\documentclass[a4paper,11pt]{article}
\pdfoutput=1
\usepackage{amssymb}
\usepackage{textcomp}
\usepackage[utf8]{inputenc}
\usepackage{latexsym}
\usepackage{epstopdf}
\usepackage{epsfig}
\usepackage{graphicx}
\usepackage{amsmath}
\usepackage{multirow}
\usepackage{subfigure}
\usepackage{a4wide}
\usepackage{cite}
\usepackage{rotating}
\usepackage{hyperref}
\newcommand{\be}{\begin{equation}}
\newcommand{\ee}{\end{equation}}
\newcommand{\ba}{\begin{eqnarray}}
\newcommand{\ea}{\end{eqnarray}}
\newcommand{\bac}{\begin{array}{c}}
\newcommand{\eaa}{\end{array}}
\newcommand{\baz}{\begin{array}{cc}}
\newcommand{\mathsym}[1]{{}}

\newcommand{\bad}{\begin{array}{ccc}}

\newcommand{\bi}{\begin{itemize}}
\newcommand{\ei}{\end{itemize}}
\newcommand{\bmt}{\begin{pmatrix}}
\newcommand{\emt}{\end{pmatrix}}
\newcommand{\bt}{\begin{tabular}}
\newcommand{\et}{\end{tabular}}

\newcommand{\benu}{\begin{enumerate}}
\newcommand{\eenu}{\end{enumerate}}
\newcommand{\bav}{\begin{array}{cccc}}

\begin{document}

\title{\Large\bf{ 
Matter Parity Violating Dark Matter Decay in Minimal SO(10), Unification,
Vacuum Stability and Verifiable Proton Decay}}

\author{\bf Biswonath Sahoo$^{*}$, M. K Parida $^{\dagger}$, Mainak Chakraborty$^{**}$\\
 Centre of Excellence in Theoretical and Mathematical Sciences,\\
Siksha 'O' Anusandhan, Deemed to be University, Khandagiri Square, Bhubaneswar 751030,\\
Odisha,  India}

%\begin{document}

\maketitle

\begin{abstract}
\noindent In direct breaking of
non-supersymmetric SO(10) to the standard model,  we investigate the possibility that dark matter (DM) decaying
 through its  mixing with right-handed neutrino (RH$\nu$) produces  
  high energy IceCube neutrinos having type-I seesaw masses. Instead
  of one universal mixing  and one common heavy RH$\nu$ mass proposed
  in a recent standard model extension, we find that underlying
  quark-lepton symmetry resulting in naturally
  hierarchical  RH$\nu$ masses  predict 
a separate mixing with each of them. We determine these mixings from the seesaw
prediction of the DM
decay rates into the light neutrino flavors. We further  show that
these mixings originate from
   Planck-scale assisted spontaneously broken matter
 parity  needed to resolve the associated cosmological
 domain wall problem. This leads to the prediction of a new LHC accessible matter-parity
 odd Higgs scalar which  also completes  vacuum
 stability in the Higgs potential for its mass $M_{\chi_S}\simeq 177$
 GeV. We have also discussed realization of relic density
 of decaying dark matter in relation to flux of IceCube neutrinos. 
Two separate
 minimal SO(10) models are further noted to predict such dark matter dynamics where a single scalar submultiplet from ${126}^{\dagger}_H$ or ${210}_H$ of intermediate mass achieves precision gauge coupling unification. Despite the presence of 
 two large Higgs representations and the fermionic dark matter host, 
 ${45}_F$, experimentally accessible proton lifetimes are also predicted with 
 reduced uncertainties.   \\
\end{abstract}
\noindent{${}^{\dagger}$email:minaparida@soa.ac.in}\\
\noindent{${}^*$email:sahoobiswonath@gmail.com}\\
\noindent{${}^{**}$email:mainak.chakraborty2@gmail.com}

\section{Introduction}\label{sec:intr}
Prominent drawbacks of the standard model (SM) are
absence of neutrino mass \cite{Salas:2017,schwetz,forero,fogli,gonzalez,nurev},   dark matter
(DM)\cite{zwicky,spergel,einasto,blumenthal,angle,strigari,Rubin:1970,Clowe:2006,XENON,LUX:2013,LUX:2016,IceCube:2013,IceCube:2014,IceCube:2016,IceCube-ANTA:2015,LHC:2017,Kadastik:2009,Hambye:2010,Frig-Ham:2009,mkp:2011,psb:DM,Mambrini:2013,Dong-Valle:2018,pnsa:2016},
 baryon asymmetry of the
universe (BAU) \cite{spergel:2003,komatsu,hindshaw,Planck15}, gauge
coupling unification \cite{Langacker:1993,Hagedorn:2016,Stella:2016,das-mkp:2001}  and vacuum
stability  of the 
Higgs potential\cite{Elias-Miro:2012,Lebedev:2013} \footnote {For  recent reviews and prospects of dark
    matter in  GUTs see
    Ref.\cite{Kadastik:2009,Hambye:2010,Frig-Ham:2009,mkp:2011,Mambrini:2013,Dong-Valle:2018,pnsa:2016}. For 
 reviews on neutrino masses see \cite{nurev}. For lack of unification in
  non-supersymmetric (non-SUSY)  
 grand desert models  see
 Ref.\cite{Langacker:1993,Hagedorn:2016,Stella:2016,das-mkp:2001}.}. 
 It is well known that supersymmetric (SUSY) grand unified theories (GUTs)
\cite{Marciano-gs:1982}
can resolve most of the issues
confronting SM but, 
in the absence of any experimental evidence of SUSY,
  non-SUSY SO(10) \cite{Georgi:1974}  without any flavor symmetry is known to fulfill most of the 
  SM limitations  except for a natural resolution of
  the gauge hierarchy problem which may be reconciled by
  resorting to fine-tuning to every loop order
  \cite{Weinberg:2007,Barr:2010}. An understanding of different
   fermion families and flavor problem is expected to emerge through a recent realization of
   comprehensive  unification in SO(18) \cite{Valle-Wilczek:2017}
   which is beyond the scope of this work.\\

  Remnants of gauged discrete symmetries \cite{Giddings:1988,Krauss:1989,Preskill:1991}
  such as R-parity in SUSY   and matter parity in non-SUSY theories
  have played crucial roles in determining stability and
  phenomenology of dark matter
  \cite{Kadastik:2009,Hambye:2010,Frig-Ham:2009,mkp:2011,pnsa:2016,Mambrini:2013,Dong-Valle:2018}.  
In general if a higher rank gauge theory containing  $U(1)_{B-L}$
as one of its subgroups leads to SM gauge theory $SU(2)_L\times
U(1)_Y\times SU(3)_C \equiv G_{213}$ , then matter parity (MP) of the SM is conserved as
a gauged discrete symmetry 
\be
Z_{MP}= {(-1)}^{3(B-L)}, \label{eq:MP}
\ee
 if the
Higgs scalar driving the symmetry breaking has $B-L={\rm even}$. Here 
$ B(L)$ stands for baryon
  (lepton) number
\cite{Kadastik:2009,Hambye:2010,Frig-Ham:2009,mkp:2011,Mambrini:2013,pnsa:2016}.
%%%%%%%%%%%%%%%%%%%%%%%%%%%%%%%%%%%%%%%%%%%%%%%%%%%%%%%%%%%%%%%%%%%555
Quite interestingly, the underlying mechanisms of type-I and type-II seesaw 
generation of  neutrino masses at tree level  in SO(10) necessarily
predict matter parity conservation of the residual gauge theory. This
is realized \cite{Gell-Mann:1978,Babu-RNM:1993} through the popular
rank-5 Higgs representation ${126}_H^{\dagger}\subset SO(10)$ which has
the following decompositions under respective sub-groups \cite{Slansky:1981}
\ba 
 {126}^{\dagger}_H&\supset& \Delta_R(1,3,{\bar 10})+\Delta_L(3,1,10)+\xi(2,2,15)...(\rm {\bf G_{224}}), \nonumber\\
&\supset&
 \Delta_R(1,3,-2,1)+\Delta_L(3,1,-2,1)+\Delta_R^{c1}(1,3,2/3,{\bar 6})+\Delta_R^{c2}(1,3,-2/3,3)+...(\rm
 {\bf G_{2213}}), \nonumber\\
&\supset&\Delta_R^0(1,0,1)+ \Delta_R^{(1)}(1,-1 ,1)+\Delta_R^{(2)}(1,-2,
 1)+\Delta_L(3,-1,1)+...(\rm {\bf G_{213}}), \label{eq:126dag}
\ea
where $G_{2213}$ denotes left-right gauge theory  $SU(2)\times
SU(2)_R\times U(1)_{B-L}\times SU(3)_C$ and the second line in
eq.(\ref{eq:126dag}) results using $Y/2=T_{3R}+ (B-L)/2$ \cite{LRS:1975}. All the
components carry $|3(B-L)|=$ even number reflecting the fact that
${126}_H$ is an even matter parity representation. Vacuum
expectation value (VEV) $V_R=\langle \Delta_R^0(1,0,1)\rangle$
carrying  $B-L=-2$ generates heavy right-handed neutrino (RH$\nu$)
Majorana mass term that
drives type-I seesaw. Similarly the left-handed (LH) triplet scalar
$\Delta_L(3,-1,1)$ that also carries  $B-L=-2$ mediates type-II
seesaw. Then direct breaking of $SO(10) \to SM$ or through
intermediate  gauge symmetries like $G_{2213}$ or Pati-Salam symmetry $SU(2)\times
SU(2)_R\times SU(4)_C$ ($\equiv G_{224}$), SM gauge theory $G_{213}$
conserves matter parity as long as ${126}_H^{\dagger}$ is
used to break $SU(2)_R\times U(1)_{(B-L)}$, or $U(1)_R\times
U(1)_{(B-L)}$, or simply $U(1)_{X}$ in $SU(5)\times U(1)_X
(X=4T_{3R}+3(B-L))$. We discuss  SO(10)
  breaking further in Sec. \ref{sec:real}.

Application of matter parity conserving non-SUSY SO(10)
 has been extensively exploited \cite{Kadastik:2009,Hambye:2010,Frig-Ham:2009,mkp:2011,pnsa:2016,Mambrini:2013} to predict fermions or scalars as
WIMP DM candidates \cite{Lee-Weinberg:1977,Griest:1989wd}. 
Gravity induced small violation of R-parity as the corresponding DM stabilising gauged discrete symmetry in
SUSY theories has been extensively investigated to predict new
interesting physical phenomena \cite{Dong-Valle:2018}.

In this work we apply
the idea of intrinsic matter
 parity and its gravity assisted spontaneous breaking to predict the
 dynamics of decaying dark matter in the minimal chain
  of non-SUSY SO(10) $\to$ SM   where 
neutrino mass is given by the popular type-I seesaw mechanism\cite{type-I}.\\
 
Recent data from IceCube 
has led to the suggestion that the observed PeV energy neutrinos
could be the decay product of
 massive dark matter \cite{IceCube-Models} which may be fermions or
 scalars  \cite{IceCube-Models,Mambrini-IceCube}.
 In one such interesting minimal
standard model extension with  Majorana fermion singlets proposed
quite recently by
Rott, Kohri, and Park (RKP) \cite{Rott:2014}, the high
scale canonical seesaw mechanism \cite{type-I}  explains the neutrino masses while
the massive dark matter $\Sigma_F$ through its extremely small mixing ( 
 $m^{(\rm mix)}\sim 10^{-5}$ eV )
with heavy degenerate right-handed neutrino (RH$\nu$) (N) decays to
produce  the  Higgs boson ($h$) and the high energy neutrino
($\nu$). The extreme smallness of $m^{(\rm mix)}$ needed to fit the
long dark matter lifetime ( $\tau_{\Sigma} > 10^{28}$ s )
 calls for a deeper theoretical explanation which is not possible
 within the simple SM extension \cite{Rott:2014}.
 As we observe, because of underlying quark-lepton symmetry \cite{JCP:1974}, the bench mark
model proposal of type-I seesaw and single universal RH$\nu$-DM mixing with one degenerate
heavy RH$\nu$ mass can not hold in the   SO(10) GUT framework without
imposing some hitherto unknown flavor symmetry. Also it is well known that one or more intermediate
gauge symmetry breakings in SO(10) gives rise to large number of paths and models leading to standard gauge theory.
As such,  without using any flavor symmetry or intermediate
gauge symmetry,  in this work we investigate
how far the underlying idea of RH$\nu$-DM mixing as origin of PeV
energy IceCube neutrinos can
be realised in the popular non-SUSY SO(10) GUT framework  including an understanding of the dynamical origin of
 mixing, DM mass, vacuum stability of the scalar potential, precision coupling
 unification and  proton lifetime prediction  which are not within the
 purview of the bench mark model \cite{Rott:2014}.

When the SM is extended with heavy right-handed neutrino(s) to
implement Type-I seesaw mechanism, the arbitrary nature of Dirac
neutrino Yukawa couplings may lead to various hierarchical or
degenerate form of RH$\nu$ mass spectra. Although no flavor symmetry has
been explicitly mentioned, one common RH$\nu$ mass $M_{N_i}=M_N
(i=1,2,3)=10^{14}$ GeV for
all three flavors and identically equal  mixing, $m^{(\rm
  mix)}=10^{-5}$ eV, has been used in the SM extension under the natural
 constraint that
the DM decays with equal branching ratios to each of three light neutrino
flavors \cite{Rott:2014}. The
equal  branching ratio hypothesis for each neutrino flavor in the
benchmark model further needs specific constraints  on the equality of
relevant Dirac neutrino Yukawa matrix elements for different flavors
$Y^{\nu}_{\alpha i}(\alpha=e,\mu, \tau),(i=1,2,3)$. Such constrained structures
of RH$\nu$ and Dirac neutrino mass matrices may be possible in the SM
extension under some imposed external flavor symmetry unspecified in
the model \cite{Rott:2014}. But in the  non-SUSY SO(10)
framework where no flavor symmetry is usually used, the type-I seesaw
based degenerate RH$\nu$ and the single universal mixing hypothesis 
with decaying dark matter are not realisable. On the other hand type-I
seesaw mechanism in SO(10) predicts the heavy RH$\nu$ masses to be
predominantly hierarchical. The basic underlying reason for
such hierarchical heavy RH$\nu$ masses, as against the
degenerate assumption of the RKP model \cite{Rott:2014}, is the quark-lepton
symmetry in SO(10) that predicts Dirac neutrino mass matrix similar to
the up-quark mass matrix \cite{JCP:1974}.
Even with such sharp contrast with the bench mark model, in this work
we show how the  complete dynamics of a massive decaying dark matter
that explains PeV energy IceCube neutrinos is predicted by intrinsic matter
parity conserving SO(10) with naturally dominant
Type-I seesaw  mechanism for neutrino masses originating from radically different 
hierarchical
ansatz for Dirac neutrino masses and RH$\nu$ mass spectra. In
particular, we show how the equality of branching ratios of DM ($=\Sigma_F$)
decay to three
different neutrino flavors determines a separate distinct mixing of
decaying dark matter (DDM) with each heavy
RH$\nu$ flavor. On the other hand utilization of one universal mixing with
naturally predicted Dirac neutrino mass and hierarchical RH$\nu$
masses  tends to drastically reduce the DM lifetime 
because of more rapid decay in the other two channels: $\Sigma_F \to
\nu_e h$, and $\Sigma_F \to \nu_{\mu} h$.

 We further show that the derived values of these three
mixings can be predicted for different light neutrino mass patterns such as
normally hierarchical (NH), invertedly hierarchical (IH), and
quasi-degenerate (QD), consistent with oscillation data such that the
$N_i-\Sigma_F$ mixings 
can be uniquely fixed once the light neutrino mass hierarchy is known.\\

Over the years implementation of gravitational effect has been found to be a very efficient mechanism in 
breaking all kinds of global symmetries, continuous or discrete
\cite{Giddings:1988,Krauss:1989,Preskill:1991,Berezinsky:1998,Boucenna:2014,Mishra-Yajnik:2009,Akhmedov-RNM:1993,Hamagushi:1998},
without causing cosmological domain wall problem
\cite{Krauss:1989,Preskill:1991,Zeldovich:1974} while at the same
time  predicting new interesting physical phenomena such as Majoron
as scalar dark matter \cite {Berezinsky:1998,Boucenna:2014,Akhmedov-RNM:1993}. 
Whereas no dynamical
explanation has been provided for the origin of  N-$\Sigma_F$  mixing
\cite{Rott:2014}, in this work  we further show how the  three small  mixings in our case
originate from  gravity assisted matter parity discrete symmetry breaking 
\cite{Berezinsky:1998,Boucenna:2014}
 without causing cosmological domain wall problem
 \cite{Zeldovich:1974,Krauss:1989,Preskill:1991}.

 We exploit Planck
 scale or gravity assisted spontaneous breaking of intrinsic matter parity
in a manner analogous to R-Parity (RP) breaking  \cite{Berezinsky:1998} in MSSM.
%%%%%%%%%%%%%%%%%%%%%%%%%%%%%%%%%%%%%%%%%%%%%%%%%%%%%%%%%%%%%%%%%%%%%%
This would require a matter parity odd singlet scalar   which indeed originates from the smallest spinorial scalar representation ${16}_H\subset$ SO(10). Analogous to eq.(\ref{eq:126dag}), the contents of ${16}_H$ under respective gauge groups are \cite{Slansky:1981}

\ba 
 {16}_H&\supset& \chi_R(1,2,{\bar 4})+\chi_L(2,1,4)\,\,\,\,\,\,\,\,\,\,\,\,\,\,
\,\,{\hfill {(\rm {\bf G_{224}})}}, \nonumber\\
&\supset& \chi_R(1,2,-1,1)+\chi^{c}_R(1,2,-1/3,{\bar 3})+\chi_L(2,1,-1,1)+\chi^{c}_L(2,1,-1/3, 3) \,\,{\hfill{(\rm {\bf G_{2213}})}}, \nonumber\\
&\supset&\chi_R^0(1,0,1)(\equiv \chi_S)+ \chi_R^{(1)}(1,1,1)+\chi_R^{(c1)}(1,1/3,{\bar 3})
+\chi_R^{(c2)}(1,-2/3,{\bar 3}) \nonumber\\
&+&\chi_L(2,-1/2,1)+\chi_L(2,1/6,3)\,\,\,\,\,\,\,
{\hfill{(\rm {\bf G_{213}})}}. \label{eq:16H}
\ea
Thus all the components of ${16}_H\,{\rm or}\,{16}_H^{\dagger}$ carry $|3(B-L)|=$ odd signifying  odd matter parity of these representations. Further, each of the two  has only one SM singlet: $\chi_S(1,0,1)\subset {16}_H$ and $\Delta_R^0\subset {126}_H$.
 But unlike the even matter parity of the singlet $\Delta_R(1,0,1)$, the singlet  $\chi_S(1,0,1)$ has odd matter parity \footnote{The analogous fermionic singlet is the RH$\nu$ $N(1,0,1)\subset {16}_F$ which has odd matter parity like all standard fermions.} that can acquire a  VEV
 ($V_{\chi}=\langle \chi_S(1,0,1)\rangle$ without breaking SM gauge symmetry but breaking matter parity spontaneously.
%%%%%%%%%%%%%%%%%%%%%%%%%%%%%%%%%%%%%%%%%%%%%%%%%%%%%%%%%%%%%%%%%%%% 

As noted above matter parity of the SM gauge theory originating as
intrinsic gauged discrete symmetry of non-SUSY SO(10) has played a crucial role in safeguarding stability of
WIMP dark matter without requiring any adhoc discrete symmetry to be imposed by hand. In this work while explaining the dynamical origin of DDM-RH$\nu$ mixings through the matter-parity violating VEV of $\chi_S$, we preserve this ability
of the SM gauge theory keeping its  possibility open for WIMP DM
embedding in the models (Model-I and Model-II) proposed in this work.
This quality of the SM gauge theory as remnant of non-SUSY SO(10) is protected by taking care to see that matter parity as gauged discrete symmetry is left unbroken as long as the SM gauge symmetry survives till the electroweak scale. This imposes the natural upper bound on the matter parity VEV $V_{\chi}=\langle \chi_S\rangle \le v_{\rm ew}= 246$ GeV.
As the  VEV of this  singlet Higgs is noted to be
 naturally bounded from above
 by the electroweak VEV,  $v_{\rm ew}= 246$ GeV, due to the
 survival of matter parity in the SM down to the electroweak scale, this SO(10) theory
 leads to the prediction of a new light Higgs scalar singlet with 
 perturbative upper bound on its mass
 $M_{\chi_S} \le 860$ GeV.  
In the process, this SO(10) theory of DDM provides its experimental testability or most desirable quality of falsifiability at LHC.   
 
Having only the standard Higgs doublet that defines the
Higgs potential, the RKP model \cite{Rott:2014} is also affected by the vacuum instability problem that
needs new  physics below $\sim 2\times 10^{9}$ GeV
\cite{Elias-Miro:2012,Lebedev:2013}. 
Interestingly, this new Higgs scalar singlet $\chi_S(1,0,1)$ that drives the Planck
scale assisted matter parity breaking completes the vacuum
stability  in the present SO(10) models (Model-I and Model-II). The
resolution of vacuum stability then predicts the $\chi_S(1,0,1)$ mass
to be $M_{\chi_S}=177$ GeV which is clearly detectable by accelerator
searches as an evidence of this model building for decaying dark matter.

After the standard model Higgs discovery at LHC, spontaneous breaking origin of
all masses in the Universe has turned out to  be a very attractive universal
hypothesis. While the standard fermions and gauge bosons get their masses
via  respective gauge and Yukawa interactions with the standard Higgs
doublet, the Higgs origins of nonstandard heavy fermions including RH$\nu$, N, and dark
matter, $\Sigma_F$, are not
possible within the simple standard model extension alone
\cite{Rott:2014} unless a number of scalar singlets are added. Also
like the SM, the  RKP model fails to unify gauge couplings.
These shortcomings are fulfilled through the present SO(10) model by
minimal modification of the grand desert.
%--------------------------------------------\\
%--------------------------------------------\\

On the realization of precision gauge coupling unification, 
 we note interesting roles of the
Higgs representations  ${210}_H$ and ${126}_H$, which have been
used to break the GUT symmetry and generate neutrino
masses. We point out that each of these two are capable of
completing the desired gauge coupling unification of the SM separately by
minimally populating the grand desert through only one of their
respective
scalar submultiplets leading to Model-I and Model-II.  
A specifically new finding of this work is the possibility of new
minimally modified grand desert in Model-II where the popular Higgs
representation ${126}_H\supset  \eta (3,-1/3,6)$   completes
precision coupling unification with its mass $M_{\eta}=10^{10.7}$ GeV. The other minimally modified grand
desert model that accommodates this decaying DM phenomenology is
Model-I where the intermediate mass of $\kappa (3,0,8)\subset {210}_H$
achieves precision coupling unification for $M_{\kappa}=10^{9.2}$
GeV. Although similar coupling unification in Model-I was noted
earlier, its connection with decaying DM and generation of $\Sigma_F$
mass and $N_i-\Sigma_F$ mixings are new. Furthermore, the vacuum stability of the SM Higgs
potential which was absent in the original suggestion
 is now complete in the presence of new scalar
singlet both in Model-I and Model-II. 

%%%%%%%%%%%%%%%%%%%%%%%%%%%%%%%%%%%%%%%%%%%%%%%%%%%%%%%%%%%%%%%%%%%

When GUT threshold effects are ignored, all single step breaking grand desert
models $[E_6, SO(10), SU(5)] \to {\rm SM}$ are affected by the
non-unification of gauge couplings in the same fashion. 
This problem has been resolved by the introduction of a number of non-standard Higgs or
fermion submultiplets at different mass scales in 
non-SUSY SU(5)
\cite{Bajc-gs,Bajc-nem-gs:DM,Perez-Imi-Rodrigo:DM,Ma-Suematsu:2007,Aizawa:2014}
with low-scale seesaw mechanisms in some cases. In
these models \cite{Bajc-nem-gs:DM,Perez-Imi-Rodrigo:DM,Ma-Suematsu:2007,Aizawa:2014} no stable dark
matter candidates can exist unless an external symmetry from outside
the GUT framework is added by hand. Similarly the phenomenological suggestion of scalar
or fermionic DM candidates as SM extension \cite{Strumia:2006}
requires the imposition of an adhoc stabilising  discrete
symmetry externally. A very profound and attractive  implementation in
the popular non-SUSY SO(10)\cite{Frig-Ham:2009} with minimal
modification of the grand desert by a TeV scale fermionic triplet dark matter
$F_T(3,0,1)$ and  an octet fermion $O_F(1,0,8)$ at higher mass
has led to a self sufficient SO(10)  model of precision coupling
unification where the traditionally used external discrete
symmetries has been replaced by intrinsic matter parity
underlying  the GUT symmetry breaking mechanism through ${126}_H$ that naturally
generates heavy Majorana neutrinos driving canonical seesaw. The model
not only predicts verifiable proton lifetime by ongoing experiments
but also it predicts the WIMP fermionic triplet DM mass $M_T\simeq 2.75\pm 0.15$ TeV which could be
detected at LHC with upgraded luminosity and at
future $e^+e^-$ colliders with energy $ > 2M_T$. Recently, out of many WIMP DM 
candidates suggested, the fermion triplet $F_T(3,0,1)$ has attracted much attention with
well documented phenomenology for direct and indirect detection
prospects \cite{Hryzuk:2014}. In the context of matter parity
conserving $SO(10)$ and dominant radiative seesaw \cite{Ma:2006} at
TeV scale, precision unification with verifiable proton lifetime has
been realized in the presence of wino like and Higgsino like DM and
gluino like fermion near the TeV scale \cite{mkp:2011}.
%%%%%%%%%%%%%%%%%%%%%%%%%%%%%%%%%%%%%%%%%%%%%%%%%%%%%%%%%%%%%%%%%%%%%%%%%%%%%

On the question of economic choice of particle degrees of freedom in
populating the grand desert for achieving precision coupling
unification in matter parity conserving SO(10) with canonical seesaw, 
each  of the two decaying dark matter (DDM) 
models  discussed here
has only one Higgs scalar submultiplet $\eta (3,-1/3, 6) \subset
{126}_H$ (Model-II) or
$\kappa (3,0,8) \subset {210}_H$ (Model-I).
 Specific roles of these
two representations ${210}_H, {126}_H$ in determining minimal SUSY SO(10) model with $26$ parameters in the Lagrangian was
 pointed out earlier \cite{aulakh-gs-vissani}. Although we are dealing with non-SUSY SO(10), the pivotal roles played by  each member of the pair is a new property of these two as noted here.
%%%%%%%%%%%%%%%%%%%%%%%%%%%%%%%%%%%%%%%%%%%%%%%%%%%%%%%%%%%%%%%%%%%%%%%%%%%%
 Despite the presence of these two large Higgs representations
  and the nonstandard fermionic representation
 ${45}_F$ hosting the dark matter, we further predict experimentally accessible proton
 lifetime predictions in both models  with substantially reduced
 threshold uncertainties compared to many earlier investigations.

In one out of several mixing solutions discussed here, we have
emphasized that the corresponding SO(10) model
is a self sufficient dynamical theory of decaying dark matter  and precision gauge
coupling unification as it does not need any externally imposed
discrete symmetry for DM stability, or additional Planck-mass fermion
singlets for the dynamical explanation of mixings, or additional
representation beyond those used for GUT symmetry breaking to complete
coupling unification and generate heavy DM mass. We have thus
concluded that if DM decays only through its mixings with heavy
RH$\nu$s mediating type-I seesaw, then matter parity violation has been observed at IceCube.

Highlights of new contributions of this work are
\begin{itemize}
\item{First realization of
decaying fermionic dark matter dynamics in non-SUSY SO(10) with type-I seesaw
and naturally hierarchical right-handed neutrinos for all types of
light neutrino mass hierarchies.}
\item{First determination of  RH$\nu$-DM mixings from Type-I seesaw,
  neutrino oscillation data  and IceCube neutrino data.}
\item{Explanation of dynamical origin of mixings through Planck-scale assisted spontaneous
breaking of intrinsic matter parity and derivation of
couplings in the associated renormalizable and non-renormalizable Lagrangians.}
\item{ Prediction of experimentally verifiable new Higgs 
scalar origin of  dark matter decay.}
\item{Resolution of the  vacuum
instability problem  of the SM scalar potential through the new light
Higgs scalar underlying dark matter dynamics.}
\item{Identification of a completely new minimal
grand desert modification model
(Model-II) for precision gauge coupling unification through
the lone intermediate mass scalar submultiplet $\eta (3,
-1/3,6)$ rooted in  
the popular SO(10) representation ${126}_H^{\dagger}$}
\item{ Implementation of Dark matter dynamics and resolution of vacuum
  stability problem in the minimal model (Model-I) achieving precision
  coupling unification through the intermediate mass scalar
  submultiplet $\kappa (3,0,8) \subset {210}_H$.}
\item{Realization of desired relic density of decaying dark matter
  through a heavy scalar exchange between the DM and the SM Higgs scalar
  suitable for generating the expected flux of PeV energy IceCube neutrinos.} 
\item{Experimentally verifiable precision proton lifetime prediction 
despite the larger Higgs representations ${126}_H,{210}_H$ and the
nonstandard fermions in ${45}_F$.}
\end{itemize}
This paper is organised in the following manner.
In Sec.\ref{sec:bench} we discuss benchmark model briefly. In Sec.\ref{sec:embed} we embed
decaying dark matter (DDM)  in SO(10) with specific discussions on
matter parity as intrinsic gauged discrete symmetry and derivation of
DM mass. We derive RH$\nu$ masses 
and the RH$\nu$-$\Sigma_F$ mixings numerically in Sec.\ref{sec:real} using type-I seesaw contribution
to  dark matter decay rate.  Dynamical generation of $N_i-\Sigma_F$ mixing
is discussed in Sec.\ref{sec:mix} using Planck-scale assisted spontaneous breaking
of matter parity. Derivation of renormalizable and
non-renormalizable Yukawa couplings of the new matter parity odd light Higgs
scalar $\chi_S$ is derived in Sec.\ref{sec:couplings}.
 Gauge coupling unification in Model-I and Model-II is
discussed in Sec.\ref{sec:gcu} while proton lifetime prediction is presented in
Sec.\ref{sec:taup}. Prediction of new light Higgs scalar and its impact on the
resolution of the vacuum
stability problem is discussed in Sec.\ref{sec:stab}. In this section
we also discuss advantages of low VEV of the matter parity violating
Higgs scalar singlet along  with a derivation from potential
minimisation.
 We discuss how 
proper relic density of the DDM is realized to generate the expected
flux of PeV energy IceCube neutrinos in Sec.\ref{sec:relic}. We summarize and conclude in
Sec.\ref{sec:sum}. In the Appendix A, Sec. \ref{sec:VP} we present diagonalisation of RH$\nu$ mass
matrices in the presence of mixings with decaying dark matter motivated models. In Appendix
B, Sec.\ref{sec:RGE} we present evolution of various gauge couplings,
top-quark Yukawa coupling 
and GUT threshold effects on unification mass in Model-I and
Model-II.

\section{The Bench Mark Model}\label{sec:bench}
\begin{figure}[h!]
\begin{center}
\includegraphics[scale=0.6]{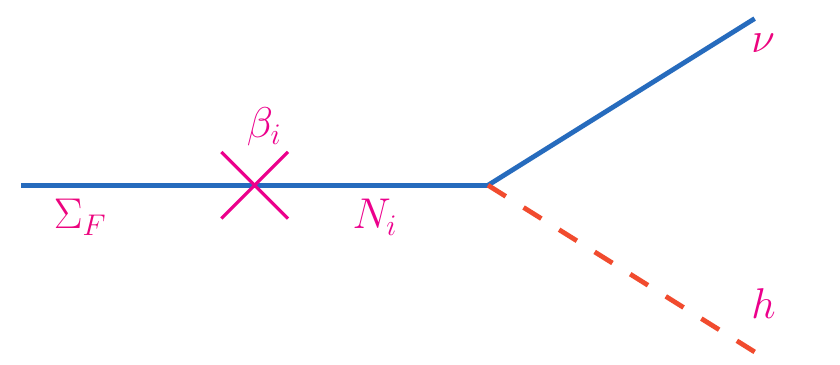}  
\caption{Feynman diagram for dark matter $\Sigma_F$ decaying to
  $\nu_L$ and standard Higgs $h$ through the DM mixing with heavy RH neutrinos $N_i$ of three generations. In the bench mark model $N_i\equiv N$ under quasidegeneracy assumption and $\beta_i=\beta$. }
\label{fig:bench}
\end{center}
\end{figure}

 For simpler
visualisation of the decay process, we present the Feynman diagram of
this model in Fig. \ref{fig:bench} where the DM $\Sigma_F$ is shown to
decay to $\nu_{L} h$. In ref.\cite{Rott:2014} all the three RH neutrino
masses have been assumed to be identical $M_{N_i}=M_N (i=1,2,3)$. 
%--------------------------
%--------------------------
%----------------Bajc-nem-gs
%---------------
Likewise, for the demonstration of the RKP model only, the mixings of all three RH neutrinos with DM $\chi $
($\equiv \Sigma_F$ in our notation) have
been assumed to be identical $\sin\beta_i=\sin\beta =\sigma/(M_N-M_{\chi})(i=1,2,3)$.
 
 The DM  decay predicts
 neutrino energy nearly equal to half the DM mass $E_{\nu}\sim M_{DM}/2$. 
Noting that the high energy neutrino flux is proportional to density
$\rho$ (square of density $\rho^2$) for decaying (self annihilating)
DM,  RKP estimate more than $50\%$ of events to be within $65^o$
($25^o$) from the galactic centre for decaying (self annihilating)
dark matter \cite{Rott:2014} . The existing
perturbative unitarity bound \cite{Griest:1989wd} already excludes annihilating WIMP DM masses larger than $100$ TeV. 
Then the observed isotropy
of IceCube neutrinos and their large energy lead to the suggestion that they
originate from decaying DM of mass $> 100$ TeV. The other attractive part
of the benchmark model
is based upon right-handed neutrino (N) extended SM for canonical seesaw which is further
extended by the addition of heavy Majorana singlet DM $\psi$ of assumed mass
$M_{\psi}$. In addition a small mixing mass term $\sigma$ between $N$ and
$\psi$ has been assumed. Also all Yukawa interactions of the DM has
been neglected.

Suppressing flavor indices of $N$, the suggested model Lagrangian is 
\be
-{\cal L}^{RKP}_{Yuk}=\lambda {{\overline \nu}}_L(h+v_{\rm ew})N +({\bar
    N}^c {\bar \psi}^c)\begin{pmatrix} M_N  & \sigma \\
 \sigma & M_{\psi}   
\end{pmatrix}
\begin{pmatrix}N \\
\psi\end{pmatrix}.  \label{RKPell}
\ee
With the assumed
 constraint on the masses
\be
M_N>M_N-M_{\psi}> M_{\psi} \gg \sigma, \label{RKPconst} 
\ee
 diagonalization
of the $4\times 4$ mass matrix in the second term in the RHS of
eq.(\ref{RKPell}) results in the heavy mass eigenvalues,
\be
M^{\pm}=\frac{1}{2}\left [M_N+M_{\psi} \pm \sqrt{{(M_N-M_{\psi})}^2+4 \sigma^2}\right]\label{eq:evndm}
\ee

The emerging canonical seesaw formula from SM extension has been assumed in its simple form
\be
m_{\nu}=-\lambda^2\frac{v_{\rm ew}^2}{M^{+}},\label{eq:RKPseesaw}
\ee
where $M_N$ represents  bare RH neutrino mass matrix and $\lambda
v_{\rm ew}$ is the assumed Dirac neutrino mass  matrix $M_D$. From
eq.(\ref{eq:evndm}) it is clear that the largest mass eigen value
$M^+\simeq M_N$ since $\sigma$ is extremely small. For the same reason
$M^- \simeq M_{\psi}$ which is the decaying DM mass observed at IceCube. 

The mass eigen states are 
\ba       
{\zeta}_{+}&=& \cos\beta\, N+\sin\beta\, \psi, \nonumber\\
{\zeta}_{-}
&=&-\sin\beta\, N + \cos \beta\, \psi  ,\nonumber\\ 
 \label{eq:esnd}
\ea
where the mixing angle is expressed as
\be
\tan 2\beta \sim 2\sigma /M_N \ll 1.\nonumber\\ 
\ee
Now rewriting the interaction Lagrangian
\begin{eqnarray}
&&-{\cal L}_{int}= \lambda h {\overline \nu}_L N \nonumber\\
&&= \lambda h {\overline \nu}_L\left[\zeta_{+}+\frac{\sigma}{(M_N-M_{\psi})}\zeta_{-}\right],\label{RKPint}
\end{eqnarray}
leads to the suppressed effective coupling constant and the suppressed
decay width
\begin{eqnarray}
&&\lambda^{eff}\equiv
\lambda^{eff}_{\alpha}(\alpha=e,\mu, \tau), \nonumber\\
&&\lambda^{eff}_{\alpha}=\sum_i \lambda_{\alpha i} \frac{{\sigma}}{(M_N-M_{\psi})},\nonumber\\
&&\Gamma_{\chi}=\Gamma^{\alpha}\simeq \frac{\lambda^2_{\rm eff}}{32\pi}M^{-} .\label{eq:Gamma}
\end{eqnarray}
Accounting for the observed PeV energy neutrino excess at IceCube requires the DM lifetime \cite{IceCube-Models}
\be
\tau_{\rm DM}=1.9N_{\nu}\times 10^{28}\,s,\label{eq:lifeDM}
\ee 
where $N_{\nu}=$ number of produced neutrinos of the same type.
Then using  $M^{-}=10^6$ GeV, eq.(\ref{eq:Gamma}), and eq.(\ref{eq:lifeDM}) gives a very small value of the mixing parameter
\ba
\lambda_{\rm eff}&\simeq & 5.3\times 10^{-29},\,\,\nonumber\\
\sigma &=& 10^{-5} eV.\,\,\label{eq:mixvalue}
\ea
In the benchmark model quasidegenerate light neutrino mass
$m_{\nu}=0.1$ eV is treated to be the result of type-I seesaw mediated
by heavy RH neutrino mass $M_N=10^{14}$ GeV which are also quasidegenerate. For the
DM decay with assumed Dirac neutrino Yukawa coupling  $\lambda \sim
1$ for all generations, the three different branching ratios have been
also assumed to be equal,
\begin{eqnarray}
 &&Br.(\Sigma_F\to \nu_e h):Br.(\Sigma_F\to \nu_{\mu} h):\nonumber\\
 &&Br.(\Sigma_F\to \nu_{\nu_{\tau}} h)::1:1:1.\label{eq:Br}
\end{eqnarray}
 
\section{Symmetry Breaking, Matter Parity and Dark Matter Mass in SO(10)}\label{sec:embed}

In the absence of any experimental evidence of non-standard gauge
bosons below the GUT scale, here we confine to the direct minimal symmetry breaking of SO(10)
to SM gauge symmetry ($G_{213}$) which is assumed to operate over the
entire range of mass scale $\mu \simeq M_W - M_{GUT}$
\be
SO(10) \to  G_{213} \to U(1)_{em}\times SU(3)_C .\label{eq:minchain}
\ee
Such a direct SO(10) breaking model is prominently well established in
SUSY GUTs \cite{aulakh-gs-vissani,Fukuyama:2005} and also has been applied in non-SUSY
SO(10)\cite{mkp:2011,pnsa:2016,Babu-RNM:1993,kynshi-mkp:1993}. In the
context of protecting non-SUSY grand desert models by GUT threshold effects
, particularly those due to superheavy gauge bosons, such a direct
breaking model of SO(10) has been also used earlier
\cite{mkp:1987,Hall:1981}. Decompositions of SO(10) representations
$10,\,16,\,{45},\, {126},\,$ and $210$ under different
subgroups are presented in eq.(\ref{eq:126dag}), eq.(\ref{eq:16H}),
eq.(\ref{eq:decall}) and also in Appendix B.   
For the  first step of symmetry breaking in eq.(\ref{eq:minchain}) we use the
SO(10) Higgs
representations ${210}_H$ and ${126}_H$ with their respective SM
singlet VEVs around the same high scale. This symmetry breaking may be
visualised in the following manner. The VEV of the $G_{2213}$ singlet
  component contained in the $G_{224}$ submultiplet ${(1,1,15)}\subset {210}_H$
breaks $SO(10)\to G_{2213}$ at $\mu \sim M_{GUT}$. At the same scale 
the SM singlet component contained in the $G_{2213}$ submultiplet $(1,3,-2,1)$
belonging to the $G_{224}$ submultiplet $\Delta_R(1,3, \overline{ 10})\subset {126}_H^{\dagger}$ acquires VEV $\sim M_R\sim M_{GUT}$ to break $SU(2)_R\times
U(1)_{(B-L)} \to U(1)_Y$. The direct breaking scenario of eq.(\ref{eq:minchain}) is thus realised when both these breaking processes,
  $SO(10) \to G_{2213}$ and $G_{2213}\to $ SM, occur at the same scale $\sim M_{GUT}$ resulting in  only twelve light gauge bosons of SM  while making all other thirty three gauge bosons of SO(10) superheavy with masses $\sim M_{GUT}$. 
Even if ${210}_H$ is replaced by ${45}_H$ and ${126}_H$ is replaced by
${16}$, the two step-breaking scenario $SO(10) \to G_{2213} \to $ SM
has been shown to be realizable. Although this would be inconsistent with matter
parity conservation because of $\langle {16}_H\rangle$ to break  $SU(2)_R\times U(1)_{B-L}$ in stead of $\langle {126}_H\rangle$, it also leads to direct breaking model as before when
the two symmetry breaking scales are identical to $\sim M_{GUT}$. The direct breaking super-grand desert scenario of all SUSY GUTs including SU(5), SO(10) and $E_6$ including MSSM (minimal supersymmetric standard model) has been noted
\cite{Marciano-gs:1982} to exhibit precision gauge coupling unification  when extrapolated to higher scales using the CERN-LEP data \cite{PDG:2012,PDG:2014,PDG:2016}. On the other hand the higher rank non-SUSY GUTs like SO(10), and $E_6$ have been shown to require at least one intermediate gauge symmetry, such as $G_{224}$ or $G_{2213}$ \cite{Babu-RNM:1993,cmgmp:1985,Kibble:1982}. But in view of a number of interesting minimal
modifications of the non-SUSY grand desert scenario
\cite{Frig-Ham:2009,mkp:2011,pnsa:2016,Hagedorn:2016,kynshi-mkp:1993}, models
without any non-standard gauge bosons below the GUT scale also appear to
be simpler choices for SO(10) as discussed in this work for predictions especially on decaying dark matter (DDM).

Alternatively, this direct symmetry breaking non-SUSY SO(10) can also be
visualised as a result of two different symmetry breakings occurring at the same
scale $\mu=M_{GUT}$. At first SO(10) breaking to $G_{224}$ can occur by the GUT scale VEV of 
Pati-Salam singlet $\xi(1,1,1)$ in ${210}_H$ or ${54}_H$ 
accompanied by the symmetry breaking $G_{224}\to $ SM driven by the VEV of $G_{224}$
sub-multiplet $\Delta_R(1,3,  \overline{10}) \subset {126}_H^{\dagger}$. As already noted, in
all the direct breaking models a grand desert is produced without
coupling unification and this problem is successfully confronted by two
different ways in this work by the introduction of just one SM scalar submultiplet $\eta (3,0,8) \subset {210}_H$ in Model-I and $\kappa (3,-1/3, 6)\subset {126}_H$ in Model-II component as discussed below in Sec.\ref{sec:gcu}.    

Defining the corresponding fine-structure constants as
$\alpha_i=\frac{g_i^2}{4\pi}, i= Y, 2L, 2R, B-L, 3C, 4C$, at
the GUT scale $\mu=M_{GUT}$ matching conditions are  $\alpha_{B-L}=\alpha_{4C}=\alpha_{3C}$
and $\alpha_{2R}=\alpha_{2L}$ leading to
\be
\frac{1}{\alpha_Y(\mu)} = \frac{3}{5}\frac{1}{\alpha_{2L}(\mu)}+\frac{2}{5}\frac{1}{\alpha_{3C}(\mu)}~.\label{eq:matching}
\ee 
Renormalisation group evolution of gauge couplings and coupling
unification have been discussed below in Sec.\ref{sec:gcu} in these two
different cases (Model-I and Model-II).

%%%%%%%%%%%%%%%%%%%%%%%%%%%%%%%%%%%%%%%%%%%%%%%%%%%%%%%%%%%%%%%%%%%%%%%%%%%

 In addition different SO(10) representations are
identified with definite values of matter parity $Z_{MP}=$ even, or
  odd
\ba
Z_{MP}&=&{\rm Even:}\,\, 10,\,45, \, 54,\, 120,\,126,\, 210,.... \nonumber\\
Z_{MP}&=&{\rm Odd:}\,\, 16,\,144,\,...... \label{mpreps}
\ea
As such all SM fermions and RH neutrinos have odd $Z_{MP}$ but the SM
Higgs scalar $h \subset {10}_H$ has even  $Z_{MP}$. Then stable DM candidates can be
assigned to be in one or more of the
 non-standard fermion representations
 ${10}_F,{45}_F,{54}_F,{120}_F,{126}_F,{210}_F,...$. Similarly scalar
 DM candidates carrying odd matter parity can be assigned spinorial representations
 ${16}_H,{144}_H, ...$.
In the following sections we will need the SM components of ${16}_H^{\dagger}$ the conjugate of the
scalar representation ${16}_H$, ${45}_H$, ${126}_H$,  ${210}_H$ and
${45}_F$ for our model descriptions out of which the components of
${126}_H$ and ${16}_H$ have been already specified in
eq.(\ref{eq:126dag}) and eq.(\ref{eq:16H}) and the remaining components are
\ba
&&{10}_H=(2,2,1)+(1,1,6).............(\bf G_{224}),\nonumber\\
&&\supset (2,2,0,1)+(1,1, 2/3,3)+(1,1,-2/3,{\bar 3})....(\bf G_{2213}),\nonumber\\
&&\supset \phi(1,1/2,1)+(2, -1/2,1)+(1, 2/3,3)+(1,-2/3,{\bar 3}),.(\bf
G_{213}),\nonumber\\
&&{210}_H=(1,1,1)+(1,1,15)+(1,3,15)+(3,1,15)+.....(\bf G_{224}),\nonumber\\
&&\supset \Phi_1 (1,1,0,1)+ \Phi_2(1,1,0,1)+ \Phi_3(1,1,0,1)+\kappa(3,1,0,8)+....(\bf G_{2213}),\nonumber\\
&& \supset
\Phi_1(1,0,1)+\Phi_2(1,0,1)+\Phi_3(1,0,1)+\kappa(3,0,8)+.....
(\bf G_{213}), \nonumber\\
&&{45}_H =(1,1,15)+(3,1,1)+(1,3,1)+(2,2,6).....(\bf G_{224}), \nonumber\\
&&\supset S_H(1,1,0,1)+(3,1,0,1)+(1,3.0.1)+ .........(\bf G_{2213}),
\nonumber\\
&&\supset S_H(1,0,1)+(3,0,1)+A_3(1,0,1)+..............(\bf G_{213})~.\nonumber\\
\label{eq:decall}
\ea
As all the components of every representation in eq.(\ref{eq:decall})
and  eq.(\ref{eq:126dag}) have been explicitly shown to possess $3(B-L)=
+2 \, {\rm or}\, -2$, they have even matter parity satisfying
$Z_{MP}=+1$. Thus VEVs in the respective SM singlet direction to be
utilsed for gauge symmetry breaking or non-standard fermion mass
generation ensures conservation of matter parity in the presence of
$G_{213}$ gauge symmetry. 
  Another representation of lower dimension that carries odd
matter parity is ${144}_H\subset$ SO(10). 

Being a SM singlet which is also a SU(5) singlet  
 $\chi_S(1,0,1)$ , in principle, can
acquire any VEV in the range ${\cal O}(M_W)-{\cal O}(M_U)$ without affecting gauge coupling evolution. But being odd under $Z_{MP}$, conservation of matter
parity in the presence of SM constrains this VEV to be at most of
order $v_{\rm ew}$, $V_{\chi} \le v_{\rm ew}=246$ GeV.

We  treat ${45}_F$ as Majorana fermionic DM
representation which contains the DDM as a Majorana fermion
singlet. This can be decomposed under $G_{224}$ and $G_{213}$ in the following
manner
\ba
{45}_F&=& {(1,1,15)}_F+{(3,1,1)}_F+{(1,3,1)}_F+{(2,2,6)}_F:\,\, {\bf G_{224}}, \nonumber\\
 &=&\Sigma_F(1,0,1)+(1,2/3,3)_F+(1,-2/3,{\bar
  3})_F+{(1,0,8)}_F+{(3,0,1)}_F+{(1,0,1)}_F+{(1,\pm 1,1)}_F \nonumber\\
&+&{(2,1/6,3)}_F+{(2,-1/6,{\bar 3})}_F+{(2,-5/6,3)}_F+{(2,5/6,{\bar
    3})}_F:\,{\bf G_{213}}.
 \label{eq:dec45}
\ea

Thus, the SM singlet fermion $\Sigma_F(1,0,1)$ as DDM is assumed to be in the $G_{224}$
submultiplet $(1,1,15)_F\subset {45}_F$. It is necessary to show how
only the DDM candidate $\Sigma_F$ can acquire PeV scale mass by Higgs mechanism while the rest $44$ components are allowed to have GUT scale masses. 

 In order to
implement such mass splitting we use the GUT scale Lagrangian
  where Yukawa interactions of the fermionic adjoint
 representation ${45}_F$ with two scalar representations ${54}_H \equiv
 E$  and ${210}_H\equiv {\bf \Phi}$
 have been included, although the presence of the former can be dispensed with
in the minimal case.
%%%%%%%%%%%%%%%%%%%%%%%%%%%%%%%%%%%%%%%%%%%%%%%%%%%%%%%%%%%%%%%%%%%%%%

\begin{eqnarray}
&&-{\cal L}^{(\rm MPC)}_{Yuk}= A_F\left(\frac{m_A}{2}+\frac {h_p}{2}{\bf \Phi}+\frac{h_e}{2}E
  \right)A_F+h.c.\,\label{Yukgut}
\end{eqnarray}
Here $m_A\sim M_{GUT}$ is the common bare mass of all the components
of ${45}_F$.
The SO(10) scalar representation  ${210}_H$  has
three SM singlets in Pati-Salam submultiplets
$\Phi_1(1,1,1), \Phi_2(1,1,15)$, and   $\Phi_3(1,3,15)$. We denote these there VEVs
by  by  $V^{(i)}_{210} =\langle \Phi^{(i)}\rangle (i=1,2,3)$. We also denote the singlet VEV
$\langle E \rangle \subset {54}_H$.
Then all the component masses of ${45}_F$ are \cite{mkp:2011,psb:DM,pnsa:2016}
\ba
m(1,-2/3,3)&=&m(1, 2/3, {\bar 3})
=m_A+\frac{1}{3 \sqrt 2}h_p{\bf V^{(2)}_{210}}-\frac{1}{\sqrt
  15}h_e \langle E \rangle ,\nonumber\\
m(1,0,8)&=&m_A-\frac{1}{3\sqrt{2}}h_p{\bf
  V^{(2)}_{210}}-\frac{1}{\sqrt {15}}h_e\langle E \rangle,\nonumber\\
m(1,0,1)&=&M_{\Sigma}=m_A+\frac{\sqrt{2}}{3}h_p{\bf V^{(2)}_{210}}-\frac{1}{\sqrt 15}h_e\langle E \rangle,\nonumber\\
. \label{eq:mass15}
\ea
These are the masses of all the components of $G_{224}$ submultiplet
$(1,1,15)_F\subset {45}_F$. The masses of the rest $30$ components are
\ba
m(2,1/6,3)&=&m(2,-1/6,{\bar 3})=m_A-h_p\frac{\bf V^{(3)}_{210}}{6}+h_e\frac{\langle E \rangle}{4\sqrt
    {15}},\nonumber\\
m(2, -5/6,3)&=&m(2, 5/6,{\bar 3})=m_A -h_p\frac{\bf  V^{(3)}_{210} }{6}+h_e\frac{\langle E \rangle}{4\sqrt
    {15}},\nonumber\\
m(1,1,1)&=&m(1,-1,1)=m_A+h_p\frac{\bf V^{(1)}_{210} }{\sqrt 6}+
\frac{\sqrt {3}}{2\sqrt {5}}h_e \langle E \rangle,\nonumber\\
m(3,0,1)&=&m_A-h_p\frac{\bf
  V^{(1)}_{210}}{\sqrt {6}}+\frac{\sqrt {3}}{2\sqrt {5}}h_e \langle E \rangle,\nonumber\\
m^{\prime}(1,0,1)&=&m_A+h_p\frac{\bf V^{(1)}_{210}}{\sqrt {6}}+\frac{\sqrt {3}}{2\sqrt{5}}h_e\langle E \rangle.\label{eq:compmass} 
\ea
    
These  formulas have the options of finetuning
two Yukawa couplings, four VEVs and $m_A$. In case we assume all other
VEVs except ${\bf V^{(2)}_{210}}$ to vanish, only one fine tuning among this VEV, $h_p$ and
$m_A$ is sufficient to make $M_{\Sigma}\simeq 1$ PeV while keeping
all other $44$ component masses near the GUT scale. It is interesting
to note that switching on other VEVs (e.g. $\langle E \rangle\subset {54}_H$)  still offers the option of keeping
only $\Sigma_F$ mass near the PeV scale while having all other masses
at the GUT scale. Similar fine tunings to keep respective component masses
light have been discussed in the corresponding physical applications in \cite{mkp:2011,psb:DM,pnsa:2016}.

 Thus using ${210}_H$  matter parity conserving non-standard Higgs origin of DM mass is
 predicted leading to the  $\Sigma_F$ mass term 
\be
-{\cal L}_{\Sigma}=(1/2) M_{\Sigma}{\bar \Sigma}^C_F (1,0,1)\Sigma_F (1,0,1)+h.c.\label{Sigmass}  
\ee
In Sec.\ref{sec:relic} we will present an  application of
this model to predict desired relic density of DDM where the
 SM singlet scalar $\xi(1,0,1)\subset \Phi_{210}(1,1,15)$ is exchanged between $\Sigma_F$ and SM Higgs $h$.

\section{ Decaying Dark Matter in SO(10) with 
  Type-I Seesaw Dominance}\label{sec:real}
\subsection{Seesaw Mechanism with DM and IceCube Neutrinos from SO(10)}\label{sec:seesawmix}
We have embedded the DM $\Sigma_F$ as a Majorana fermion singlet in ${45}_F$ and
also derived its mass from non-standard Yukawa interactions of
SO(10). 
The Yukawa Lagrangian respecting the SM gauge symmetry below the GUT scale can be
 written as
\ba
-{\cal L}_{Yuk}&=&Y^{\nu}{\bar l}_LN{\tilde {\phi}} +{(1/2)}M_N{\bar {N^C}}N\nonumber\\
 &&+ {(1/2)}M_{\Sigma}{\bar{\Sigma_F^C}}\Sigma_F +{(1/2)}m^{(\rm
    mix)}{\bar {N^C}}\Sigma + f l_L^TC\tau_2 \Delta_L l_L+h.c.,
\label{yukNSIG}
\ea
 where, in our notation, $N(\Sigma_F)$ represent the RH neutrino(DM
 fermion), $\phi=$ the standard Higgs scalar field,
  ${\tilde \phi}=i\tau_2\phi*$, and $\Delta_L (3,-1,1)\subset {126}_H$. 
The mass parameter $m^{(\rm mix)}$ is the analogue of the $N-\Sigma_F$ mixing
 discussed in Sec.\ref{sec:bench}.
Using the Higgs field vacuum expectation value
$\langle \phi \rangle=v_{\rm ew}/\sqrt (2)\simeq  174.1$ GeV, the neutral fermion mass matrix
in the $(\nu, N, \Sigma_F)$ basis turns out to be
\ba
{\cal M}_{\bf O} =  
\begin{pmatrix} 0 & M_D  & 0  \\
 M_D^T  & M_N & {m^{(\rm mix)}}^T\\ 0 & m^{(\rm mix)} &
 M_{\Sigma} \end{pmatrix}.  \label{fullmnu}
\ea
where $M_D=$ the Dirac neutrino mass matrix $=Y^{\nu}\langle \phi \rangle$,
$Y^{\nu}$ being the Yukawa coupling matrix. The type-II seesaw
contribution to neutrino mass has been
  neglected assuming $\Delta_L$ mass at the
GUT scale. 
Here the $M_N$ block  is a $3\times 3$ matrix and $m^{(\rm mix)}$ is a three component row matrix 
\begin{equation}
m^{(\rm mix)}= (m_1^{\rm mix}, m_2^{\rm mix}, m_3^{\rm mix}).\label{eq:mixrow}
\end{equation}
Diagonalization of the first $2\times 2$ block leads to the canonical seesaw
formula along with heavy RH neutrino mass eigenvalues
\ba
m_{\nu}&=&-M_D\frac{1}{M_N}M_D^T, \nonumber\\ 
m_{N}&=&M_N.\label{eq:evcan}
\ea
The mass eigen-states are 
\ba       
{\hat {\nu}}&=& \cos\alpha \,\nu+\sin\alpha \, N, \nonumber\\
{\hat {N}}&=&-\sin\alpha\, \nu + \cos \alpha \, N,\nonumber\\ 
\tan 2\alpha&=& 2M_D/M_N.\label{eq:escan}
\ea
Similarly, diagonalisation of the second block in eq.(\ref{fullmnu})
under the condition $M_N>M_N-M_{\Sigma}> M_{\Sigma} \gg m_{\rm mix}$ 
gives
\be
M^{\pm}=\frac{1}{2}\left[M_N+M_{\Sigma} \pm \sqrt{({(M_N-M_{\Sigma})}^2+4{m^{(\rm
      mix)}}^2)}\right], \label{eq.evndm}
\ee
which gives the heavier ($\Sigma_{\rm Heavy}$) and the lighter
($\Sigma_{\rm Light}$) mass eigen states with respective masses
\ba
M_{\Sigma_H}&=& M^+\simeq M_N+{m^{(\rm mix)}}^2/(M_N-M_{\Sigma}), \nonumber\\
M_{\Sigma_l}&=&M^-\simeq M_{\Sigma}-{m^{(\rm mix)}}^2/(M_N-M_{\Sigma}),\label{eq:evdm}
\ea
\ba       
{\hat {\Sigma}}_{\rm Heavy}&=& \cos\beta\, N+\sin\beta\, \Sigma_F, \nonumber\\
{\hat {\Sigma}}_{\rm Light}
&=&-\sin\beta \, N + \cos \beta\, \Sigma_F,\nonumber\\ 
\tan 2\beta &\simeq& 2m^{(\rm mix)}/M_N. \label{eq:esndm}
\ea

It has been shown in the next section that, 
in terms of individual RH$\nu$ states $N_i$ and their corresponding
mixings $m_i^{\rm mix}$, eq.(\ref{eq:esndm}) can be replaced by 
\ba       
{\hat {\Sigma}}^i_{\rm Heavy}&=& \cos\beta_i\, N_i+\sin\beta_i\, \Sigma_F, \nonumber\\
{\hat {\Sigma}}_{\rm Light}^i
&=&-\sin\beta_i \, N_i + \cos \beta_i\, \Sigma_F,\nonumber\\ 
\tan 2\beta_i &\simeq& 2m_i^{(\rm mix)}/M_{N_i}. \label{eq:esndmm}
\ea

It is clear from eq.(\ref{eq:evdm}) that the conditions
$M_{N_i}>M_{N_i}-M_{\Sigma}> M_{\Sigma}$ are easily satisfied in the canonical
seesaw origin for neutrino masses with $M_{N_i} \sim 10^{9}-10^{15}$
GeV. 

Even if the lightest RH$\nu$ $N_1$ is lighter than the PeV scale DM
mass  we note that a similar formula for mixing is also valid
\ba
\tan 2\beta_1 &\simeq& 2m_1^{(\rm mix)}/M_{\Sigma_F}, \nonumber\\
 M_{\Sigma_F}& \gg & M_{N_1}, \label{eq:mixN_1}
\ea

\subsection{Determination of RH$\nu$ Masses}
 SO(10) has Pati-Salam symmetry $SU(4)_C$
at GUT scale. As such it predicts Dirac neutrino mass matrix similar
to the up-quark mass matrix. This makes Dirac neutrino Yukawa
couplings and type-I seesaw prediction of RH
neutrino masses predominantly hierarchical. This is the reason why the
simplified picture of seesaw adopted in \cite{Rott:2014} with SM extension is not
applicable in SO(10). Here we derive all the three RH$\nu$ masses by
fitting the type-I seesaw formula with most recent neutrino
oscillation data \cite{Salas:2017,schwetz,forero,fogli,gonzalez}.
%%%%%%   my additions %%%%%%%%%%%%%%%%%%%%%%%%
At first we choose a certain hierarchy (NH/IH/QD) of light neutrino masses.
Assuming numerical value of one of the mass eigenvalues, the other two are computed 
using best fit value of mass squared differences as shown in Table
\ref{tab_lhnu} of Appendix \ref{sec:VP}.
The mixing matrix $U_{\rm PMNS}$\footnote{In this context it is worthwhile to mention that the total diagonalisation matrix
is given by $U_{tot}={\rm
  diag}(e^{i\phi_1},~e^{i\phi_2},~e^{i\phi_3})U_{PMNS}$ with
$U_{PMNS}=U (\theta_{ij},\delta){\rm diag}(e^{i\alpha_M/2},~e^{i\beta_M/2},
~1)$, where $\phi_i$s are unphysical phases and $\alpha_M,\beta_M$ are
Majorana phases.}, is then constructed  according to PDG convention
\cite{PDG:2012,PDG:2014,PDG:2016} with 
best fit values of mixing angles $(\theta_{12},\theta_{23},\theta_{13})$ and Dirac CP phase $(\delta)$. 
Armed with mass eigenvalues and mixing matrix, it is now easy to obtain the effective light neutrino
mass matrix $m_\nu$ 
\begin{equation}
m_\nu=  U_{\rm PMNS} {\rm diag}(\hat{m}_{\nu_1},\hat{m}_{\nu_2},\hat{m}_{\nu_3}) U^T_{\rm PMNS}~.
\end{equation}
Again as we are dealing with Type-I seesaw dominated scenario, the effective light neutrino mass matrix
can be written as
\begin{equation}
m_\nu \simeq -M_D M_N^{-1} M_D^T,
\end{equation}
from which the RH$\nu$ mass matrix can be estimated as
\begin{equation}
M_N=\left[ M_D^{-1} m_\nu (M_D^T)^{-1}\right]^{-1}. \label{MN}
\end{equation}
Thus $M_N$ can be calculated if numerical values of $M_D$ and $m_\nu$ are already known to us.
For this purpose we determine the Dirac neutrino mass matrix  using up-quark and down-quark diagonal basis respectively
at the GUT scale  $M_U\sim 10^{15}$ GeV. They are given by  
\begin{eqnarray} 
M^{(u)}_D({\rm GeV}) =\begin{pmatrix}
0.00054 & (1.5027+0.0038i)10^{-9} &  (7.51+3.19i)10^{-6} \\
(1.5027+0.0038i)10^{-9}& 0.26302 &  9.63\times 10^{-5} \\
(7.51+3.19i)10^{-6} & 9.63\times 10^{-5} & 81.9963 \\
\end{pmatrix},\label{eq:MD_udia}\\
%%%%%%%%%%%%%%%%%%%%
M^{(d)}_D({\rm GeV})=\left( \begin{array}{ccc}
             0.01832+0.00441i & 0.08458+0.01114i & 0.65882+0.27319i\cr
             0.08458+0.01114i &  0.38538+1.56\times10^{-5} & 3.3278+0.00019i\cr
             0.65882+0.27319i & 3.3278+0.00019i & 81.8543-1.64\times10^{-5}i\cr
             \end{array}\right)\label{eq:MD_ddia}~.
\end{eqnarray}   
This predicts almost diagonal structure of Dirac neutrino Yukawa neutrino coupling
matrix $(Y^\nu=M^{(u)}_D/v_{\rm ew})$ upto a very good approximation in the  
up-quark diagonal basis. But in the down-quark diagonal
basis the off-diagonal elements are also quite significant. Using the
Dirac neutrino mass matrix in the up-quark diagonal basis 
$M_D=M_D^{ (u)}$ , or in the down-quark diagonal basis $M_D=M_D^{(d)}$
as discussed above 
 in eq.(\ref{MN}), the RH$\nu$ mass matrix $(M_N)$ is calculated. Thereafter
the complex symmetric $M_N$ matrix is diagonalised by a unitarity $V_P$ matrix as
\begin{equation}
M_N=V_P \hat{M}_N V^T_P ~~{\rm where}~\hat{M}_N={\rm diag} (\hat{M}_{N_1},\hat{M}_{N_2},\hat{M}_{N_3}). 
\end{equation}
\paragraph{}
It is to be noted that for calculation of any physical process involving RH$\nu$ (such as decay of $N$ to $\nu_\alpha,h$)
we have to go to the physical basis or mass basis of the RH$\nu$s. But even after diagonalising $M_N$ with $V_P$
matrix the resulting diagonal matrix $(\hat{M}_N)$ may contain complex entries. If all three diagonal elements are complex,
one of them can be made real  by taking out its phase which can be treated as the unphysical phase.
The remaining phase parameters in the other two elements are nothing but Majorana phases which can be absorbed in the $V_P$
matrix. In this way we can get real right handed neutrino masses and the corresponding total diagonalisation matrix $V_P$
with Majorana phases included in it.
\paragraph{}
This whole exercise is repeated for each hierarchy (NH, IH, QD1, QD2) of light neutrinos taking into account both u-quark and
d-quark diagonal basis. Thus, as a whole, we have analysed eight
cases. The diagonalising matrix and the mass eigenvalues of 
the RH$\nu$s are presented systematically in Appendix \ref{sec:VP}.
\subsection{Different $N_i-\Sigma_F$ Mixings}
We now consider the mixing of the fermionic dark matter $\Sigma_F$
with the RH$\nu$ $N_i(i=1,2,3)$. The Majorana type mixing term between
them is given by
\begin{equation}
-\mathcal{L}=\left( \overline{N^c} ~~\overline{\Sigma^c_F} \right)
\mathcal{M} \left(\begin{array}{c} N \\
                                                                               \Sigma_F
                                                                              \end{array}\right) ,
\end{equation}
where $\mathcal{M}$ is a $4\times 4$ matrix and $N$ contains three RH$\nu$ fields given by $N \equiv (N_1,~N_2,~N_3)^T$.
The explicit form of $\mathcal{M}$ is given by
\begin{equation}
\mathcal{M}=\left(\begin{array}{cc}
                   (M_N)_{3\times 3} & m^{(mix)} \\
                   (m^{(mix)})^T & M_\Sigma
                  \end{array} \right)
\end{equation}
where $m^{(mix)}$ is a column matrix with three entries: $m^{(mix)}=(m_1^{mix},~m_2^{mix},~m_3^{mix})^T$. As discussed
in the previous section, complex symmetric $M_N$ matrix, can be diagonalised by $3\times3$ unitary $V_P$ matrix.
Then we assume the $4\times4$ $\mathcal{M}$ matrix to be block diagonalised by unitary $4\times4$
$V^\prime_P$ matrix 
\begin{equation}
V^\prime_P=\left(\begin{array}{cc}
                  (V_P)_{3\times3} & O \\
                  O^T &  1
                 \end{array}\right) ~~{\rm where} ~ O\equiv (0,~0,~0)^T.
\end{equation}
Thus, after block diagonalisation, we are left with
\begin{equation}
\mathcal{M}_{block-dia}= \left(\begin{array}{cc}
                  \hat{M}_N & m^{(mix)} \\
                  (m^{(mix)})^T &  M_\Sigma
                 \end{array}\right).\label{b_dia}
\end{equation}
For full diagonalisation the above matrix (\ref{b_dia}) has to be rotated again by a matrix $V_S$ which can be represented as 
a combination of four rotation matrices as $V_S=R_{14}(\beta_1).R_{24}(\beta_2).R_{34}(\beta_3)$. For small values of $\beta_i$,
$V_S$ can be presented to a good approximation as 
\begin{equation}
{V_S}\simeq
\begin{pmatrix}
1 & 0 & 0 & \beta_1\\
0 & 1& 0 &  \beta_2\\
0 & 0 & 1 & \beta_{3}\\
-\beta_1& -\beta_2 & -\beta_3 & 1\\
\end{pmatrix} ~~{\rm where}~\beta_i \simeq\frac{m_i^{mix}}{M_{Ni}-M_\Sigma}~(i=1,2,3) .\label{eq:vs}
\end{equation}
Thus we can say that the $\mathcal{M}$ matrix is fully diagonalised through a two step rotation (first by $V^\prime_P$ 
followed by $V_S$) and the total diagonalisation matrix is 
\begin{equation}
V_N=V^\prime_P V_S .\label{eq:VN}
\end{equation}
 It should be noted that the physical or 
diagonal basis of these heavy fermionic fields are obtained by multiplying $(V_N)^\ast$\footnote{The complex conjugation comes
because the mass matrix is written in the Majorana basis.} matrix with the flavor basis states $(N~\Sigma_F)$.

\subsection{Determination of Mixing Parameters}
Using the three constraint equations for the three partial branching ratios
of the DM decay $\Sigma_F \to \nu_{\alpha}+h (\alpha=e,\mu, \tau)$, we now determine the mixing parameters $m^{\rm mix}_i (i=1,2,3)$. 
Equality of three branching ratios imply
\begin{eqnarray}
\Gamma (\Sigma_F \rightarrow \nu_\alpha~h) =
 \frac{M_\Sigma}{32\pi} \sum_{i=1,2,3} |Y_{\alpha i} (V_{N_{i4}})^\ast|^2=\Gamma~,~
(\alpha=e,\mu,\tau). \label{mmix}
\end{eqnarray}
This is actually a set of three equations (for $\alpha=e,\mu,\tau$) each of which contains three unknown
parameters $m_i^{mix}(i=1,2,3)$. The common decay width $\Gamma$ in the RHS of the above equation is the 
inverse of life time ($\tau_\Sigma$) of the dark matter particle $\Sigma_F$ ($\tau_\Sigma \sim 10^{28}$ sec) which is much 
greater than the life time of Universe. These three equations  in eq.(\ref{mmix}) are then solved simultaneously
to get the values of the unknown mixing parameters which in turn produces equal branching ratio of the 
decay of DM to each neutrino flavor. Following the same methodology we calculate these mixing parameters
for the previously mentioned eight cases (NH, IH, QD1, QD2) with u-quark diagonal basis and d-quark diagonal basis.
The results are presented in a concise manner in Table \ref{tab_uq} and Table \ref{tab_dq}.

Solving eq.(\ref{mmix}) we find the three mixing parameters for the normally hierarchical (NH) pattern of light
neutrino masses
\begin{eqnarray}
&&m^{\rm mix}_1=-2.724 \times 10^{-8} {\rm eV},\nonumber\\
&&m^{\rm mix}_2=-3.255 \times 10^{-8} {\rm eV},\nonumber\\
&&m^{\rm mix}_3= 2.395 \times 10^{-4} {\rm eV}.\label{eq:mmixi}
\end{eqnarray}
Out of these, the first two are of the same order but the third is $4$
orders larger than each of them and nearly $50$ times larger than the value derived in the
bench mark model. Thus we have successfully derived the three
different mixings the decaying dark matter is predicted to possess
with the three hierarchical RH$\nu$s of non-SUSY SO(10) GUT that gives
type-I seesaw ansatz for light neutrino masses. 
We have also solved for the RH$\nu$ mass and mixing parameter spectra
using neutrino oscillation data in the cases of quasi-degenerate (QD)
and the inverted hierarchical (IH) light neutrino mass patterns. 
%%%%%%%%%%%%%%%%%%%%%%%%%% 
% The
% results are presented in Table \ref{Yb_m} where we have presented
% the Majorana phases in the RH$\nu$ sector which were nessary to
% derive the desired solutions.
%%%%%%%%%%%%%%%%%%%%% 
For the QD type solutions we have
chosen one set of  light
neutrino masses, QD1, which satisfy the recent cosmological bound \cite{Planck15}
and another set, QD2, expected to be reachable by Katrin
experiment \cite{Katrin}.\\
%\par\noindent{\large\bf QD1}\\  
\begin{eqnarray}
&&{\large\bf \rm QD1}~~~(\hat {m}_1,\hat{m}_2, \hat {m}_3) 
=(0.0630079,0.0636035,0.0800)\,\,{\rm eV}.\label{eq:QD1mass}\nonumber\\
%\end{eqnarray}   
%\par\noindent{\large\bf QD2}
%\begin{eqnarray}
&&{\large\bf \rm QD2}~~~(\hat {m}_1,\hat{m}_2, \hat {m}_3) =
(0.1938,0.1940,0.2)\,\,{\rm eV}.\label{eq:QD2mass}
 \end{eqnarray}
The QD2 choice may need priors in addition to cosmological bound.
%%%%%%%%%%%%%%%%%%%%%%%%%%%%%%%%%%%%%%%%%%%%%%%%%%
%%%%%%%%%% Table for u-quark diagonal basis %%%%%
%%%%%%%%%%%%%%%%%%%%%%%%%%%%%%%%%%%%%%%%%%%%%%%%%%%%
%%%%%%% U quark db with QD1 & QD2 %%%%%%%%%%%%%%%%%%%
\begin{table}[!h]
%\caption{}
\caption{Predictions of mixing parameters of decaying dark matter
  $\Sigma_F$ with three heavy right handed neutrinos $N_i(i=1,2,3)$
  from Type-I seesaw dominance in SO(10), neutrino oscillation data
  with NH, QD, and IH type masses
  and IceCube neutrino data. Two non-vanishing Majorana phases
  $\alpha_M$ and $\beta_M$ of heavy RH$\nu$s needed for solutions
   have been indicated in each case. Dirac neutrino mass matrix $m_D$ is taken according to up-quark diagonal basis 
   (eq.(\ref{eq:MD_udia})). }
\begin{center}
\begin{tabular}{ |c|c|c|c|c|c|c| } 
\hline
 Mass& $\hat{M}_{N_1}$ (GeV) &  $\hat{M}_{N_2}$ (GeV) &
 $\hat{M}_{N_3}$ (GeV) & $m^{\rm mix}_1$ (eV) &  $m^{\rm mix}_2$ (eV)
 & $m^{\rm mix}_3$ (eV) \\ 
 ordering& $(\alpha_M)$&$(\beta_M)$ & & & & \\\hline
 NH & $8.9\times 10^5 $  & $2.28\times 10^9$ & $1.2\times 10^{15}$& $-2.724$& $-3.255$& $2.395$\\ 
 &$({147}^{\circ})$ & $({108}^{\circ})$ & $ $ & $\times10^{-8}$& $\times10^{-8}$& $\times10^{-4}$\\\hline
 QD1 & $4.75\times 10^3$  & $9.6\times 10^8$ & $9.225 \times 10^{13}$& $-2.608$& $4.282$& $-1.093$\\ 
 & $(179^{\circ})$& $(174^{\circ})$& $ $& $\times10^{-8}$& $\times10^{-8}$&$\times10^{-5}$\\\hline
 QD2 & $1.54\times 10^3$  & $3.47\times 10^8$ & $3.35 \times 10^{13}$& $2.61$& $-1.7$& $6.97$\\ 
 & $(178^{\circ})$& $(-40.18^{\circ})$& $ $& $\times10^{-8}$& $\times10^{-8}$&$\times10^{-6}$\\\hline
 IH & $2.976\times10^{15}$ & $2.5\times10^9$ & $6.1\times 10^3$& $-1.044$& $1.656$& $2.613$\\ 
  & $(-33^{\circ})$ & $(-33^{\circ})$&  & $\times10^{-3}$& $\times10^{-7}$& $\times10^{-8}$\\\hline
% cell7 & cell8 & cell9 & & &\\ \hline
\end{tabular}
\label{tab_uq}
\end{center}
\end{table}
%%%%%%%%%%%%%%%%%%%%%%%%%%%%%%%%%%%%%%%%%%%%%%%%%%%%%%%%%%%%%
%%%%% Table for down quark diagonal basis %%%%%%%%%%%%%%%%%%%
%%%%%%%%%%%%%%%%%%%%%%%%%%%%%%%%%%%%%%%%%%%%%%%%%%%%%%%%%%%%%
%%% For d quark DB  with QD1 & QD2 %%%%%%%%%%%%%%%%%%%%%%%%%%%%%%%%%%%%%%%%%
\begin{table}[!h]
%\caption{}
\caption{Predictions of mixing parameters of decaying dark matter
  $\Sigma_F$ with three heavy right handed neutrinos $N_i(i=1,2,3)$
  from Type-I seesaw dominance in SO(10), neutrino oscillation data
  with NH, QD, and IH type masses
  and IceCube neutrino data. Two non-vanishing Majorana phases
  $\alpha_M$ and $\beta_M$ of heavy RH$\nu$s needed for solutions
   have been indicated in each case. Dirac neutrino mass matrix $m_D$ is taken according to 
   the down-quark diagonal basis (eq.(\ref{eq:MD_ddia}))}
\begin{center}
\begin{tabular}{ |c|c|c|c|c|c|c| } 
\hline
 Mass& $\hat{M}_{N_1}$ (GeV) &  $\hat{M}_{N_2}$ (GeV) &
 $\hat{M}_{N_3}$ (GeV) & $m^{\rm mix}_1$ (eV) &  $m^{\rm mix}_2$ (eV)
 & $m^{\rm mix}_3$ (eV) \\ 
 ordering& $(\alpha_M)$&$(\beta_M)$ & & & & \\\hline
 NH & $5.85\times 10^4 $  & $3.69\times 10^9$ & $1.14\times 10^{15}$& $2.948$& $-1.045$& $6.463$\\ 
 &$({-177.08}^{\circ})$ & $({-3.6}^{\circ})$ &  & $\times10^{-8}$& $\times10^{-7}$& $\times10^{-4}$\\\hline
 QD1 & $4.71\times 10^3$  & $9.9\times 10^8$ & $9.14 \times 10^{13}$& $3.1$& $3.78$& $1.75$\\ 
 & $(145.5^{\circ})$& $(-5.3^{\circ})$&  & $\times10^{-8}$& $\times10^{-8}$&$\times10^{-5}$\\\hline
 QD2 & $1.55\times 10^3$  & $3.5\times 10^8$ & $3.35 \times 10^{13}$& $-1.96$& $2.2$& $5.81$\\ 
 & $(9.27^{\circ})$& $(-3.35^{\circ})$&  & $\times10^{-8}$& $\times10^{-8}$&$\times10^{-6}$\\\hline
 IH & $3.177\times10^{15}$ & $2.23\times10^9$ & $6.59\times 10^3$& $-1.03$& $1.045$& $-3.1$\\ 
  & $(-141.25^{\circ})$ & $(-144.9^{\circ})$& & $\times10^{-3}$& $\times10^{-7}$& $\times10^{-8}$\\\hline
% cell7 & cell8 & cell9 & & &\\ \hline
\end{tabular}
\label{tab_dq}
\end{center}
\end{table}

Thus we have found that realistic Type-I seesaw dominance in SO(10) that fits  
the neutrino oscillation data and the IceCube data results in
substantially different predictions on the RH$\nu$ mass  and
 mixing spectra  compared to the simplistic assumptions of the SM extension
 \cite{Rott:2014}. This holds true for Dirac neutrino masses evaluated in both the up-quark or  the
 down-quark diagonal basis. For a given light neutrino mass pattern,
 NH, IH, or QD, clearly there are three distinct values of
 $N_i$-$\Sigma_F$ mixings consistent with IceCube neutrino data and
 the natural hypothesis that $\Sigma_F$ decays with equal probability
 to each light neutrino flavor. Whereas benchmark model holds for QD
 type neutrino mass hierarchy in the SM extension with the stated value ${\hat
   m}_{\nu_i}=0.1$ eV with a universal heavy mass $M_N=10^{14}$ GeV, our SO(10) ansatz matches with all types of
 light neutrino mass hierarchies and predominantly hierarchical
 $M_{N_i}$ values covering the range $10^{4} - 10^{15}$ GeV. 

\section{Dynamical Generation of RH$\nu$-DM Mixing}\label{sec:mix}
In the following section we explore theoretical origin of
$N_i-\Sigma_F$ mixings derived in the previous section using neutrino oscillation data and
IceCube neutrino data. For convenience we choose solutions derived in the up-quark
diagonal basis and for other cases similar derivations apply.
 As the intrinsic matter parities of  $\Sigma_F\subset {45}_F$ and
 RH$\nu$ $N_i\subset {16}_{F_i}$ are even and odd, respectively, their
 mixing is possible if this gauged discrete symmetry is broken
 explicitly or spontaneously. 
 The usual mechanisms of breaking a discrete symmetry, which
might be an intrinsic gauged discrete symmetry of the theory or
externally imposed upon it,  are known to result in cosmological
domain wall problem. A natural resolution of the domain wall problem emerges if
the discrete symmetry breaking is assisted by gravity or Planck
scale \cite{Giddings:1988,Krauss:1989,Preskill:1991,Berezinsky:1998,Boucenna:2014,Mishra-Yajnik:2009}. In particular, because of the redundancy of parameters of 
local gauge transformation, the Planck-scale assisted symmetry breaking
has been noted to be more effective  if the discrete symmetry is an intrinsic gauged
discrete symmetry of the theory \cite{Giddings:1988,Krauss:1989,Preskill:1991,Berezinsky:1998} and
the matter parity  in our model being a gauged
discrete symmetry ideally matches with this situation.
 The purpose of this section is to
discuss the possibility of  different renormalizable and
nonrenormalizable interactions for the Planck scale assisted matter
parity breaking that
gives rise to the extremely small values of the mixings.
 
\subsection{Planck-Scale Assisted RH$\nu$-DM Mixing}
In Sec. \ref{sec:embed} we have shown how the non-standard Yukawa interaction in SO(10) has the capability to predict the DM ($\Sigma_F$) mass. We show how the matter parity conserving SO(10) model that predicts
 type-I seesaw dominance as well as its high scale, also predicts the
 extremely  small value of
 $N_i-\Sigma_F$ mixing through Planck-scale assisted matter parity breaking.
 We assign  the decaying singlet fermion DM 
$\Sigma_F (1,0,1)$  to the
 nonstandard fermionic representation ${45}_F$  which has even 
 matter parity. Similarly the RH$\nu$ being
 in the spinorial representation ${16}_F$ possesses odd matter parity. Therefore, as the $N\Sigma_F$ fermion bilinear has
 odd matter parity, any mass-dimensioned coefficient of
 this term can not be generated without breaking matter parity. 
As the  generation of this discrete symmetry breaking, either softly or spontaneously, leads to the well known domain wall problem, in this work we follow the idea that the cosmologically safe matter parity breaking can be achieved by Planck scale effects
\cite{Giddings:1988,Krauss:1989,Preskill:1991,Berezinsky:1998}. 
We assume the presence of a
SO(10) singlet fermion $N^{\prime}$ having Planck mass $\sim 10^{19}$
GeV. Its mixing $m_i^{(\rm Br)}$ with the RH$\nu$ $N_i$ which is a SU(5)-singlet and carries
odd matter parity acts  as a source of the matter parity breaking
term. This  $m_i^{(\rm Br)}$ can emerge from the VEV $V_{\chi}$ of a SM scalar singlet $\chi_S(1,0,1)\subset {16}_H^{\dagger}$
\ba 
-{\cal L}_{\rm Plmix}&=&m_i^{(\rm Br)}N_i N^{\prime} \nonumber\\
&&\subset Y_i^{\chi}V_{\chi}N_i N^{\prime} \nonumber\\
&&\subset Y_i^{\chi}{16}^i_F.1_F.{16}_H^{\dagger}, \nonumber\\  
m_i^{(\rm Br)}&=& Y_i^{\chi}\langle \chi_S(1,0,1)\rangle
= Y_i^{\chi}V_{\chi}.\label{eq:Plmix}
\ea

\begin{figure}[h!]
\begin{center}
\includegraphics[scale=0.4]{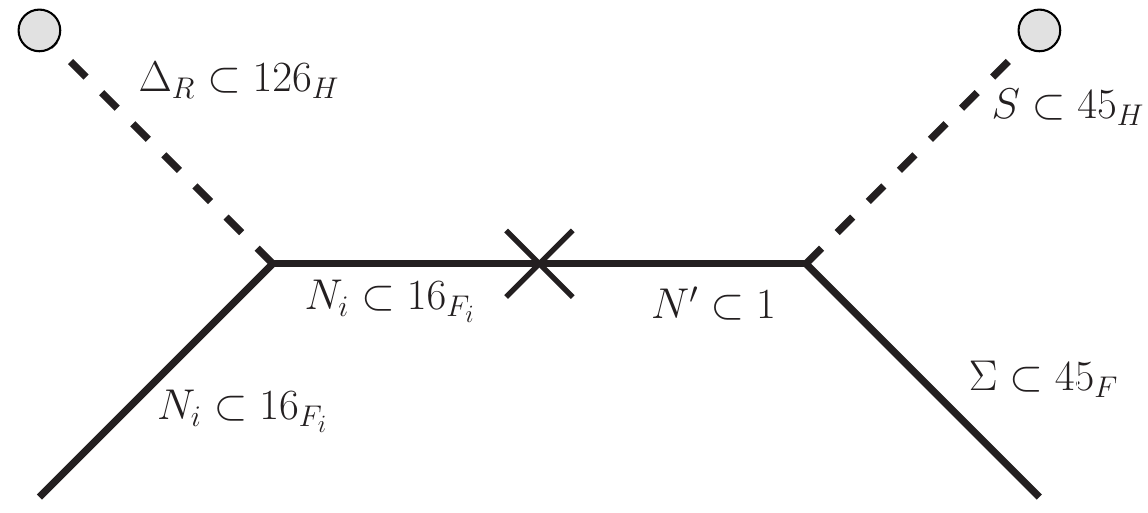}
\caption{Feynman diagram generating $N_i-\Sigma_F$ mixing in SO(10). Here
  $N^{\prime}$ is the SO(10) singlet fermion of Planck mass and
  $S_H$ is the SM singlet Higgs component of ${45}_H$ that acquires
  VEV $V_H$.}
\label{fig:mixint}
\end{center}
\end{figure}

In the present model, the added presence of the SO(10)
singlet $N^{\prime}$ of Planck mass also permits the SO(10)
invariant Higgs fermion interaction  
\ba
-{\cal L}^{(Pl)}_{Yuk}&=&y_{45}N^{\prime}{45}_F{45}_H \,\,\nonumber\\
&& \to y_{45}N^{\prime}{\Sigma}_F S_H\,\, ,\label{eq:Plyuk}
\ea
where the SM scalar singlet $S_H \subset {45}_H$ that has even matter
parity can acquire VEV 
$\langle S_H \rangle=V_H \sim {\cal O}(M_W)-{\cal O}(M_{\rm GUT})$.
Thus the Planck scale assisted matter parity breaking mechanism can be
visualised to originate from a
  seesaw type Feynman diagram shown in
Fig.\ref{fig:mixint} \footnote{When N$^{\prime}$ is integrated out, 
  Fig. \ref{fig:mixint} leads to the effective
  5-dim. operator scaled by Planck mass: ${(\eta
    {16}_F{45}_F{16}_H^{\dagger}{45}_H)}(1/ M_{Planck}) \subset {(\eta
    N.\Sigma_F.\chi_S.S_H)}(1/ M_{Planck})$ where $\eta$ represents 
product of relevant couplings. This is shown in Fig. \ref{fig:NRSigN} below.}.   
%----------------------------------------------------
%----------------------------------------------------
In Fig.\ref{fig:mixint} we have also used the Yukawa interaction $f
{16}_F.{16}_F.{126}^{\dagger}_H$ .
Then using VEV~~ $ \langle \Delta_R\rangle=V_R \simeq  M_U$, with $M_N=f V_R$ gives the
$N_i-\Sigma_F$ mixing 
\be
-{\cal L}_{(\rm seesaw)}=y_{45}\frac{m_i^{(\rm Br)}V_H}{M_{N^{\prime}}}N\Sigma_F, \label{eq:seesawmix}  
\ee
leading to
\ba
m_i^{(\rm mix)}&=&y_{45}\frac{m_i^{(\rm Br)}V_H}{M_{N^{\prime}}},\nonumber\\
&&\simeq y_{45}\frac{m_i^{(\rm Br)}V_H}{M_{\rm Planck}}. \label{eq:mmix}
\ea
It is clear that wide range of values of the explicit  matter parity breaking parameter $m_i^{\rm Br}$ in
$m_i^{\rm Br}N_i N^{\prime}$ triggers the $N_i-\Sigma_F$ mixings
reported in Sec. \ref{sec:real}.
%-------------------------------
%-------------------------------
%----------------------------------
We further note that keeping matter parity conservation of the SM
gauge symmetry, it is possible to assign any VEV to $\chi_S$ under the
constraint $V_{\chi_S}
\le {\cal O} (M_W)$. 

\par In what follows
we show how the extremely small value of $N_i-\Sigma_F$ mixings are
  realized by Planck-scale assisted spontaneous breaking of matter
  parity via renormalizable and nonrenormalizable interactions.  

\subsubsection{Through VEV of Scalar Singlet in ${16}_H^{\dagger}$}
\begin{figure}[h!]
\begin{center}
\includegraphics[scale=0.5]{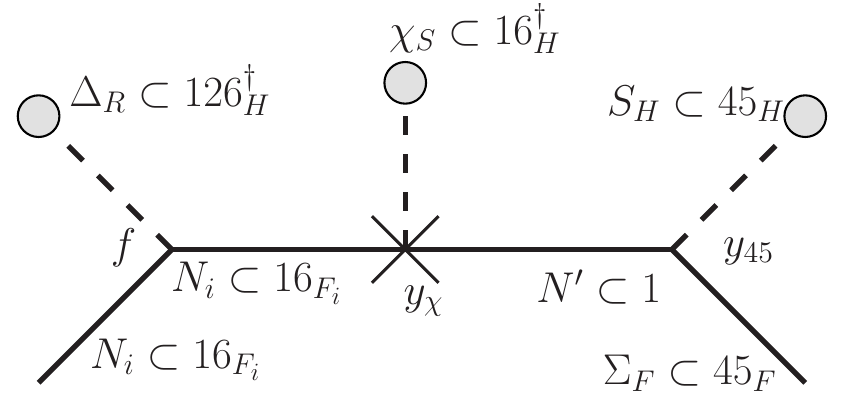}
\caption{Feynman diagram generating N-N$^{\prime}$ mixing that leads to Planck
  scale assisted matter parity breaking and extremely small value of 
N-$\Sigma_F$
mixing. Here
  N$^{\prime}$ is the SO(10) singlet fermion  having Planck mass,
  $S_H$ is the SM singlet Higgs component of ${45}_H$ that acquires
  VEV $V_H$ , and $\chi_S$ is the singlet scalar component of
  ${16}_H^{\dagger}$ with allowed VEV $V_{\chi}\le 
{\cal O}(M_{W})$.}
\label{fig:chisvev}
\end{center}
\end{figure}

The solutions for $m_i^{(\rm Br)}$ stated above can be  understood
further on the basis of spontaneous breaking of matter parity
elucidated through Feynman diagram of Fig. \ref{fig:chisvev}.\\ 
In eq.(\ref{eq:Plmix}), $m_i^{(\rm Br)}$ can be assigned its spontaneous
symmetry breaking origin through the VEV $V_{\chi}$ of the  Higgs scalar
singlet $\chi_S(1,0,1) \subset {16}_H^{\dagger}$ which is also a singlet
under SU(5) and SM gauge theories. As it carries odd matter parity, the
conservation  of matter parity of SM down to the electroweak scale is ensured only if 
 $V_{\chi} \le {\cal O}(v_{\rm ew})=246$ GeV. For the purpose of this
work the scalar singlet $\chi_S(1, 0,1)$  is treated to be
real \footnote{ With the SO(10) invariant piece of the Higgs potential
  $V_{\rm split}= M_U^2 {16}_H^{\dagger} {16}_H +
  [\mu_{\Delta}{126}_H^{\dagger}{16}_H{16}_H+ h.c.]+{(\rm other\,\, terms)}$,
  where the trilinear coupling
  $\mu_{\Delta} \sim M_U$, a straight forward derivation shows that
  any one of the
  real or the imaginary components of $\chi_S(1,0,1)\subset {16}_H^{\dagger}$ can
  be fine tuned to remain as light as possible  while the other
  component can acquire mass  near the
  GUT scale. As a result the imaginary part decouples from
  contributing to any physical quantity below the GUT scale and 
  self-consistency of  model predictions of this work is guaranteed. This result has been discussed further in more
  detail in Sec.\ref{sec:imchi}.}. \\
 
In Fig. \ref{fig:chisvev} we have shown how
the N-N$^{\prime}$ mixing is generated through the electroweak scale VEV
$\langle \chi_S\rangle=V_{\chi} \sim {\cal O}( M_w)$ which breaks matter parity spontaneously
alone via the SO(10) invariant gauge interaction term
$Y^{\chi}{16}_F.1_F.{16}_H^{\dagger}$. Here $Y^{\chi}$ is the
associated Yukawa coupling. The SM gauge symmetry breaks in the usual
manner
through the  VEV of the standard Higgs doublet $\phi \subset {10}_H$
that carries even matter parity with VEV $v_{\phi}=v_{\rm ew}= 246$ GeV.
%----------------------------------------------------
%----------------------------------------------------
This gives 
\be
m_i^{\rm Br}=Y_i^{\chi}V_{\chi} \le Y_i^{\chi} v_{\rm ew}.\label{eq.vevMw}  
\ee

%--------------------------------------------------------------
\subsubsection{Through Nonrenormalizable N-N$^{\prime}$ Interaction}
\begin{figure}[h!]
\begin{center}
\includegraphics[scale=0.4]{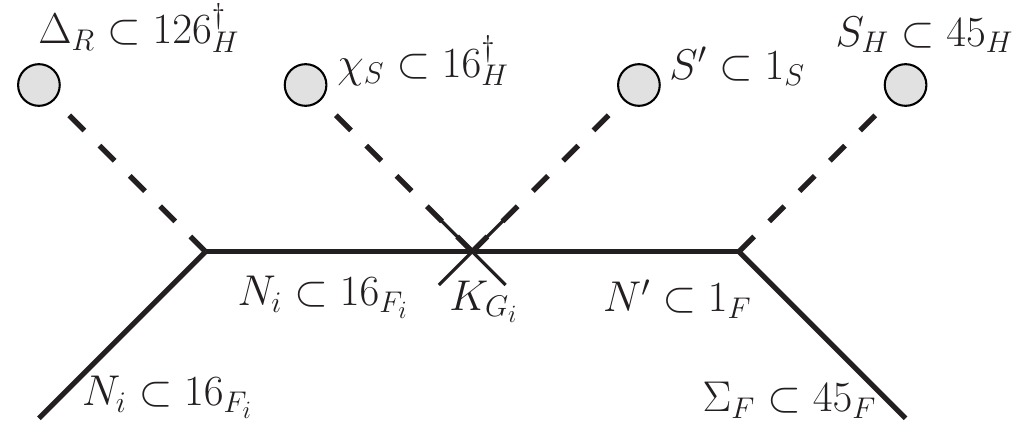}
\caption{Nonrenormalizable N-N$^{\prime}$ interaction contributing to
  Planck-scale assisted matter parity breaking. Here S$^{\prime}$ is a SO(10)
  singlet scalar.} 
\label{fig:NR-I}
\end{center}
\end{figure}

The RH$\nu$ and the SO(10) singlet fermion N$^{\prime}$ may also have a Planck
scale mediated nonrenormalizable interaction through a dim.5 operator
which requires the introduction of a SO(10) singlet scalar S$^{\prime}$
\be
 -{\cal L}_{NR-I}=K_{G_i} {16}_{F_i}.1_F.{16}_H^{\dagger}S'/M_{Planck}. \label{eq:dim5-1} 
\ee
 Feynman diagram for this interaction and the corresponding seesaw
mechanism is shown in Fig.\ref{fig:NR-I}.
%----------------------------------------------------------------
%--------------------------------------------------------------

Noting that $V_{S'}$, the VEV of S$^{\prime}$, can be
anywhere above the electroweak scale, eq.(\ref{eq:dim5-1}) gives
\be
m_i^{\rm Br}=K_{G_i} \frac{V_{\chi}V_{S'}}{M_{Planck}}.\label{eq:nrmBr}
\ee
Using $V_{\chi}\sim 100 $ GeV, $M_{Planck}=10^{19}$ GeV, and
$V_{S'}=10^8-10^{19}$ GeV this predicts a wide range of values of
$m_i^{\rm Br}=1 \,{\rm eV}-100 \,{\rm GeV}$. Through this mechanism we have
shown that small values of matter parity breaking parameter $m_i^{\rm
  Br} \sim 1 $ eV are also realizable even though  $V_{\chi}\sim
v_{\rm ew}$.
The smallness of  $m_i^{\rm Br}$ in this case is a result of
Planck-scale suppression as well as the SM matter parity restricted
smaller value of $V_{\chi}$. 
\subsubsection{Direct Nonrenormalizable $N-\Sigma_F$ interaction}
\begin{figure}[h!]
\begin{center}
\includegraphics[scale=0.6]{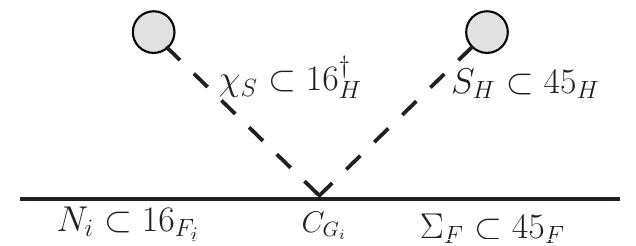}
\caption{Nonrenormalizable N-$\Sigma_F$ interaction contributing to
  Planck-scale assisted matter parity breaking. Here $S_H$ is a SM
  singlet scalar in ${45}_H$.} 
\label{fig:NRSigN}
\end{center}
\end{figure}

Noting that breaking matter parity as gauged discrete symmetry
essentially needs assistance from gravity or Planck scale effects
\cite{Giddings:1988,Krauss:1989,Preskill:1991,Berezinsky:1998} we introduce the following, 
 dim.5 operator scaled by
\ba
-{\cal L}_{NR-II}&=&C_{G_i}{16}_F^i.{45}_F.{16}_H^{\dagger}.{45}_H/M_{Planck} \nonumber\\
             &\to&C_{G_i} N_i\Sigma_F\chi_S S_H/M_{Planck}.\label{eq:dim5-2}
\ea
where the constant $C_{G_i} \simeq 1$. Even though this nonrenormalizable interaction results by integrating out  $N^{\prime}$ that mediates the
  Feynman diagram of Fig.\ref{fig:mixint}, it is possible to write
down eq.(\ref{eq:dim5-2}) independently as the Planck scale assisted dim.5
operator in the absence of $N^{\prime}$.
With $V_{\chi_S} \sim {\cal O}( M_w)$, the small N-$\Sigma_F$ mixing $m_{\rm mix}=10^{-5}$ results for $V_H=\langle S_H \rangle\sim {\cal O}(10^{3})$ GeV. 
Thus, with matter parity
 as its intrinsic gauge discrete symmetry, this SO(10) model
without the introduction of the additional fermion singlet or external discrete
symmetry becomes
a self sufficient theory of unified matter and decaying dark matter.
For the sake of completeness, this interaction is shown through the
Feynman diagram of Fig. \ref{fig:NRSigN}.
%------------------------------------------------------------
%-------------------------------------------------------------

We emphasize that the underlying mechanism of matter parity breaking
that requires Planck scale assistance \cite{Berezinsky:1998} plays a crucial role in
providing a natural explanation of N-$\Sigma_F$ mixing. Without such
gravity or Planck scale effects \cite{Berezinsky:1998},
direct breaking of matter parity  would give rise to cosmological domain
wall problem \cite{Krauss:1989,Preskill:1991}. 
\section{Determination of Renormalizable and Non-Renormalizable
  Couplings}\label{sec:couplings}
Our dynamical explanation of $N_i-\Sigma_F$ mixings would be complete
by estimating the relevant couplings of different interaction
Lagrangians discussed above which is undertaken in this section.
We have shown that single and double Planck-scale suppression can
occur leading to extremely small mixings between each of the three
heavy RH neutrinos  $N_i\subset {16}_F^{(i)}(i=1,2,3)$ and the decaying
DM $\Sigma_F\subset {45}_F$. We now estimate different Yukawa couplings
of RH$\nu$ of  three generations which give such mixings. 
\subsection{Renormalizable Solutions with Domain Wall Problem}
 Denoting the renormalizable 
Yukawa interaction of $\chi_S$ with  $N_i (i=1,2,3)$ through 
\begin{eqnarray}
&&-{\cal L}^R= h^{R}_{i} {45}_F {16}^{i}_F {16}_H^{\dagger} \nonumber\\
&&\to   h^{R}_{i} V_{\chi}N_i\Sigma_F,\nonumber\\
&&m^{\rm mix}_i= h^{R}_{i}V_{\chi}, \label{eq:YukR}
\end{eqnarray}
and using $V_{\chi}\le v_{\rm ew}$ we can estimate the limiting values of
Yukawa couplings for three generations for each case of solutions
given in eq.(\ref{eq:RDW}) and the first line of Table \ref{tab_uq}
. Thus corresponding to up-quark diagonal basis and NH, we have 
\begin{equation}
|h_i^R|\ge (1.10\times
10^{-19},1.32\times 10^{-19}, 9.70\times 10^{-16}) (i=1,2,3).\label{eq:RDW}
\end{equation}
\subsection{Solutions Without Domain Wall Problem}
 Planck scale assisted solutions expected to be free
of the domain wall problem predict different values of Yukawa
couplings noted below.

\par\noindent{\large \bf (a) Renormalizable Yukawa Coupling from Single Planck-Scale Suppression:-}\\
In order to deal with three different RH$\nu$s with hierarchical
masses we replace our notations in all relevant figures and equations
with $N\to N_i$, $Y^{\chi}\to Y^{\chi}_i$, $m^{\rm Br}
\to m^{\rm Br}_i$, and $m^{\rm mix} \to m^{\rm mix}_i$ for different
RH$\nu$ flavors $N_i (i=1,2,3)$.  
Integrating out all the relevant heavy fields in Fig.\ref{fig:mixint}, and
Fig.\ref{fig:chisvev} we derive expression for effective mixing and Yukawa
couplings due to  Planck-scale suppression.

\ba
&&m^{\rm mix}_i\simeq y_{45}\frac{m^{\rm Br}_iV_H}{M_{\rm Planck}},  \nonumber\\ 
&&=y_{45}\frac{Y^{\chi}_iV_HV_{\chi}}{M_{\rm Planck}},\nonumber\\
&&=h^{\rm eff}_{i}V_{\chi}, \nonumber\\
&&h^{\rm eff}_{i}=y_{45}\frac{Y^{\chi}_iV_H}{M_{\rm Planck}}.\label{eq:singlePl}
\ea 
 In the NH case this leads to
\begin{eqnarray}
&&h^{\rm eff}_{i}\ge m^{\rm mix}_i/v_{\rm ew},\nonumber\\
%&&|h_i^{\rm eff}|\ge (1.14\times 10^{-18},1.38\times 10^{-18},
%  1.0\times 10^{-13}) (i=1,2,3),\nonumber\\
&&|Y_i^{\chi}|\ge (1.10\times 10^{-8},1.32\times 10^{-8},
  9.7\times 10^{-5}),
\label{eq:Yukeff}
\end{eqnarray}
where in deriving the last step we have used $y_{45}=1$ and $V_H=10^8$
GeV in eq.(\ref{eq:singlePl}). The Yukawa couplings $Y^{\chi}_i$ are
renormalizable. With such intermediate value of scalar singlet VEV of
even matter parity there will be an additional singlet scalar of mass
$M_{SH}\simeq 10^8$ GeV.\\

\par\noindent{\large \bf (b) Effective Nonrenormalizable Coupling from Double Planck Suppression:-}\\

Integrating out all relevant fields in Fig.\ref{fig:NR-I} leads to
corresponding expressions originating from double Planck suppression.

\ba
&&m^{\rm mix}_i\simeq y_{45}\frac{K_{G_i}V_HV_{S^{\prime}}V_{\chi}}{M^2_{\rm Planck}},\nonumber\\
&&\equiv h^{( eff, NR )}_{i}V_{\chi}, \nonumber\\
&&h^{( eff, NR)}_{i}=y_{45}\frac{K_{G_i}V_HV_{S^{\prime}}}{M^2_{\rm
    Planck}}.\label{eq:doublePl}
\ea
Then for the NH case using $y_{45}\simeq 1$,
$V_H=V_{S^{\prime}}=10^{15}$ GeV we get
\begin{equation}
|K_{G_i}| \ge (1.10\times 10^{-11},1.32\times 10^{-11},
  9.7\times 10^{-8}).\label{eq:YukdoublePl}
\end{equation}
 In this case there will be no additional Higgs scalar singlets of
 intermediate mass except for the light $\chi_s$ discussed in Sec.\ref{sec:stab}.
 \par\noindent{\large \bf (c) Effective Nonrenormalizable Coupling Without  Fermion Singlet $N^{\prime}$}\\

We have shown in Sec.\ref{sec:mix} that extremely small mixings can be generated by
non-renormalisable ${\rm dim}.5$ interaction. As discussed above, replacing $N\to
N_i$ in Fig. \ref{fig:mixint} and correspondingly $C_{G}\to C_{G_i}$
as in 
eq. (\ref{eq:dim5-2}) we get
in the NH case  
\begin{equation}
|C_{G_i}|\ge (1.14\times 10^{-4},1.38\times 10^{-4},
  1.0).\label{eq:nonprime}
\end{equation}
where the equality holds for $V_{\chi}=v_{\rm ew}$. 
As already noted, an interesting outcome of this estimation is that
the present SO(10) theory  is not only
free from invoking any external discrete symmetry for DM stability, but also it does
not  need any Planck mass singlet fermion $N^{\prime}$ as in the case
(a) and case (b) discussed above to achieve
domain-wall free cosmologically acceptable matter parity breaking
through Planck-scale suppressed ${\rm dim.} 5$ operator 
of eq.(\ref{eq:dim5-2}).

Different allowed values of couplings without the cosmological domain
wall problem are summarized in Table \ref{tab_coup} as the class of
``NO'' solutions. 
\begin{table}[!h]
%\caption{}
\caption{Renormalizable ($h_i^R, Y_i^{\chi}$) and nonrenormalizable
  ($h_i^{(eff,NR)},C_{G_i}$) solutions for $\chi_S$ Yukawa couplings
  in the presence (labeled as ``YES'') and
 absence 
(labeled as `` NO'')
 of cosmological domain wall problem all of which predict the three
 $N_i-\Sigma_F$ mixings for the NH type light neutrino masses in the
 up-quark diagonal basis given in Table \ref{tab_uq}. The VEV of the
 scalar singlet has been fixed at its upper limit $V_{\chi}=v_{\rm
   ew}=246$ GeV. For lower allowed values of this VEV $V_{\chi}<
 v_{\rm ew}$, these couplings
 are enhanced by the scaling factor $v_{rm ew}/V_{\chi}$.}
\begin{center}
\begin{tabular}{ |c|c|c|c|c| } 
\hline
Domain Wall& Coupling& $i=1$  &  $i=2$  & $i=3$ \\ 
 \hline
YES &$h_i^R$ & $1.1\times 10^{-18} $  & $1.4\times 10^{-18}$ & $1.0\times 10^{-13}$\\ 
 \hline
 NO&$|Y_i^{\chi}|$ & $1.1\times 10^{-7}$  & $1.4\times 10^{-7}$ & $1.0
 \times 10^{-2}$ \\\hline 
NO&$h_i^{(eff,NR)}$& $1.1 \times 10^{-7}$& $1.1\times 10^{-7}$&$1.0\times
10^{-2}$ \\\hline
 NO & $|C_{G_i}|$&$1.1\times 10^{-4}$  & $1.4\times 10^{-4}$ & $1.0$ \\\hline
\end{tabular}
\label{tab_coup}
\end{center}
\end{table}

It is interesting to note that Type-I seesaw dominance in matter
parity conserving SO(10)
predicts decaying dark matter dynamics most generally by accommodating
all the three different types of light neutrino mass hierarchies and even
satisfying cosmological bound without or with priors in the QD cases. The QD2 type
solutions predict neutrinoless double beta decay rate close to the
current experimental limits.   
In the case of mixings solutions presented for IH, QD1, and QD2 type
of mass hierarchies and also for down-quark diagonal
basis, the corresponding Yukawa couplings can be estimated following
the same procedure. Once the light neutrino mass hierarchy is
determined in future, most of these Yukawa coupling predictions would converge
only to two alternative sets corresponding to up-quark or down-quark
diagonal bases. On the other hand if NH or IH type hierarchy is
confirmed, then the benchmark model solutions would be either ruled
out or revised.

%%%%%%%%%%%%%%%%%%%%%%%%%%%%%%%%%%%%%%%%%%%%%%%%%%%%%%%%%%%%%%%%
\section{Gauge Coupling Unification}\label{sec:gcu}
As pointed out in previous sections, two Higgs representations ${210}_H,{126}_H$ have been shown to define a SUSY SO(10) GUT model \cite{aulakh-gs-vissani} with minimal number of $26$ parameters. In this section we show that in the direct
non-SUSY SO(10) breaking to SM driven by these two representations, either ${210}_H$ or ${126}_H$ is capable of supplying just one scalar submultiplet of intermediate mass to complete minimal modification of the grand desert as discussed below in case of Model-I and Model-II.

\subsection{Unification in the Minimal Model-I}
It is interesting to see how the present minimal model generating DM
mass gives rise to gauge coupling unification of the standard gauge
theory by exercising utmost economy on the choice of lighter fields to
populate the grand desert by just one nonstandard scalar submultiplet $\kappa (3,0,8)
\subset {210}_H$. Although such type of model was suggested briefly for
coupling unification \cite{kynshi-mkp:1993},
the input parameters used
at that time were not as accurate as available now
\cite{PDG:2012,PDG:2014,PDG:2016}
\ba   
\alpha_S(M_Z)&=& 0.1182\pm 0.0005, \, \nonumber\\
\sin^2\theta_W (M_Z)&=&0.23129\pm 0.00005,\,\nonumber\\
\alpha^{-1}(M_Z)&=& 127.94 \pm 0.02.\, \label{eq:inputpara}
\ea
 Further there were no neutrino oscillation data or information on dark
matter available at that
time to establish natural high scale type-I seesaw dominance in this model.
 Also no connection with matter parity conservation, or DM candidates or their masses were discussed in the GUT
framework. Neither was there any justification in favour of
 ${210}_H$ as the mediator of non-standard Yukawa origin of decaying DM mass.
  The choice of $\kappa
(3,0,8)$ at intermediate scale was purely from curiosity to achieve
unification. The experimental bound on proton lifetime has increased almost
by more than one order over the years from 1993 till date that calls for estimation of
uncertainties in the model predictions  accurately. Apart from embedding the
IceCube DM, our other motivation is to see how far the present grand
unification framework can be constrained by the ongoing search
experiments on proton lifetime  in near future
\cite{JCP:2017}.   It is quite interesting to note that the DM fermion
representation ${45}_F$ which has given prominent threshold effects
elsewhere \cite{pnsa:2016} has exactly vanishing contribution in the
present case (Model-I).

In addition, in this
work we have
estimated GUT threshold effects under partially degenerate assumption
which states that the superheavy masses belonging to the same SO(10)
representation are degenerate in masses \cite{rnm-mkp:1993,lmpr:1995}. We find
that under this constraint, an attractive region of parameter
space requires the inclusion of threshold effects of superheavy gauge
bosons in the adjoint representation ${45}_V$ with their masses only few
times different from $M_U^0$ \cite{mkp:1987}. 
  
 We found in Sec.\ref{sec:embed} that the Higgs
representation ${210}_H$ plays two crucial roles in the GUT symmetry breaking as well as generating the desired
dark matter mass that decays to produce PeV energy IceCube
neutrinos.  
We fine tune the parameters of the GUT scale Lagrangian in such a way
that only the mass of component $\kappa (3,0,8)\subset {210}_H$  is substantially lighter than the GUT scale
while keeping all other superheavy component masses of 
${210}_H$ near the GUT scale.

\begin{figure}[h!]
\begin{center}
\includegraphics[scale=0.55]{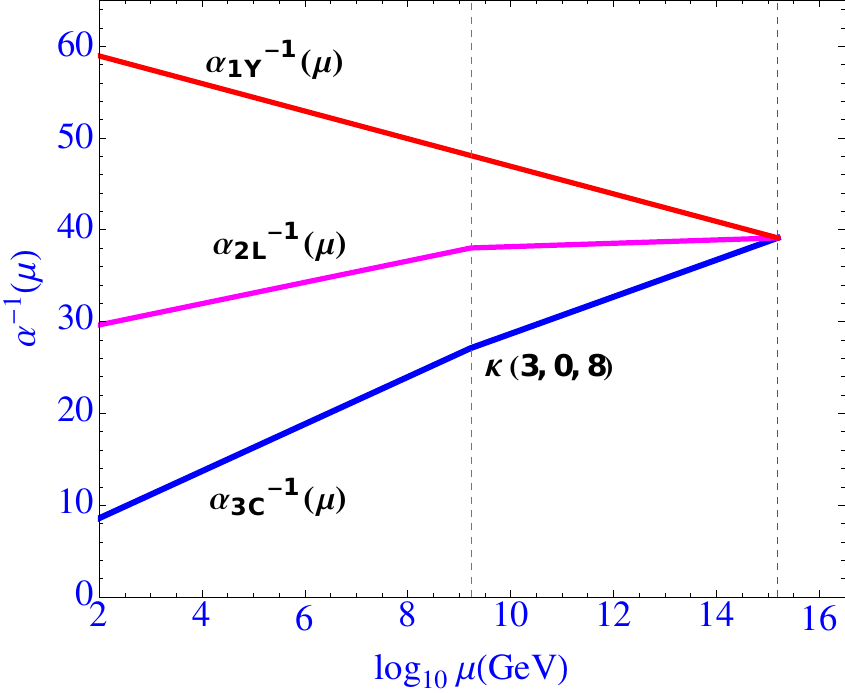}  
\caption{Unification of SM gauge couplings in the presence of $\kappa(3,0,8)$ as discussed in the text. The vertical dashed lines represent the mass scales
$M_{\kappa}=10^{9.23}$GeV and $M_{U}=10^{15.2446}$GeV, respectively. }
\label{fig:gcu1}
\end{center}
\end{figure}

Using the contributions of SM particles and the scalar $\kappa (3,
0,8)$ though the renormalization group (RG)
equations \cite{GQW}, the unification of 
gauge couplings is shown in Fig. \ref{fig:gcu1}.
%------------------------------\\
%------------------------------\\
%-------------------------------\\
%-------------------------------\\

 The unification of three  gauge couplings of SM at one-loop level
 is consistent with
\ba
M^0_U  &=& 10^{15.2+0.0446 }\, {\rm GeV},  \nonumber\\
M^0_{\kappa} &=& 10^{9.23}\, {\rm GeV}, \nonumber\\
\alpha^{-1}_G &=& 41.77. \, \label{eq:massscaleskappa}  
\ea
The quantity $+0.0446$ in the exponent is due to the matching of the
SM coupling constants at the GUT scale that occurs even if all the
superheavy particle masses are exactly degenerate with $M_U^0$
\cite{mkp:1987,Hall:1981,Weinberg:1980,Ovrut:1982}.  
It is clear from the Fig. \ref{fig:gcu1} that excellent unification of
 gauge couplings of the standard gauge theory below the GUT
scale without the assumption of any other
intermediate gauge symmetry has been achieved  with only one
non-standard Higgs field ($\kappa(3,0,8)$) having intermediate mass
$M_{\kappa}=10^{9.23}$ GeV. 
\subsection{Unification in  Minimal Model-II}
In this case while all nonsinglet scalar components of ${210}_H$ and all other
component masses in ${126}_H$ have masses at the GUT scale, only the component
$\eta (3,-1/3, 6)\subset {126}_H$ is at $M_{\eta}=10^{10.73}$ GeV. Similar to
eq.(\ref{eq:massscaleskappa}), the RG solutions in this case are
\ba
M^0_U  &=& 10^{15.24+0.0445 }\, {\rm GeV},  \nonumber\\
M^0_{\eta} &=& 10^{10.73}\, {\rm GeV}, \nonumber\\
\alpha^{-1}_G &=& 38.397 \, \label{eq:massscaleseta}  
\ea

\begin{figure}[h!]
\begin{center}
\includegraphics[scale=0.55]{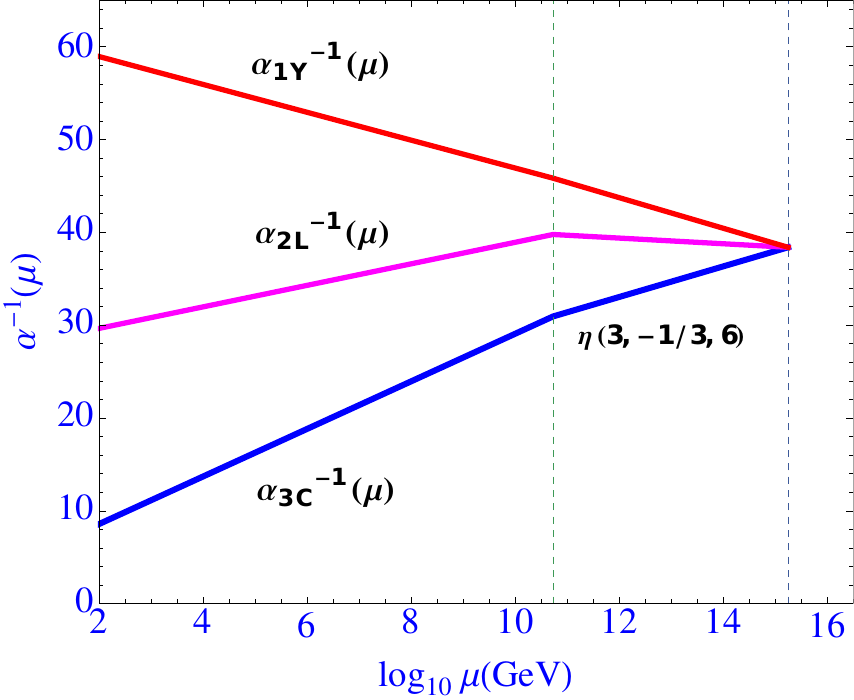}  
\caption{Unification of SM gauge couplings in the presence of $\eta(3,-1/3,6)$ as discussed in the text. The vertical dashed lines represent the mass scales
$M_{\eta}=10^{10.73}$GeV and $M_{U}=10^{15.28}$GeV, respectively. }
\label{fig:gcu2}
\end{center}
\end{figure}

Excellent unification of SM gauge couplings is shown in Fig.\ref{fig:gcu2}.    
%------------------------------\\
%------------------------------\\
%-------------------------------\\
%-------------------------------\\

\section{Proton Lifetime Prediction}\label{sec:taup}
\subsection{Decay Rate}
Currently the measured value on the lower limit of the proton  life
time for the decay modes $p\to e^+\pi^0$ and $p\to \mu^+\pi^0$ are
 \cite{Shiozawa:2014,Abe:2017}
\begin{eqnarray}
&&\tau_p^{expt.}(p\to e^+\pi^0)~\ge ~1.6\times 10^{34}~~{\rm
    yrs.},\nonumber\\
&& \tau_p^{expt.}(p\to \mu^+\pi^0)~\ge ~7.7\times 10^{33}~~{\rm
    yrs.}.\label{taupexpt}        
\end{eqnarray}
We investigate our model capabilities to account for this lower limit.
\subsection{Analytic Formulas for Decay Width}
 Including strong and electroweak renormalization
effects on the ${\rm d}=6$ operator and taking into account quark mixing, chiral symmetry breaking
effects, and lattice gauge theory estimations, the decay rates
  are \cite{Babu-Pati:2010,Buras:1978,Bajc,Nath-Perez}, 
\be  
\Gamma(p\rightarrow e^+\pi^0)
=\left(\frac{m_p}{64\pi f_{\pi}^2}
\frac{{g_G}^4}{{M_U}^4}\right)|A_L|^2|\bar{\alpha_H}|^2(1+D'+F)^2\times R,
\label{width}
\ee
where $ R=[A_{SR}^2+A_{SL}^2 (1+ |{V_{ud}}|^2)^2]$ for $SU(5)$, but $R=
[(A_{SR}^2+A_{SL}^2) (1+ |{V_{ud}}|^2)^2]$ for $SO(10)$, $V_{ud}=0.974=$ 
 the  $(1,1)$ element of $V_{CKM}$ for quark mixings, and
$A_{SL}(A_{SR})$ is the short-distance renormalization factor in the
left (right) sectors.  In eq.(\ref{width}) $A_L=1.25=$
long distance renormalization factor but  
$A_{SL}\simeq A_{SR}=2.542$. These are numerically estimated by
evolving the ${\rm dim.} 6$ operator for proton decay by using the
anomalous dimensions of ref.\cite{Buras:1978} and the beta function
coefficients for gauge couplings of this model. In eq.(\ref{width})  
 $M_U=$ degenerate mass of  superheavy gauge bosons, $\bar\alpha_H =$
hadronic matrix elements, $m_p =$proton mass
$=938.3$ MeV, $f_{\pi}=$ pion decay 
constant $=139$ MeV, and the chiral Lagrangian parameters are $D=0.81$ and
$F=0.47$. With $\alpha_H= \bar{\alpha_H}(1+D'+F)=0.012$ GeV$^3$ estimated from 
lattice
gauge theory computations  \cite{Aoki:2007,Munoz:1986}, we obtain  $A_R \simeq A_LA_{SL}\simeq
A_LA_{SR}\simeq 2.726$ and the expression for the
 inverse
decay rate is
\begin{equation}
\Gamma^{-1}(p\rightarrow e^+\pi^0) 
 =
  \frac{4}{\pi}\frac{f_{\pi}^2}{m_p}\frac{M_U^4}{\alpha_G^2}\frac{1}{\alpha_H^2
    A_R^2}\frac{1}{F_q},\label{taup}
\end{equation}
where the GUT-fine structure constant $\alpha_G=0.0263$ and the
factor  $F_q=2(1+|V_{ud}|^2)^2\simeq 7.6$ have been used for $SO(10)$. 
This formula has 
the same form as given in \cite{Babu-Pati:2010}.
\subsection{Analytic Formula for Threshold Effects}
In the single step breaking models discussed in this work, GUT threshold effects
due to superheavy degrees of freedom in different SO(10) representations  are
expected sources of  major
 uncertainties on unification scale and  proton
lifetime prediction.  The underlying origin of threshold
effects due to smaller quantum corrections proposed in
\cite{Weinberg:1980,Hall:1981,Ovrut:1982}   has been also
addressed in SO(10)
\cite{rnm-mkp:1993,lmpr:1995,mkp:1987},
and more recently in \cite{pnsa:2016}. Details have been also given in
the Appendix which yield the following corrections arising from
different superheavy particles in the loops.
 Further we have estimated the threshold uncertainties following the
 partially degenerate assumption  introduced in \cite{rnm-mkp:1993,lmpr:1995,RNM-PLB:1993}
 which states that the superheavy components belonging to the same GUT
 representation are degenerate with  the same superheavy scale
 around $M_U$. A new expected source of threshold uncertainty is due
 to  fermion representation ${45}_F$. In Model-I and Model-II
 discussed here, we investigate possible cancellations in  reducing
 threshold uncertainties. \\

Noting that the superheavy scalars, fermions, and gauge bosons
contribute through small log evolutions and defining $\eta_j^{\prime}=\log_{10}(M_j/M_U)$, we have the following formula for GUT threshold effects

\ba 
\frac{M_U}{M_U^0}&=& 10^{\bf\pm C_{\rm input}} \, ({\bf{\rm input\, parameters}})\,\nonumber\\
&&\times 10^{\bf\pm C_{(126)}\eta_{(126)}^{\prime}}\,({\bf{\rm superheavy\, scalars\, in}\, {126}_H})\,\nonumber\\
&&\times 10^{\bf\pm C_{(210)}\eta_{(210)}^{\prime} }\,({\bf{\rm superheavy\, scalars\, in}\, {210}_H})\,\nonumber\\
&&\times 10^{\bf\pm C_{(45)}\eta_{(45)}^{\prime} }\,({\bf{\rm superheavy\, scalars\, in}\, {45}_H})\, \nonumber\\
&&\times 10^{\bf\pm C_{(16)}\eta_{(16)}^{\prime} }\,({\bf{\rm superheavy\, scalars\, in}\, {16}_H})\, \nonumber\\
&&\times 10^{\bf \pm C_{10}\eta_{(10)}^{\prime} }\,({\bf {\rm superheavy\, scalars\, in}\, {10}_H})\,\nonumber\\ 
&&\times 10^{\bf \pm C_{V}\eta_{V}^{\prime} }\,({\bf{\rm superheavy\, gauge\, bosons\,in }\, {45}_V})\,\nonumber\\
&&\times  10^{\bf \pm C_{F}\eta_{F}^{\prime} }\,({\bf{\rm superheavy\, fermions\, in}\, {45}_F}),
\label{eq:errors}
\ea
where $M_i(i={10}_H,{16}_H, {45}_H,.....) $ is the respective degenerate superheavy particle mass in the  SO(10) representation  \cite{rnm-mkp:1993}. 
In  eq.(\ref{eq:errors}) the first line represents $3\sigma$ uncertainty of input values of
 $\sin^2\theta_W (M_Z)$, $\alpha_S (M_Z)$ and $\alpha (M_Z)$ \cite{PDG:2012,PDG:2014,PDG:2016} given in eq.(\ref{eq:inputpara}). Other  contributions
represent superheavy particle contributions from respective representations associated with  spontaneous symmetry breaking of gauge
symmetries and also in the generation of DM mass including its mixing with RH$\nu$.

As shown in Appendix we have derived threshold
corrections by providing the longitudinal modes of superheavy gauge
bosons from ${210}_H$ and ${126}_H$. The relevant superheavy components of scalars can be found from representations given in Table \ref{tab:decomp}. The decompositions of ${54}_H$ whose VEV contributes to DM or color octet fermion mass through eq.(\ref{eq:compmass}) has been skipped from Table  \ref{tab:decomp} as its nonsinglet components give vanishing threshold effects  in Model-I and Model-II. GUT threshold effects due to ${45}_H$ whose SM singlet component $S_H$ enters into all the Feynman diagrams of Sec.4 can be estimated from the decomposition given for ${45}_F$ but using the small log evolution formula for scalars given in the Appendix. Similarly superheavy gauge boson threshold effects have been estimated in Model-I and Model-II.  

\subsection{Predictions in the Minimal Model-I} 

In this case our estimated values of coefficients occurring in eq.(\ref{eq:errors}) are
\ba
C_{\rm input}&=&0.110, \nonumber\\
 C_{(10)}&=& -0.0196, \nonumber\\
C_{(126)}&=&-0.03743,\, C_{(210)}=0.0322,\, \nonumber\\
C_{V}&=&-0.9358,\, C_{(16)}=0.0,\,   \nonumber\\
C_{(45)}&=&0.0,\,    C_{(54)}=0.0.\, \nonumber\\
C_{F}&\simeq &0.0.                  \label{eq:thcoef1} 
\ea
The coefficients $C_{(16)},C_{(45)}$, and $C_{F}$ due to superheavy
components of Higgs representations ${16}_H,{45}_H$ and the fermion
representation ${45}_F$ in the partially degenerate case vanish 
\cite{RNM-PLB:1993,rnm-mkp:1993,Weinberg:1980,Hall:1981,Ovrut:1982,mkp:1987}. 
  
 The estimated GUT scale in the partially degenerate case turns out to
 be
\be
M_U=10^{15.2446\pm 0.1100 \pm 0.089275|\eta_S|\pm 0.9358|\eta_V|}.
\label{eq:degMU}
\ee
This leads to the proton lifetime 
\begin{eqnarray}
&&\tau_p^{SO(10)}\simeq  10^{32.940\pm 0.440 \pm 0.3571|\eta_S|\pm
    3.743|\eta_V|}~~{\rm yrs}.\label{taupnum}
\end{eqnarray}
It is interesting to note that, despite the two large Higgs representations ${210}_H, {126}_H$, the nonstandard fermion representation ${45}_F$, and other Higgs representations ${45}_H,{16}_H, {10}_H$,  major contribution to threshold uncertainty in Model-I is only due to superheavy gauge bosons. The superheavy Higgs boson threshold effect that acts as a major source of uncertainty
in intermediate scale models \cite{lmpr:1995} is much smaller   and
the fermion threshold contribution is absent in this direct symmetry
breaking model.

Numerical estimations on proton lifetime  for Model-I are shown in Table
\ref{tab:taupdeg1} for different splitting factors of superheavy masses. 
%--------------------------------------
%-------------------------------------
\begin{table}[h!]
\caption{ Upper limits on predicted proton lifetime in Model-I as a function of superheavy scalar (S) and gauge boson
  (V) mass splittings as defined in the text and Appendix . The factor
  $10^{\pm 0.44}$ represents  uncertainty due to input parameters.}
\centering
\begin{tabular} {|p{1.5 cm} |p{1.5 cm }|p{2.5 cm}|p{1.5 cm }|p{1.5 cm}|p{2.5 cm}|}  
 \hline
 ${M_S \over M_U} $ & ${M_V \over M_U}$& $\tau_P (yrs)$ & ${M_S \over M_U} $
 & ${M_V \over M_U} $ & $\tau_P (yrs)$   \\ \hline
 $10$ & $2$ & $2.65\times10^{34\pm 0.44}$ & $5$ & $5$ & $6.4\times10^{35\pm 0.44}$ \\
\hline
 $10$ & $3$ & $1.21\times10^{35\pm 0.44}$ & $2$ & $5$ & $4.61\times10^{35\pm 0.44}$ \\ 
\hline
 $5$ & $3$ & $9.45\times10^{34\pm 0.44}$ & $2$ & $4$ & $2\times10^{35\pm 0.44}$ \\ 
\hline
 $5$ & $4$ & $2.77\times10^{35\pm 0.44}$ & $2$ & $3$ & $6.81\times10^{34\pm 0.44}$ \\
\hline
% $10$ & $2$ & $1.5\times10^{35}$ & $3$ & $4$ & $7.5\times10^{35}$ \\ 
%\hline
\end{tabular} 
\label{tab:taupdeg1}
\end{table}
%-------------------------------------------

\subsection{Predictions in the Minimal Model-II}

In this case the coefficients of eq.(\ref{eq:errors}) are
\ba
C_{\rm input}&=&0.1334, \nonumber\\
C_{(10)}&=& -0.02024,~~C_{(126)}=-0.044534, \nonumber\\
C_{(16)}&=&0.0,~~ C_{(210)}=0.0809717,  \nonumber\\
C_{V}&=&-0.9352, ~~C_{(45)}=0.0. \nonumber\\
C_{(54)}&=&0.0,~~C_{F}\simeq 0.0.\,~~ \label{eq:thcoef}
\ea

  With the partially degenerate assumption that the superheavy masses of different components of a given SO(10) representation have separately degenerate masses, or with the complete degeneracy assumption of identical masses for all superheavy scalars,  maximising the threshold uncertainties gives the following results.\\
\subsection{Lifetime with Partially Degenerate Superheavy Scalar}

\be
M_U=10^{15.28\pm 0.1334 \pm 0.1457|\eta_S^{\prime}|\pm 0.9352|\eta_V^{\prime}|},\label{eq:pdegMU2}
\ee
\begin{eqnarray}
&&\tau_p^{SO(10)}\simeq  10^{33.11\pm 0.5335 \pm
    0.5828|\eta_S^{\prime}|\pm 3.74|\eta_V^{\prime}|}~~{\rm yrs}.\label{taupnum-pdg}
\end{eqnarray}
In this  case 
predictions on proton lifetime  are presented in Table \ref{tab:taupdeg2}.
%----------------------------------
%----------------------------------
\begin{table}[h!]
\caption{Upper limits on predicted proton lifetime in Model-II as a function of
  superheavy scalar (S) and gauge boson (V) mass splittings  defined
  in the text and Appendix.The factor $10^{\pm 0.53}$ represents  uncertainty due to input parameters.}
\centering
\begin{tabular} {|p{1.5 cm} |p{1.5 cm }|p{3 cm}|p{1.5 cm }|p{1.5 cm}|p{3 cm}|}  
 \hline
 ${M_S \over M_U} $ & ${M_V \over M_U}$& $\tau_P (yrs)$ & ${M_S \over M_U} $
 & ${M_V \over M_U} $ & $\tau_P (yrs)$   \\ \hline
 $5$ & $2$ & $4.469\times10^{34\pm 0.53}$ & $2$ & $3$ & $1.19\times10^{35\pm 0.53}$ \\
\hline
 $8$ & $2$ & $5.87\times10^{34\pm 0.53}$ & $3$ & $3$ & $1.51\times10^{34\pm 0.53}$ \\ 
\hline
 $10$ & $2$ & $6.693\times10^{34\pm 0.53}$ & $5$ & $3$ & $2.03\times10^{35\pm 0.53}$ \\ 
\hline
 $1$ & $3$ & $7.97\times10^{34\pm 0.53}$ & $1$ & $4$ & $2.33\times10^{35\pm 0.53}$ \\
\hline
% $10$ & $2$ & $1.5\times10^{35}$ & $3$ & $4$ & $7.5\times10^{35}$ \\ 
%\hline
\end{tabular} 
\label{tab:taupdeg2}
\end{table}
%------------------------------------\\
%------------------------------------\\ 

Including uncertainty due to input parameters and assuming all
superheavy fermion and gauge boson masses identical to $M_U$, we find that this model predicts proton lifetime up to $3\times 10^{34\pm 0.58}$ yrs
($ 8\times 10^{34\pm 0.58}$ yrs) for degenerate (partially degenerate) superheavy
scalar masses  which is accessible to ongoing proton decay searches.\\

\section{Vacuum Stability of the Scalar Potential}\label{sec:stab} 
While explaining dynamical origin of $N_i-\Sigma_F$ mixings, we have
found that Planck-scale assisted spontaneous symmetry breaking
provides an attractive mechanism through the smaller VEV $V_{\chi}\leq v_{\rm ew}$ 
of a matter parity odd Higgs singlet naturally present in the Higgs
representation ${16}_H^{\dagger}$ of SO(10). Because of its order
$v_{\rm ew}$ VEV, this Higgs scalar is predicted to be light with
perturbative upper bound on its mass
$\leq 860$ GeV. We discuss below how such a nonstandard 
Higgs singlet scalar resolves the issue of vacuum instability in
Model-I and Model-II discussed in the previous sections. 
It is well known that the SM Higgs potential\footnote{It is to be noted that adding a constant term to the potential and using the minimization condition
$\mu^2=\lambda_\phi v_{\rm ew}^2$,  the potential can be rewritten in a
  convenient form as $V_{sm}=\lambda_\phi[(\phi^\dagger
    \phi)-\frac{v_{\rm ew}^2}{2}]^2$ where we 
have omitted the constant term which does not affect the equation motions. It is evident from the above expression of potential that the minimum is at 
$\langle \phi^\dagger \phi \rangle=\frac{v_{\rm ew}^2}{2}$.}
\begin{equation}
V_{\rm SM}=-\mu^2\phi^{\dagger}\phi+\lambda_{\phi}
\left(\phi^{\dagger}\phi\right)^2\,\, \label{eq:smpot}
\end{equation}
develops instability due to radiative corrections as the quartic
coupling becomes negative for larger values of the Higgs field
$\Lambda_I\sim |\phi| \ge 2\times 10^9$ GeV. One popular solution to this
vacuum instability has been suggested through the introduction of
additional scalar field(s) of mass below the instability scale 
$\Lambda_I$ \cite{Elias-Miro:2012,Lebedev:2013,Lebedev:2015}. These may be scalar singlet candidates
corresponding to WIMP \cite{Garg:2017,gond:2010,chen:2012,khan:2014}
or decaying scalar DM manifesting
through PeV scale IceCube neutrinos \cite{IceCube-Models,Mambrini-IceCube}. 
\vspace{5mm}
\paragraph{}
We have shown in Sec.\ref{sec:mix} that dynamical origin \cite{MKP-ARC:2010}  of
$N_i-\Sigma_F$ mixings predicts a non-standard Higgs scalar singlet
$\chi_S(1,0,1)$ with VEV $V_{\chi}\le v_{\rm ew}=246$ GeV.
To examine how the presence of this $\chi_S$ affects the evolution of
the standard Higgs quartic coupling, we consider additional
contributions to $V_{\rm SM}$ due to $\chi_S$ field
modified scalar
potential in the presence of $\phi$ and $\chi_S$
\begin{equation}
%V_{\phi\chi}=M_{S_H}^2/2S_H^2+\lambda_{S_H}S_H^4+\lambda_{\phi S_H)\phi^{\dagger}\phi S_H^2\,\,\label{eq:SHcorr}  
V_{\phi-\chi}=-\mu_s^2 \chi_S^\dagger \chi_S +\lambda_\chi (\chi_S^\dagger\chi_ S)^2 + 2 \lambda_{\phi \chi} (\phi^\dagger \phi)(\chi_S^\dagger \chi_S) 
\end{equation}
After using the minimization conditions for both scalar fields $\phi$
and  $\chi_S$, the entire potential $V= V_{\rm SM}+V_{\phi-\chi}$  can be written in a convenient
form   
\begin{equation}
V=
\lambda_\phi\left[(\phi^\dagger \phi)-\frac{v_{\rm ew}^2}{2}\right
  ]^2 +\lambda_\chi\left[(\chi_S^\dagger
    \chi_S)-\frac{V_{\chi}^2}{2}\right]^2 
+ 2\lambda_{\phi \chi}\left(\phi^\dagger \phi -\frac{v_{\rm
      ew}^2}{2}\right)
\left(\chi_S^\dagger \chi_S -\frac{V_{\chi}^2}{2}\right) \label{v_sh}
\end{equation}
where $V_{\chi}$ is the VEV of the newly added scalar singlet. For $\lambda_\phi,\lambda_\chi>0$ and $\lambda_{\phi \chi}^2<\lambda_\phi \lambda_\chi$, the minimum for the 
total potential in eq.(\ref{v_sh}) is given by $\langle \phi^\dagger \phi \rangle=\frac{v^2}{2}$ and $\langle \chi_S^\dagger \chi_S \rangle=\frac{V_{\chi}^2}{2}$.
To know about the high scale behavior of the Higgs quartic coupling $(\lambda_\phi)$ we have to solve its renormalization group (RG) equation
which has been modified due to addition of the singlet scalar field $(\chi_S)$. Actually the RG equation of $\lambda_\phi$ is 
a coupled first order differential equation which involves the quartic coupling of the singlet scalar $(\lambda_\chi)$, the coupling
of the interaction term $(\lambda_{\phi \chi})$, the gauge couplings
$(g_{2L},g_{1Y}, g_{3C})$ and dominant Yukawa coupling  $h_t$ due to
the top quark. We have considered one loop RG equations for the scalar
quartic couplings shown below. Two loop equations 
for gauge couplings and top quark Yukawa coupling are discussed in
 Appendix B of Sec.\ref{sec:RGE}. 
\begin{eqnarray}
&&\frac{d \lambda_{\phi }}{d \ln \mu}= \frac{1}{(4\pi)^2}\left[ ( 12
    y_{\rm top}^2 -3 {g_{1Y}}^2 -9g_{2L}^2) \lambda_\phi -6 y_{\rm top}^4 +
\frac{3}{8} { 2 g_{2L}^4 +(g_{1Y}^2 +g_{2L}^2)^2 } +24 \lambda_\phi^2 + 4 \lambda_{\phi \chi}^2 \right]~,\nonumber\\
%%%%%%%%%%%%%%%%%%%%%%%%%
&&\frac{d \lambda_{\phi \chi}}{d \ln \mu}=
  \frac{1}{(4\pi)^2}\left[\frac{1}{2}(12y_{\rm top}^2 -3 g_{1Y}^2 -9 g_{2L}^2)\lambda_{\phi \chi}+
4\lambda_{\phi \chi} (3 \lambda_\phi +2 \lambda_\chi) + 8 \lambda_{\phi \chi}^2 \right]~, \nonumber\\
%%%%%%%%%%%%%%%%%%%%%%%%%
&&\frac{d \lambda_{ \chi}}{d \ln \mu}= \frac{1}{(4\pi)^2}\left[ 8 \lambda_{\phi \chi}^2 + 20 \lambda_\chi^2 \right]~.\label{eq:qrge}
\end{eqnarray}
    As in our Model-I and Model-II, the Dirac neutrino mass matrix has
    been assumed to be same as that of the up quark mass matrix and
 the most dominant contribution affecting RG evolution of
 $\lambda_{\phi}(\mu)$ for mass scales $\mu > M_{N_3}$ is the element
 $Y_{33}={(M_D)}_{33}/v_{\rm eW}$.
Thus the equation governing the evolution of $\lambda_\phi$ will have
a new additional term $(4 Y_{33}^2 \lambda_\phi -2 Y_{33}^4)$ in the
RHS of of the first of eq.(\ref{eq:qrge}).
Similarly the new term to be added in the RHS of the $\lambda_{\phi \chi }$ equation is $2Y_{33}^2 \lambda_{\phi \chi}$. 

%%%%%%%%%%%%%%%%%%%%%%%%%%%%%%%%%%%%%%%%%%%%%%%%%%%%%
\subsection{Vacuum Stability with Spontaneously Broken Matter Parity} 
We now show how  the presence of the Higgs scalar singlet
$\chi_S(1,0,1)\subset{16}_H^{\dagger}$ carrying odd
matter parity and having perturbative mass upper bound  $M_{\chi_S}< 860$ GeV, which has been predicted to explain
the dynamical origin of extremely small value of the N-$\Sigma_F$ mixings, resolves the
issue of vacuum instability of the SM Higgs potential. All the results
derived so far and others to follow for a real scalar singlet which
could be  either the real or
imaginary part of $\chi_S(1,0,1)$. At the end of Sec.\ref{sec:vchi} we
show that if any one of these two component masses is made as light as the
electroweak scale by allowed fine tuning of the parameters of the relevant
scalar potential, the other component automatically acquires GUT scale
mass. As a result only the light scalar singlet component of $\chi_S(1,0,1)$ modifies the
standard Higgs quartic coupling contributing to the resolution of
vacuum stability problem while the heavy mass decouples from making any such contribution.     

\subsubsection{Higgs Doublet-Singlet Mixing and LHC Constraints}
It is clear from the expression of the potential (eq.(\ref{v_sh})) that the ordinary SM scalar doublet $(\phi)$ and the newly introduced scalar singlet $(\chi_S)$
mix through the $\lambda_{\phi \chi}$ term due to which their masses also get modified little bit. Now our primary task is to diagonalize the mass matrix 
  of the scalars which enables us to find the mass eigenvalues of the SM Higgs like state, singlet like state 
and the mixing angle between them. The mass of the SM Higgs like state
is around $125$ GeV where as that of the singlet like state is unknown. We 
have to analyze the phenomenological implications of this singlet scalar at LHC. Before using a certain value of the mixing angle and mass of the 
new scalar in our RG running analysis, we have to check the compatibility of our chosen set with present LHC data.
\paragraph{}
The mass matrix of the two scalars in $(\phi,\chi)$ basis can be
derived from the expression of the potential  in eq.(\ref{v_sh}) 
\begin{equation}
\mathcal{M}^2 =2 \begin{pmatrix}
                \lambda_\phi v_{\rm ew}^2 & \lambda_{\phi \chi} v_{\rm
                  ew}V_{\chi} \\
                \lambda_{\phi \chi} v_{\rm ew} V_{\chi} & \lambda_\chi V_{\chi}^2
                \end{pmatrix} ~.
\end{equation}
Without any approximation this matrix is diagonalized with the 2 dimensional orthogonal rotation matrix and the mixing angle turns out to be 
\begin{equation}
\theta= \frac{1}{2} \tan ^{-1}\left( \frac{2 \lambda_{\phi \chi}
  v_{\rm ew} V_{\chi}}{\lambda_\phi v_{\rm ew}^2 -\lambda_\chi V_{\chi}^2 } \right)~. 
\end{equation}
Denoting the  mass eigenvalue of the SM Higgs like state as $m_1$ and singlet scalar
like state as  $m_2$, we have
\begin{eqnarray}
&&m_1=2(\lambda_\phi v_{\rm ew} ^2 \cos^2\theta + \lambda_\chi
  V_{\chi}^2 \sin^2 \theta +\lambda_{\phi \chi} v_{\rm ew} V_{\chi} \sin 2\theta) \nonumber\\
&&m_2=2(\lambda_\phi v_{\rm ew}^2 \sin^2\theta + \lambda_\chi
  V_{\chi}^2 \cos^2 \theta -\lambda_{\phi \chi} v_{\rm ew} V_{\chi} \sin 2\theta)~.
\end{eqnarray}
Among $8$ parameters in the set
$(m_1,m_2,\lambda_\phi,\lambda_\chi,\lambda_{\phi
  \chi},v_{\rm eW},V_{\chi},\theta)$ only two are known: the SM Higgs mass $m_1 \sim 125$
GeV, and the SM VEV $v_{\rm ew}=246$ GeV.
Matter parity conservation down to the electroweak scale introduces a upper limit on VEV of $\chi_S$,
 $V_{\chi} \leq v_{\rm ew}=246$ GeV. The other parameters are constrained
from different measurements performed at LHC
\cite{Lebedev:2015}.

\begin{figure}[h!]
\begin{center}
\includegraphics[scale=0.5,angle=0]{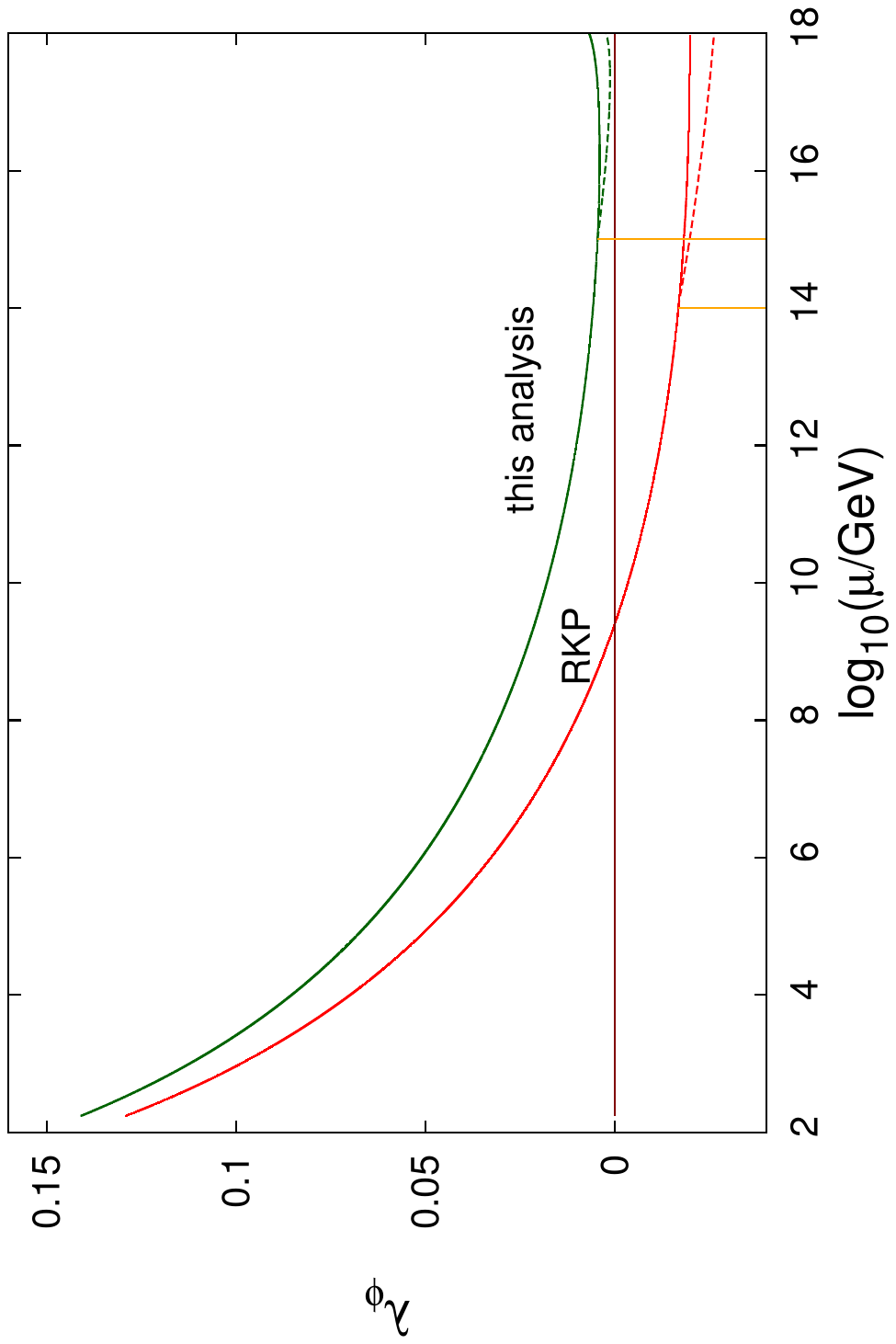}  
\caption{ Resolution of vacuum instability problem  by the predicted second Higgs scalar $\chi_{S}$ of mass $M_{\chi_S}\sim 177$ GeV carrying 
odd matter parity. The evolution of standard Higgs quartic coupling in
the presence of $\chi_S$ modifications is denoted by the upper green
curve and the dotted line shows the
effect of Dirac neutrino Yukawa modification for $M_{N_3}=10^{15}$ GeV. The lower red curve
marked ``RKP'' represents the quartic coupling evolution in the
benchmark model for $M_N=10^{14}$ GeV.} 
\label{fig:chis}
\end{center}
\end{figure}

We now discuss  briefly about the experimental constraints on 
singlet scalar mixing parameters quoted above. In this model the constraints come from three different kinds of measurements: (i) Electroweak precision data,
(ii) Higgs coupling measurements, and  (iii) Different searches for a Higgs like scalar.
\paragraph{}
In the model under consideration, the electroweak observables are mainly modified due to the new one loop contributions to the $W$ and $Z$ propagators.
These new contributions arise due to (i) loop diagram with the singlet scalar, (ii) modification of the coupling of the SM Higgs with the gauge bosons.
Incorporating those corrections the shifts in the electroweak parameters are calculated. A global $\chi^2$ \footnote{The $\chi^2$ is a function of scalar
singlet mass, mixing angle between scalar singlet and SM doublet and other known parameters of SM} analysis can be carried out to get a exclusion plot for 
$m_2$ vs $\theta$. Sizable constraint comes for $m_2 \leq 60$ GeV and $m_2 \geq 170 $ GeV and strongest limit is obtained when $m_2 \geq 450$ GeV.
\paragraph{}
The presence of the singlet scalar or in other words the mixing between the newly added singlet scalar with the ordinary SM Higgs, modifies the coupling
of the SM Higgs with other fermions and gauge bosons. Again these couplings are involved in computations of several decay widths which are observable 
at LHC. Taking only $\gamma \gamma$ or $4l$ as final states and with the consideration that singlet scalar mass is outside $[120-130]$ GeV a combined
constraint on the singlet mass and mixing angle $\theta$ is obtained as 
\begin{equation}
\sin \theta < 0.44 ~~{\rm at}~ 95\% ~{\rm CL} 
\end{equation}
for 
\begin{equation}
m_2 \geq \frac{m_1}{2} ~(62.5)~(\rm GeV) ~{\rm and}~ m_2 \notin[120,130]~{\rm GeV}.
\end{equation}
When $m_2$ becomes smaller than $m_1/2$, then the decay channel $\phi\rightarrow \chi_S\chi_S$ opens up leading to a stringent bound on $\theta$ which is
strongly dependent on the coupling responsible for $\phi,~\chi_S$ mixing i.e, $\lambda_{\phi \chi}$. It has been observed that for larger values of $\lambda_{\phi \chi}$
almost whole $m_2-\theta$ parameter space is ruled out.
\paragraph{}
Similar kinds of constraints on $m_2-\theta$ parameter space can be obtained from search of different possible direct decay channels of the singlet state.
\paragraph{}
A combined exclusion plot for $m_2-\theta$ parameter space can be drawn taking into account all three types of measurements discussed above. A complete
analysis and the corresponding exclusion plot can be found in Ref.\cite{Lebedev:2015}. It is to be noted that for $m_2=(10-250)$ GeV \footnote{In our case we need not go beyond
$250$ GeV since the maximum value of the VEV of the scalar singlet is $246$ GeV. This in turn allows us to take $m_2$ at most $\sim$ $250$ GeV} $|\sin \theta| <0.1$
is allowed for any value of $m_2$ within these limits whereas higher values of the mixing angle $(\theta)$ are permissible in few pockets of mass ranges of which the regime 
relevant to our analysis is $m_2=(160-180)$ GeV with $|\sin \theta| <0.4$. 
\paragraph{}
In our actual numerical analysis we vary the unknown quartic couplings $(\lambda_\phi,\lambda_\chi,\lambda_{\phi \chi})$ over a wide range of values $(0.001-0.1)$. The 
VEV of $\chi_S$ ($V_{\chi}$) is varied from a small value upto $246$
GeV, whereas that of $\phi$ ($v_{\rm ew}$) is kept fixed at $246$ GeV and mass of the SM Higgs like state is 
taken to be around $125$ GeV. The highest allowed value of $\theta$ is different for different value of the mass eigenvalue $m_2$. This upper bound on $\theta$
is chosen from Fig.3 of Ref\cite{Lebedev:2015} and utilised appropriately for different values of the scalar singlet mass in our numerical analysis. For each allowed set of parameters
$(m_1,m_2,\lambda_\phi,\lambda_\chi,\lambda_{\phi \chi},v,V_{\chi_S},\theta)$ we analyze vacuum stability and perturbativity of the quartic couplings upto the Planck scale.
After repeating this exercise for replica of many such sets, it is found that the couplings loose their perturbativity much before the Planck scale if we take the 
initial value of $\lambda_\chi  \geq 0.3$. The problem of vacuum stability is not cured unless we take electroweak scale value of $\lambda_{\phi \chi}\geq 0.034$.
\begin{table}[!h]
\caption{Values of quartic couplings, VEVs, mixing angle,  SM  Higgs
  mass $(m_1)$ and scalar singlet $\chi_S$ mass $(m_2)$ at electroweak
  scale consistent with experimental constraints which predict vacuum stability. }
\begin{center}
\begin{tabular}{ |c|c|c|c|c|c|c|c| } 
\hline
 $\lambda_\phi$ & $\lambda_\chi$  & $\lambda_{\phi \chi}$ & $|\sin \theta|$ & $v$ (GeV)& $V_{\chi}$ (GeV)& $ m_1$ (GeV)& $m_2$ (GeV)   \\ \hline
 $0.141$ & $0.251$  & $0.037$ & $0.3$ & $246$ & $245$ & $125.25$ & $177.5$ \\ \hline
% cell7 & cell8 & cell9 & & &\\ \hline
\end{tabular}
%\end{table}
\label{para_sp}
\end{center}
\end{table}
To satisfy these two conditions simultaneously we have to allow $|\sin
\theta| \geq 0.3$ and from Fig.3 of \cite{Lebedev:2015}, it is clear that this value of mixing angle is only
allowed in the mass range $(160<m_2<180)$ GeV. The resolution for vacuum stability problem using one such set of parameters (as given in the Table \ref{para_sp}) is
shown in Fig.\ref{fig:chis} where we have also included the contribution of the Dirac neutrino Yukawa matrix which is shown by dotted lines.
For  SM  extension with type-I seesaw extension Dirac neutrino Yukawa
effect is shown at $10^{14}$ GeV as would be applicable to the bench
mark model \cite{Rott:2014} which has been identified as the curve RKP
in Fig.\ref{fig:chis}. Our model predictions of quartic coupling shown
as the upper curve in this figure naturally includes the Dirac
neutrino Yukawa affecting the RG evolution for $\mu > 10^{15}$ GeV. 
As shown in Fig.\ref{fig:chis}, the SM vacuum  is indeed stable upto
Planck scale when supplemented by modifications due to $h-\chi_S$
mixing even after Dirac neutrino Yukawa corrections are included. The
desired quartic coupling also lies well below the perturbative limit.

%----------------------------------\\
%-----------------------------------\\
%------------------------------\\
%------------------------------\\

It is pertinent to point out that vacuum stability in SO(10) with
$SU(2)_L\times SU(2)_R\times U(1)_{B-L}\times SU(3)_C$  intermediate
symmetry and matter parity conservation has been carried out through a
WIMP scalar DM near TeV scale \cite{Mambrini:2013}. In a different
interesting approach, the imaginary part of a complex scalar field
whose real part generates heavy RH$\nu$ mass for type-I seesaw has
been treated as the source of IceCube neutrinos \cite{Mambrini-IceCube}. A number of other
models have been also suggested for IceCube neutrino events \cite{IceCube-Models}.
In our present model WIMP DM as matter parity even non-standard Majorana fermionic
singlet originating from ${45}_F$ of SO(10) is also easily accommodated in
addition to the PeV scale decaying DM discussed here ensuring
coupling unification and verifiable proton lifetime.
%%%%%%%%%%%%%%%%%%%%%%%%%%%%%%%%%%%%%%%%%%%%%%%%%%%%%%%%%%%%%%%%%%%%%

\subsection{Low $V_{\chi}$ from Potential Minimisation}\label{sec:vchi}

In this section we show that potential minimisation of the full Higgs
potential can predict a low VEV $V_{\chi}=\langle Re(\chi_S) \rangle\le v_{ew}$
where we treat only the real part of $\chi_S$ to be as light as the
electroweak scale while keeping its imaginary part at the GUT
scale. In the last part of this section we have proved how such a wide gap can
be realized between the two component masses of $\chi_S$.\\

Instead of low VEV $V_{\chi}\le v_{ew}$, we have checked implications of
high $V_{\chi}\sim M_{GUT}$ assumption:(i) This violates matter parity of Lagrangian
predominantly by introducing additional large contribution to heavy
RH$\nu$ mass via Planck scale effects, thus contradicting the basic
assumption of this model that matter parity violating part is a small
perturbation over matter parity conserving part. (ii) Explanation of
$\Sigma_F-N_i$ mixings with $V_{\chi}=M_{GUT}$ not only requires much
smaller values of couplings than reported here, but also it opens up
at least one different induced contribution  for the dark matter decay
$\Sigma_F \to \nu h$  that destabilizes the RKP model
\cite{Rott:2014}.(iii) The mass of the scalar singlet $\chi_S$ being
 as high as the GUT scale does not solve the vacuum stability
problem through its Higgs portal interaction with the standard
Higgs. We thus search for the prediction of low $V_{\chi}$ in
 such direct SO(10) breaking to SM.
  
We have shown that the VEV $V_{\chi}\le v_{ew}=246$ GeV is necessary
for  the only SM singlet scalar component Re($\chi_S(1,0,1)$) contained in
${16}_H^{\dagger}$ in order to allow matter parity as gauged discrete
symmetry to coexist with SM all the way down to the electroweak scale.  
Now we address the question whether spontaneous symmetry breaking of
SO(10) can allow such small VEV through fine tuning of the model
parameters in the GUT Lagrangian.

It is well known  that this question is related to the stability of Higgs vacuum in the SM
or the Higgs mass at the electroweak scale.
While SUSY GUTs have a natural resolution to the well known gauge
hierarchy problem, in non-SUSY GUTs the scalar mass is kept at the
electroweak scale by fine tuning of parameters in the GUT Lagrangian
to every loop order.
Following the discussion of Sec.\ref{sec:RGE} and notations of Table
\ref{tab:decomp}   we write the full SO(10) invariant Higgs potential 
including ${16}_H$,
  ${10}_H$,${45}_H$,${126}_H$,${210}_H$ and their conjugates noting
  that ${10}_H$,${45}_H$ and ${210}_H$ are self-conjugates. In writing
  such scalar potentials we have suppressed tensor indices \cite{aulakh-gs-vissani,Fukuyama:2005}
for the sake of simplicity
\ba
V_{16}&=&M_{16}^2{16}_H^{\dagger}{16}_H+\lambda_{16}{({16}_H^{\dagger}{16}_H)}^2
+\lambda_{(16,10)}{16}_H^{\dagger}{16}_H{10}_H{10}_H \nonumber\\
&+&\lambda_{(16,45)}{45}_H{45}_H{16}_H^{\dagger}{16}_H+\lambda_{(16,126)}{126}_H^{\dagger}{126}_H{16}_H^{\dagger}{16}_H
\nonumber\\
&+&\lambda_{(16,210)}{210}_H{210}_H{16}_H^{\dagger}{16}_H 
+m_{(16,126)}\left
[{126}_H{16}_{H}^{\dagger}{16}_{H}^{\dagger}+{126}_H^{\dagger}{16}_{H}{16}_{H}\right]
\nonumber\\
&+& m_{(16,10)}\left[{16}_{H}{16}_{H}{10}_H+{16}_{H}^{\dagger}{16}_{H}^{\dagger}{10}_H\right]
+m_{(16,45)}{45}_H{16}_H^{\dagger}{16}_H\nonumber\\
&+&m_{(16,210)}{210}_H{16}_H^{\dagger}{16}_H.\label{eq:pot16} 
\ea

\ba
V_{10}&=& M_{10}^2{10}_H^2+\lambda_{10}{10}_H^4+\lambda_{(16,10)}{16}_H^{\dagger}{16}_H{10}{10}_H+
\lambda_{(126,10)}{126}_H^{\dagger}{126}_H{10}_H{10}_H \nonumber\\
&+&\lambda_{(45,10)}{45}_H{45}_H{10}_H {10}_H+m_{(45,10)}{45}_H{10}_H^2
+\lambda_{(210,10)}{210}_H{210}_H{10}_H{10}_H~. \label{eq:pot10} 
\ea
As noted before we identify $\chi_S(1,0,1)$ as the real part of the SM 
scalar singlet in $\chi_S(1,0,1)^{\dagger}\subset {16}_H^{\dagger}$
\ba
\chi_S^{\dagger}&=&\chi_{S_R}-i\chi_{S_I}, \nonumber\\
\chi_{S_R}&=&Re(\chi_S^{\dagger}),\nonumber\\
\chi_{S_I}&=&-Im(\chi_S^{\dagger}), \label{eq:chicomp}
\ea
We use decompositions given in eq.(\ref{eq:126dag}),
eq.(\ref{eq:16H}), eq.(\ref{eq:decall}) and Appendix B. In the
generalised case at first we use GUT scale VEVs in all SM singlet directions
and denote them as  $(A_0,A_3) \subset {45}_H$ and $V_{210}^{(i)}(i=1,2,3) \subset {210}_H$
which are along the singlet component directions of respective  Pati-Salam
submultiplets $G_{224}$ 
\ba
A_0&=&V_H=\sqrt{2} \langle (1,1,15)\rangle \subset {45}_H, \nonumber\\
A_3&=& \sqrt{2}\langle (1,3,1)\rangle \subset {45}_H , \nonumber\\
V_{210}^{(1)}&=&\sqrt{2} \langle (1,1,1)\rangle \subset {210}_H, \nonumber\\
V_{210}^{(2)}&=&\sqrt{2} \langle (1,1,15)\rangle \subset {210}_H, \nonumber\\
V_{210}^{(3)}&=&\sqrt{2}  \langle (1,3,15)\rangle \subset {210}_H, \nonumber\\
V_R&=&\sqrt{2}  \langle (1,3,{\bar 10})\rangle \subset{126}_H,\nonumber\\
 \label{eq:vevdef}
\ea
In addition we  use 
\ba
V_{\chi}&=&\sqrt{2}\langle Re(\chi_S)\rangle=\sqrt{2} \langle (1,2,{\bar
  4})\rangle \subset {16}_H,\nonumber\\
V_{\phi}&=&\sqrt {2} \langle\phi\rangle\subset \langle (2,2,1)\rangle \subset {10}_H.\label{eq:vevchiphi}
\ea
The VEVs of $Im(\chi_S)$ and the left handed doublet VEV in ${16}_H$ or
${16}_H^{\dagger}$  are taken to be vanishing wherever necessary
\be
\langle\chi_L(2,1, 4)\rangle=\langle Im(\chi_S)\rangle=0.\label{eq:vevzer}
\ee 
Although all our derivations discussed here can apply in the presence
of all GUT scale VEVs defined in eq.(\ref{eq:vevdef}), no generality
is lost by confining to the minimal case of the two models (Model-I
and Model-II)  for which we fix
\be
 V^{(1)}_{210}=V^{(3)}_{210}=A_3=0.\label{eq:minimal}
\ee
Now identifying $V^{(2)}_{210}=V_{210}, \lambda_{210}^{(2)}=\lambda_{210}$ and
 using eq.(\ref{eq:126dag}), eq.(\ref{eq:16H}),
eq.(\ref{eq:decall}), eq.(\ref{eq:vevdef}),
eq.(\ref{eq:vevchiphi}) and eq.(\ref{eq:minimal}), we 
minimise the potentials $V_{16}$ and $V_{10}$ to obtain
\ba
\lambda_{(16)}V_{\chi}^2&=&-[\,M_{16}^2+\lambda_{(16,10)}V_{\phi}^2/2+P],\nonumber\\
P&=&\lambda_{(16,45)}A_0^2/2+(1/2)\lambda_{(16,126)}V_{R}^2+(1/2)\lambda_{(16,210)}V_{210}^2 \nonumber\\
&+&\sqrt{2}m_{(16,126)}V_R+m_{(16,45)}A_0/\sqrt{2}+m_{(16,210)}V_{210}/\sqrt{2}. \label{eq:vchi1}
\ea  
\ba
\lambda_{(10)}V_{\phi}^2&=&-[\,M_{10}^2+\lambda_{(16,10)}V_{\chi}^2/2+Q], \nonumber\\
Q&=&\lambda_{(10,45)}A_0^2/2 +
\lambda_{(10,126)}V_R^2/2+(1/2)\lambda_{(10,210)}
V_{210})^2+m_{(10,45)}A_0/\sqrt{2}. \label{eq:vphi1}
\ea 
Now solving the two equations eq.(\ref{eq:vchi1}) and eq.(\ref{eq:vphi1})  for
 simultaneously gives
\be
V_{\chi}^2= -\omega_2(M_{16}^2+P)+\omega_1(M_{10}^2+Q).\label{eq:vchisq}
\ee
\be
V_{\phi}^2= -\frac{\lambda_{16}}{\lambda_{10}}\omega_2(M_{10}^2+Q)+
\omega_1(M_{16}^2+P).\label{eq:vphisq}
\ee
 where 
\ba
\omega_1&=&\frac{2\lambda_{(10,16)}}{(4\lambda_{10}\lambda_{16}-\lambda_{(10,16)}^2)},
\nonumber\\
\omega_2&=&\frac{4\lambda_{10}}{(4\lambda_{10}\lambda_{16}-\lambda_{(10,16)}^2)}. \label{eq:w12}
\ea
These equations retain the same form in the more generalised cases 
when, instead of eq.(\ref{eq:minimal}), all GUT scale VEVs defined in
eq.(\ref{eq:vevdef}) are included. In such a generalised case
the following replacements are made  in the definitions of P and Q given in eq.(\ref{eq:vchi1}) and eq.(\ref{eq:vphi1}):
$\lambda_{(16,210)}V_{210}^2\to \sum_{i=1}^3\lambda_{(16,210)}^{(i)}{(V_{210}^{(i)})}^2,\,\,
\lambda_{(10,210)}V_{210}^2\to \sum_{i=1}^3\lambda_{(10,210)}^{(i)}{(V_{210}^{(i)})}^2,\,\,
m_{(16,210}V_{210}\to\sum_{i=1}^3m_{(16,210)}^{(i)}V_{210}^{(i)},\,\,$
and $A_{0}^2\to A_{0}^2+A_{3}^2$ \footnote{ In the same fashion additional effect
 due to GUT scale VEV of the $G_{224}-$ singlet scalar $\langle E \rangle
  \subset {54}_H$ discussed in Sec.\ref{sec:embed} and occuring as
  alternative solution to DM relic density noted in Sec.\ref{sec:relic}
  can be included without affecting self-consistency of model
  predictions discussed in Sec.\ref{sec:imchi}.}. Similarly  self-consistency of model predictions
discussed below in  Sec.\ref{sec:imchi}
is guaranteed in the
generalised case with the corresponding replacements in  eq.(\ref{eq:potcomp}) and  eq.(\ref{eq:mtheta}).\\

Besides the mass squared terms $M_{16}^2$ and
$M_{10}^2$ being of order $M_{GUT}^2$, each of the 
three VEVs $V_{210}$, $V_R$, $A_0$ can be of order $M_{GUT}$. In
addition each of the trilinear couplings
$m_{(16,10)}$, $m_{(16,45)}$, $m_{(16,126)}$, $m_{(16,210)}$, and $m_{(10,45)}$
is also of order $M_{GUT}$. Then each of the terms in $P$ and $Q$ is proportional to $M_{GUT}^2$ having unknown coefficients. This suggests that
our models can satisfy the low scale matter parity violating condition  $V_{\chi}\le V_{\phi}\simeq {\cal O}(M_{\rm W})$ by fine tuning the mass dimensionful parameters and
nine quartic couplings in eq.(\ref{eq:vchisq})
and eq.(\ref{eq:vphisq}). In the general non-minimal case of eq.(\ref{eq:vphi1}), compared to eq.(\ref{eq:vchi1}), the number of such parameters  is less but they are enough  
to yield the well known result $V_{\phi}^2 \sim M_{\rm W}^2$ through
fine tuned cancellations among different terms involving GUT scale
parameters. In the sense of extended survival hypothesis \cite{Aguila:1981,rnm-gs:1983}, the fine
tunings needed to make $V_{\phi}\sim v_{ew}$ which is associated with
electroweak gauge symmetry breaking belong to the category of minimal
fine tuning. Other fine tunings needed to keep $\kappa (3,0,8)$ at
$M_{\kappa}=10^{9.2}$ GeV and $Re(\chi_S)$ light are among the
category of additional fine tunings.  

Common to both minimal and non-minimal cases we note some interesting
possibilities of cancellations which could be exploited for
physical applications.
 Even if the input VEV $V_H\simeq
10^3-10^8$ TeV as utilised in some cases of Sec.\ref{sec:couplings},
there is provision for mutual fine-tuned cancellation of corresponding
terms. For example cancellation can occur between $\lambda_{(16,15)}V_H^2$ and $m_{(16,45)}V_H$ in the
expressions for $P$ and $Q$ in eq.(\ref{eq:vchisq}) and eq.(\ref{eq:vphisq}) in the minimal case
and similarly among corresponding terms in the nonminimal case.    
In our analysis we have not included radiative corrections which can be also
significant in estimating the minimum of a Higgs potential more accurately 
\cite{Bertolini:2010}.

\subsubsection{GUT Scale Mass of  Im($\chi_S$) and
    Self-Consistency of Model Predictions}\label{sec:imchi}
All our discussions presented in this section on minimisation of Higgs
potential and derivation of low $V_{\chi}$ apply if the real part of
$\chi_S(1,0,1)$ has mass $\sim M_{W}$ while its  imaginary  part does
not contribute to any physical quantity below the GUT scale. Further all the analyses  in
  Sec. \ref{sec:stab} for vacuum stability  have been carried out under the assumption that 
  $\chi_S(1,0,1)=\chi_{S_R}$ is a real singlet Higgs scalar and resolution of
  vacuum stability has predicted its mass to be $M_{\chi_S}\simeq 177$
  GeV. There is  the possibility that the imaginary part of
  $\chi_S(1,0,1)$, if light, can also contribute to relevant quantities affecting the self consistency of predictions discussed
  so far. On the other hand if the imaginary part  has GUT scale mass,
  it would decouple from making any such contributions at lower scales
  which would
  guarantee self-consistency. In  this section we discuss how this
  latter possibility is naturally realised in both of our models
  (Model-I and Model-II) without needing any additional fine tuning. \\

Using $V_{16}$ from eq.(\ref{eq:pot16}) and eq.(\ref{eq:chicomp}) we at first
extract the potential contributing to the  mass terms for the real and
imaginary parts of $\chi$. Although the following proof can be applied
in the most general case of both the models, for the sake of simplicity we confine only
to the minimal case in which all GUT scale VEVs vanish along with
$\langle\chi_L(2,-1/2,1)\rangle=\langle\chi_{S_I}\rangle=0$
 except
$V_R, V_{210}^{(2)}\equiv V_{210}$ and $A_0=V_H$ 
\ba
V_{\rm mass}&=& [M_{16}^2+\lambda_{(16,10)}V_{\phi}^2/2+
(1/2)\lambda_{(16,45)} A_0^2 +\lambda_{(16,210)}V_{210}^2/2\nonumber\\
&+&\lambda_{(16,126)}V_R^2/2+m_{(16,45)}A_0/\sqrt{2}+
m_{(16,210)}V_{210}/\sqrt{2}](\chi_{S_R}^2+\chi_{S_I}^2)\nonumber\\
&+&\sqrt{2}m_{(16,126)}V_R(\chi_{S_R}^2-\chi_{S_I}^2)
+(1/2)\lambda_{16}V_{\chi}^2\chi_{S_R}^2+\lambda_{16}V_{\chi}^2\chi_{S_I}^2,
\label{eq:potcomp} 
\ea 
where the last two terms follow from
$\lambda_{16}{({16}_H^{\dagger}{16}_H)}^2$ under the 
assumption that $\langle \chi_{S_I} \rangle =0$. 
Then the masses of real and imaginary parts can be  separately written as   
\ba
M_{\chi_{S_R}}^2&=&M_{\theta}^2+\lambda_{16}V_{\chi}^2/2+\sqrt
{2}m_{(16,126)}V_R ,\nonumber\\
&=&-\lambda_{16}V_{\chi}^2/2. \label{eq:remass}
\ea

\be
M_{\chi_{S_I}}^2=M_{\theta}^2+\lambda_{16}V_{\chi}^2-\sqrt{2}m_{(16,126)}V_R, \label{eq:immass}
\ee
where, in the minimal cases,
\ba
M_{\theta}^2&=&M_{16}^2+\lambda_{(16,10)}V_{\phi}^2/2+m_{(16,210)}V_{210}^2/2+\lambda_{(16,45)}A_0^2/2\nonumber\\
&+&\lambda_{(16,126)}V_R^2/2+m_{(16,45)}A_0/\sqrt{2}+m_{(16,210)}V_{210}/\sqrt{2}.
\label{eq:mtheta}
\ea
The second line of eq.(\ref{eq:remass}) follows from
eq.(\ref{eq:vchi1}). Now using eq.(\ref{eq:remass}) we have
\be
M_{\theta}^2=-\lambda_{16}V_{\chi}^2-\sqrt{2}m_{(16,126)}V_R,\label{eq:mthsol}
\ee
which through eq.(\ref{eq:immass}) gives GUT scale mass to the
imaginary part of the scalar singlet
\be
M_{\chi_{S_I}}^2=-2\sqrt{2}m_{(16,126)}V_R\sim M_{GUT}^2.\label{eq:imsol}
\ee
Thus, once we realize $V_{\chi}\sim M_W$ by fine tuning or,
equivalently,  
Re$(\chi_S)$ is made to acquire mass of ${\cal O}(M_W)$  to implement
low scale matter parity breaking and resolve vacuum stability issue of the
scalar potential, the GUT scale mass of $Im(\chi_S)$ automatically
follows. Because of such high value of its mass, Im($\chi_S$) decouples from
making any contribution  to
the standard Higgs quartic couplings below the GUT scale reported in 
Sec.\ref{sec:stab}
 and also all other relevant quantities
predicted in Sec.\ref{sec:mix} and 
Sec.\ref{sec:couplings}. Thus self-consistency of model predictions is
guaranteed. As already noted in Sec.\ref{sec:vchi}, the self
consistency of model predictions is maintained in the generalised case
even when the effects of 
all additional
GUT scale VEVs defined in eq.(\ref{eq:vevdef}) are included.  
    
%%%%%%%%%%%%%%%%%%%%%%%%%%%%%%%%%%%%%%%%%%%%%%%%%%%%%%%%%%%%%%%%%%%%%
\section{Relic Density of Decaying Dark Matter and Flux of IceCube Neutrinos}\label{sec:relic}
In this section we discuss how the proper relic density of the decaying dark matter candidate $\Sigma_F$ is realized within the present SO(10) framework to generate the expected flux of PeV energy IceCube neutrinos. 
It was shown by Griest and Kaminkowski \cite{Griest:1989wd} that any elementary particle having mass greater than $340$ TeV can not have a thermal origin while the  upper limit on  thermal dark matter mass has been  derived to be $\sim 200$ TeV
\cite{vonHarling:2014kha}. As we are dealing 
with a very massive ($\sim$ PeV) Majorana type decaying dark matter (DDM),
its relic density must have a non-thermal origin. In the following sections we shall discuss one such
suitable mechanism compatible with our theoretical framework which can successfully address the observed relic density of the DDM.

\subsection{Relic Density through the Exchange of a Heavy Scalar Singlet}
As discussed earlier, the SO(10) representation ${45}_F$ has two singlets under SM. We have used the singlet fermion contained in the Pati-Salam sub-multiplet $(1,1,15)_F\subset {45}_F$ as the Majorana fermion DM component. 
The present mechanism of generating relic density  is similar to the
non-equilibrium thermal dark matter (NETDM) mechanism discussed in a number of recent works in matter parity conserving non-SUSY SO(10) with high scale intermediate gauge symmetries
\cite{Mambrini:2013,Mambrini:2015}. In the present work in the absence of any intermediate gauge symmetry, the desired heavy scalar singlet mediating the Higgs exchange process is naturally present at the GUT scale. 
At some early epoch during the evolution of the Universe, when the temperature of the thermal bath containing the SM particles is very high,  the Majorana type fermionic DM $(\Sigma_F \subset 45_F)$
can be produced due to the scattering of SM particles by  the exchange
of a very heavy scalar $(\xi \subset {210}_H)$ of mass $M_{\xi} \gg
M_{\Sigma}$. This exchanged scalar can be the scalar singlet in
${(1,1,15)}_H \subset {210}_H$. Alternatively, this exchanged scalar
could be the $G_{224}$ singlet ${(1,1,1)}_H\subset {54}_H$. Therefore
the production rate of DM particles is kept at extremely small level
with negligible self interaction probability among them. One notable
aspect of this mechanism is that the DDM fermion $\Sigma_F$ having no renormalisable interaction with the SM particles is prevented from being in the thermal bath. Thus, while the DM particles are produced by SM particles in the thermal bath via heavy scalar exchange, they do not annihilate due to mutual interactions. Further they do not attain thermal equilibrium justifying their nomenclature as NETDM. 
In \cite{Mambrini:2013,Mambrini:2015}, one main reason for the need of
$G_{224}$ as intermediate gauge symmetry  has been gauge coupling
unification \cite{Babu-RNM:1993,cmgmp:1985} and the heavy scalar mass
exchanged to achieve NETDM has been fine tuned to be at the $G_{224}$ breaking
intermediate scale. Now that we achieve precision gauge coupling unification in Model-I and Model-II with minimally modified grand desert in each case, we need not adopt additional fine tuning to assign the mass of $\xi$ at the intermediate scale. Its
natural presence around the GUT scale successfully drives the DDM generation process. As discussed in Sec.\ref{sec:embed}, $\Sigma_F$ possesses  nonstandard Yukawa couplings $h_p$ and $h_e$ with ${210}_H$ and ${54}_H$, respectively. But only $h_p$ with ${210}_H$ of our minimal models has been shown to be enough to keep
$\Sigma_F$ mass at PeV scale.

\begin{figure}[h!]
\begin{center}
\includegraphics[scale=0.4]{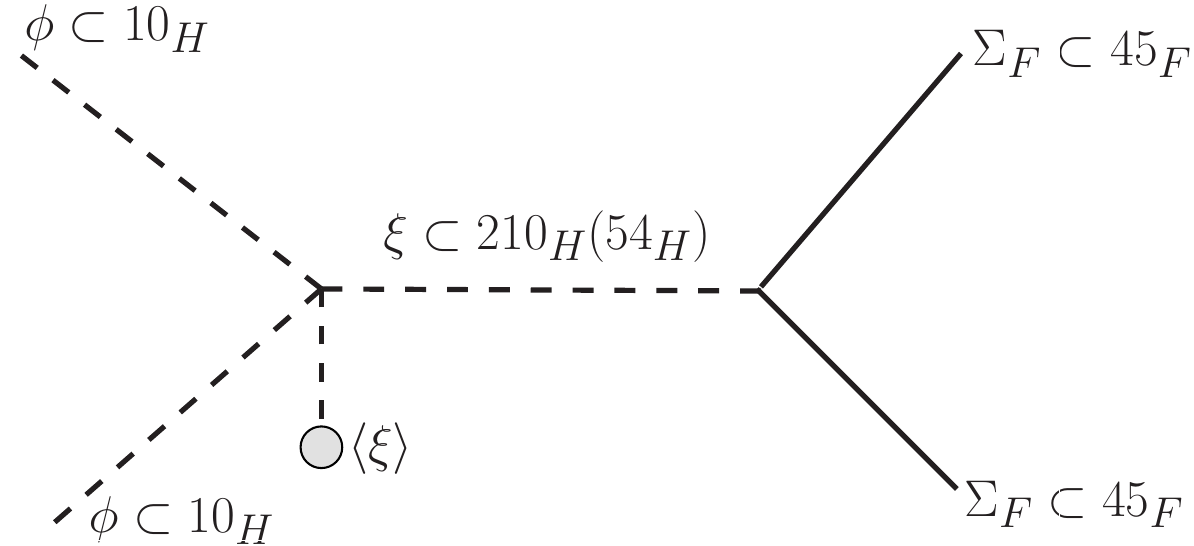}  
\caption{Feynman diagram for the decaying dark matter ($\Sigma_F$)
  production through the exchange of heavy scalar singlet $\xi$
  contained in ${210}_H\subset$ SO(10). The mechanism also operates if
  the Higgs representation ${210}_H$ is replaced by ${54}_H\subset$
  SO(10) as indicated in the parenthesis.}
\label{scl_exg}
\end{center}
\end{figure}

 Thus, although the decaying dark matter $\Sigma_F\subset {45}_F$ can not have
 renormalisable interaction directly with SM particles, it does possess
 such Yukawa interaction of the type $f_{\Sigma}{\Sigma}_F{\Sigma}_F
 \xi \subset f_{\Sigma}{45}_F{45}_F X_H$ where $X_H ={210}_H$ 
   or ${54}_H$. Thus $f_{\Sigma}\simeq {\cal O} (h_p)$,  or  $\simeq {\cal O} (h_e)$ of
 eq.(\ref{eq:mass15}) and  eq.(\ref{eq:compmass}) of Sec.\ref{sec:embed}.  At the other end, the exchanged DM does interact via renormalisable quartic interaction $\lambda_{\xi} \phi^{\dagger}\phi \xi^{\dagger}\xi \subset \lambda_{\xi}{10}_H{10}_H X_H^{\dagger}X_H $ where $\phi=$ SM Higgs doublet and $\lambda_{\xi} \sim \lambda_{(10,210)}$ which has been also defined in Sec.\ref{sec:vchi}.
Alternatively, when  the $G_{224}$ singlet of ${54}_H$ is used instead
of the singlet in ${210}_H$, this quartic
coupling can be replaced correspondingly. No generality is lost even if
this quartic coupling is treated to be small  in the cases
of the two SO(10) models. 
 From these two interactions, we can easily construct the Feynman diagram  shown in Fig.\ref{scl_exg} which elucidates  
the underlying NETDM type mechanism without any intermediate gauge symmetry for the production of the
DDM. As explained above the DM interacts with the SM particle (the SM
Higgs $h$) at high temperature bath whereas the DM itself can not be in
the thermal bath.

%%%%%%%%%%%%%%%%%%%%%%%%%%%%%%%%%%%%%%%%%%%%%%%%%%%%%%%%%%%%%%%%%%%%%%%%%%%%
%%%%%%%%%%%%%%%%%%%%%%%%%%%%%%%%%%%%%%%%%%%%%%%%%%%%%%%%%%%%%%%%%%%%%%%%%%%%%

 To find out the relic density of the DM produced in this mechanism we have to solve the Boltzmann equation 
\begin{equation}
\frac{d Y_\Sigma}{d z} =\sqrt{\frac{\pi}{45}}\sqrt{g_\ast}M_\Sigma M_{pl}\frac{\langle \sigma v \rangle}{z^2} Y_{eq}^2, \label{boltz}
\end{equation}
where $Y_\Sigma=n_\Sigma/s$, $Y_{eq}=n_{eq}/s$, $n_\Sigma$ is the number density of the dark matter particle, $n_{eq}$ is the equilibrium number density of
the initial state standard model particle and $s$ is the entropy density.
In eq.(\ref{boltz})   $z=M_\Sigma/T$ where $T$ is  the temperature of the Universe, $M_{pl}$ is the Planck mass
and $g_\ast=107$ is the effective number of massless degrees of freedom. The thermally averaged cross section of the above process multiplied by the equilibrium 
density squared is given by
\begin{equation}
\langle \sigma v \rangle n_{eq}^2 \simeq \frac{T}{512 \pi^5}\int_{4M_\Sigma^2}^\infty d \hat{s} \sqrt{\hat{s}-4M_\Sigma^2}~ K_1\left(\frac{\sqrt{\hat{s}}}{T}\right)\sum | \mathcal{M} |^2,
\label{sv}
\end{equation}
where $\hat{s}$ represents the square of centre of mass energy, $K_1(x)$ is the Bessel function of second kind and $\mathcal{M}$ denotes the Feynman amplitude of the process under consideration.
The summation is over all possible initial states as well as the spin of the final states. The sum of the modulus squared amplitude is calculated as
\begin{equation}
\sum | \mathcal{M} |^2 \simeq F_{\xi} \left(\frac{\hat{s}-4M_\Sigma^2}{M_\xi^2}\right) \label{mfi}
\end{equation}
where $F_{\xi}$ is a numerical factor which takes into account the
couplings of the two vertices of the Feynman diagram of
 Fig. \ref{scl_exg} i.e approximately 
\be
F_{\xi}\sim f_{\Sigma}\lambda_{\xi}.\label{eq:Fxi}  
\ee
%%%%%%%%%%%%%%%%%%%%%%%%%%%%%%%%%%%%%%%%%%%%%%%%%%%%%%
Substituting the expressions of thermally averaged annihilation
cross section  from eq.(\ref{sv}) and modulus squared amplitude from eq.(\ref{mfi}) in the 
Boltzmann equation (eq.(\ref{boltz})), we integrate it to very high value of $z$ (or equivalently from $T_{RH}$ to temperature of present epoch), to get the present day dark matter
density as
\begin{equation}
Y_{\Sigma}^{(\rm inf)}= \frac{F_\xi}{1024\pi^7}\left( \frac{45}{\pi g_\ast}\right)^{3/2}\frac{M_{pl}M_\Sigma}{M_\xi^2}\int_{\small \frac{M_\Sigma}{T_{RH}}}^{\frac{M_\Sigma}{T_0}}
\frac{1}{x^2} \left(\int_{2x}^\infty t (t^2 -4x^2)^{3/2}K_1(t)dt\right) dx
\label{boltz_sol}
\end{equation}
where $T_0$ is the temperature at the present epoch. So the quantity $M_\Sigma/T_0$ tends to infinity and the dark matter density (scaled by entropy density) at present epoch
is termed as $Y_{\Sigma}^{(\rm inf)}$.
%%%%%%%%%%%%%%%%%%%%%%%%%%%%%%%%%%%%%%%%%%%%%%%%%%%%%%%%%%%
Under the plausible  assumption that the
dark matter mass is much smaller than the reheating temperature,
$M_{\Sigma}\ll T_{RH}$.
Then  the abundance of dark matter particle at present epoch  comes out to be
\begin{equation}
{ Y}_\Sigma^{(\rm inf)} \simeq \left( \frac{45}{\pi g_\ast}
\right)^{3/2}\frac{F_{\xi} M_{Pl}T_{RH}}{64 \pi^7 M_\xi^2}. \label{eq:YM0} 
\end{equation}
Denoting values at the present epoch by subscript ``${(\rm inf)}$'',
${ Y}_\Sigma^{(\rm inf)}$ is also related to known
quantities
\be
{ Y}_\Sigma^{(\rm inf)}s^{(\rm inf)}M_{\Sigma}=(\Omega_{\rm DM}
h^2)\left[\frac{\rho^{(\rm inf)}_c}{h^2}\right],\label{eq:y0}
\ee
where $h$ denotes the Hubble parameter and
\ba
(\Omega_{\rm DM}h^2)&=& 0.12, \nonumber\\
\frac{\rho^{(\rm inf)}_c}{h^2}&=& 1.05 \times 10^{-5} {\rm GeV}{\rm
  cm}^{-3}, \nonumber\\
s^{(\rm inf)}&=& 2.89\times 10^{-3}\,{\rm cm}^{-3}.\label{eq:cdata} 
\ea
Then using eq.(\ref{eq:cdata}) and eq.(\ref{eq:y0}) in
eq.(\ref{eq:YM0}) we have
\ba
T_{RH} = 1.3\times 10^7 \times \left[\frac{\Omega_{\rm
    DM}h^2}{0.12}\right]\left[\frac{F_{\xi}^{-1}g_{*}^{3/2}}{10^6}\right] %\nonumber\\
  \left[\frac{M_{\xi}}{10^{16}}\right]^2\left[\frac{M_{\Sigma}}{10^6}\right]^{-1}\label{eq:TRHSig}
\ea
where $M_{\xi}, M_{\Sigma}$ are in GeV. As an illustration of resulting analytic
solutions for the heavy scalar mass $M_{\xi}(\xi \subset {210}_H)$  and reheating
temperature $T_{RH}$ under this approximation we get for $\Omega_{\rm
    DM}h^2=0.12$ and $M_{\Sigma}=10^6$ GeV
\ba
 T_{RH}&=& 1.3\times 10^7 {\rm \,GeV}, M_{\xi}=3.1 \times 10^{15}{\rm
   \,GeV}, F_{\xi}^{-1}g_{*}^{3/2}=10^7,\nonumber\\
 T_{RH}&=& 1.3\times 10^7 {\rm \,GeV}, M_{\xi}=10^{16}{\rm
   \,GeV}, F_{\xi}^{-1}g_{*}^{3/2}=10^6,  \nonumber\\
 T_{RH}&=& 1.3\times 10^9 {\rm \,GeV}, M_{\xi}=10^{17}{\rm \,GeV},
 F_{\xi}^{-1}g_{*}^{3/2}=10^6,\nonumber\\
 T_{RH}&=& 1.3\times 10^8 {\rm \,GeV}, M_{\xi}=10^{17}{\rm \,GeV},
 F_{\xi}^{-1}g_{*}^{3/2}=10^5.\label{eq:anasol}
\ea
As discussed in Sec.\ref{sec:gcu} the threshold corrected GUT scale
in our model can easily acquire the range of values $M_{GUT} \simeq
10^{15.5}-10^{16.5}$ GeV. Therefore  the desired superheavy Higgs scalar
component of $\Phi^{(2)}(1,1,15)$ left over after spontaneous symmetry
breaking naturally acts as the $\xi$ particle which is exchanged
between the DM and the SM Higgs as shown in the Feynman diagram of
Fig.\ref{scl_exg} to generate the desired relic density. Then this
mass can be easily of the same order as $M_{GUT}$ 
Since $g_\ast \sim 100$, and $f_{\Sigma}\sim 0.1$, the desired quartic
coupling of the exchanged scalar $\xi$ with SM Higgs scalar are
$\lambda_{\xi}\sim 0.01$ when $F_{\xi}^{-1}g_{*}^{3/2}=10^6$,
$\lambda_{\xi}\sim 0.01-0.1$ when $F_{\xi}^{-1}g_{*}^{3/2}=10^5$ and
$\lambda_{\xi}\sim 0.0001-0.001$ when
$F_{\xi}^{-1}g_{*}^{3/2}=10^7$.  These results are presented in Table
\ref{tab:TRHMxi}.
%%%%%%%%%%%%%%%%%%%%%%%%%%%%%%%%%%%%%%%%%%%%%%%%%%%%%%%%%%%%%%%%%
\begin{table}[h!]
\caption{Allowed analytic solutions for exchanged heavy scalar mass $M_{\xi}$
  and reheating temperature $T_{RH}$ as a function of quartic coupling
  $\lambda_{\xi}$ consistent with relic density $\Omega h^2=0.12$.}
\begin{center}
\begin{tabular}{|c|c|c|c|} 
\hline
$M_{\xi}$ & $F_{\xi}^{-1}g_\ast^{3/2}$& $\lambda_{\xi}$ & $T_{\rm RH}$ \\ 
$(GeV)$ & & &$(GeV)$\\
\hline
$3.1\times 10^{15}$&$10^7$ & $\sim 0.001$& $1.3\times 10^{7}$\\ 
 \hline
 $10^{16}$&$10^{6}$&$\sim 0.01$ & $1.3\times 10^7$\\
 \hline 
$10^{17}$&$10^6$& $\sim 0.01$& $1.3\times 10^{9}$ \\
\hline
$10^{17}$&$10^5$& $\sim 0.1$& $1.3\times 10^{8}$ \\
\hline
\end{tabular}
\label{tab:TRHMxi}
\end{center}
\end{table}
%%%%%%%%%%%%%%%%%%%%%%%%%%%%%%%%%%%%%%%%%%%%%%%%%%%%%%%%%%%%%%%

\begin{figure}[h!]
\begin{center}
\includegraphics[scale=0.8]{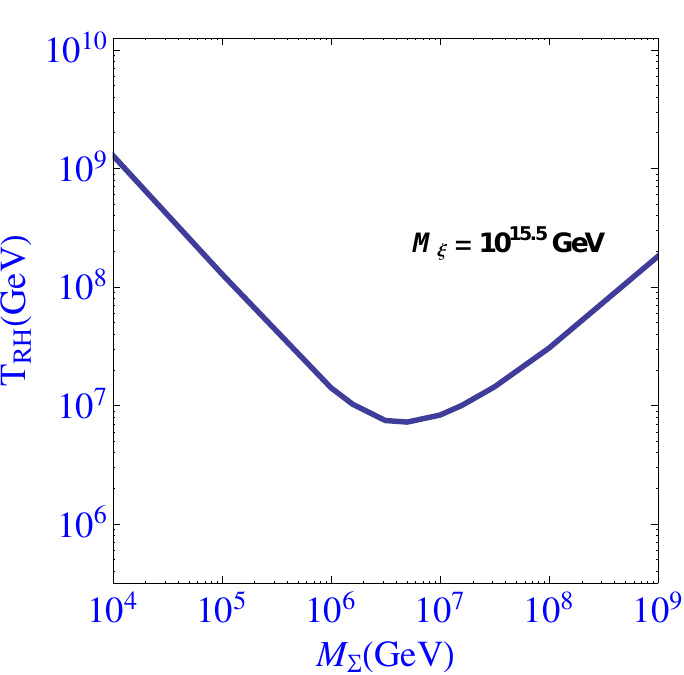}  
\caption{Variation  of reheating temperature as a function of
  decaying dark matter mass for  exchanged scalar mass close to the
  threshold uncorrected GUT scale
  $M_{\xi}= 3\times10^{15}$ GeV  consistent with relic density $\Omega_{DM}
  h^2 \sim 0.12$ and for $\lambda_{\xi} \simeq 0.001$.}
\label{fig:xi15p5}
\end{center}
\end{figure}

Alternatively, without using any approximation we have also carried out numerical integration of eq.(\ref{boltz_sol}) to
express $T_{RH}$ as a function of DDM mass as shown in
Fig.\ref{fig:xi15p5} for $M_{\xi}=10^{15.5}$ GeV.  
%%%%%%%%%%%%%%%%%%%%%%%%%%%%%%%%%%%%%%%%%%%%%%%%%%%%%%%%%%%%%%%%%%%%%%%%%%%%%%%%%%%%%%%%%%%%%
%%%%%%%%%%%%%%%%%%%%%%%%%%%%%%%%%%%%%%%%%%%%%%%%%%%%%%%%%%%%%%%%%%%%%%%%%%%

\begin{figure}[h!]
\begin{center}
\includegraphics[scale=0.8]{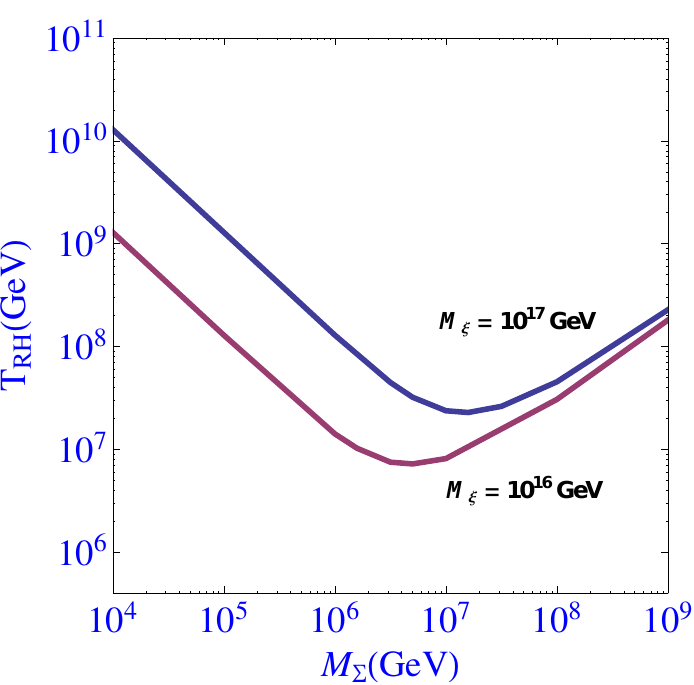}  
\caption{Variation  of reheating temperature as a function of
  decaying dark matter mass for two values of exchanged scalar masses
  $M_{\xi}=10^{17}$ GeV with $\lambda_\xi=0.1$ (upper curve) and $M_{\xi}=10^{16}$ GeV with $\lambda_\xi=0.01$ (lower
  curve) consistent with $\Omega_{DM} h^2 \sim 0.12$ }
\label{fig:TRH1617}
\end{center}
\end{figure}

In Fig. \ref{fig:TRH1617} numerical solutions  for two different sets of $M_{\xi},F_{\xi}$
combinations  are presented showing variation of reheating temperature
with DM mass.

%%%%%%%%%%%%%%%%%%%%%%%%%%%%%%%%%%%%%%%%%%%%%%%%%%%%%%%%%%%%%%%%%%%%%%%%%%% 
%%%%%%%%%%%%%%%%%%%%%%%%%%%%%%%%%%%%%%%%%%%%%%%%%%%%%%%%%%%%%%%%%%%%%%%%
In the region of interest the solutions presented in
Fig.\ref{fig:TRH1617} are found to be similar to the analytic
solutions given in eq.(\ref{eq:anasol}). Thus,
it is clear that $T_{RH}\simeq 10^7-10^9$ GeV could reproduce the
desired relic density for DDM mass $M_{DM}=M_{\Sigma}=10^6$ GeV with
 the exchanged heavy particle masses $M_{\xi}\sim
10^{15}-10^{17}$ GeV predicted  within this non-SUSY SO(10) GUT
framework.

\subsection{Flux of  IceCube Neutrinos}
In this subsection we  try to explain the two PeV energy neutrino events \cite{IceCube:2013} detected by IceCube through the decay of heavy Majorana type fermionic
DM whose mass is also in the PeV range. We will calculate the flux of neutrinos due to DM decay in the galactic halo and compare it to the flux required for 
two PeV energy neutrino events at IceCube. It is to be noted that the flux may have an extragalactic contribution too. However for the present analysis we 
 neglect this extragalactic contribution since it has been found in some previous works\cite{Beacom:2006tt,Yuksel:2007ac,Yuksel:2007dr} that this contribution is smaller than the galactic contribution by an 
order of magnitude. 
%\paragraph{}
\\\\
The differential flux generated due to the decay of the DM in the Milky Way halo is given by
\begin{equation}
\frac{d \Phi}{d E_\nu} = \frac{1}{4\pi M_\Sigma \tau_\Sigma} \frac{d N_\nu}{d E_\nu} \int_0^\infty ds ~\rho [r(s,l,b)]
\label{lsi}
\end{equation}
where $M_\Sigma$ and $\tau_\Sigma$ are mass and decay life time of the DDM particle $\Sigma_F$  and
$\frac{d N_\nu}{d E_\nu}$ is the energy spectrum of the neutrinos produced due to the DM decay.
Since in the present work we are concentrating only on one decay channel of the DM, i.e $\Sigma_F \rightarrow \nu~h$,
the spectrum would be a delta function $\frac{d N_\nu}{d E_\nu}=\delta(E_\nu-\frac{M_\Sigma}{2})$. In eq.(\ref{lsi})
$\rho(r)$ is the density profile of the dark matter particle in our Galaxy where $r$ is the distance 
from the galactic centre. For all numerical estimations we have used NFW profile for the density $\rho(r)$,
which is expressed as
\begin{equation}
\rho(r)=\frac{\rho_h}{\frac{r}{r_c}(1+\frac{r}{r_c})^2}. 
\end{equation}
Here the critical radius $r_c \simeq 20$ kpc and $\rho_h \simeq 0.33$ {$\rm GeVcm^{-3}$}. 
The integral of eq.(\ref{lsi}) over $s$ variable is known as line of sight integral which has to be evaluated
to obtain the flux at earth and $s$ is related to $r$ through the relation
\begin{equation}
r(s,l,b)=\sqrt{s^2+R^2+2sR\cos b \cos l} 
\end{equation}
where $R\simeq 8.5$ kpc is the distance of sun from the galactic centre and $l,b$ are galactic coordinates 
known as longitude and latitude, respectively. Now upon evaluating the integral eq.(\ref{lsi}) can be written
in a simpler form as
\begin{equation}
 \frac{d \Phi}{d E_\nu}= D_h \delta(E_\nu-M_\Sigma/2)
\end{equation}
where 
\begin{equation}
 D_h=1.7\times10^{-13}\left(\frac{1{\rm PeV}}{M_\Sigma}\right) \left( \frac{10^{28} s}{\tau_\Sigma} \right) {\rm cm^{-2}s^{-1}sr^{-1}}~.
\end{equation}
The number of expected neutrino events at Icecube in small energy bin $\Delta E_\nu$ can be obtained by evaluating 
the following integral over the differential flux:
\begin{equation}
N=\int_{\Delta{E_\nu}} \frac{d \Phi}{d E_\nu} A (E_\nu) d E_\nu \label{event}
\end{equation}
where $A (E_\nu)$ is the exposure at energy $E_\nu$. From the plot provided by Icecube collaboration\cite{IceCube:2013}, it is clear that 
the number of events near PeV energy for data accumulated during 662 days is nearly two, or more specifically at $E_\nu=1.3$ PeV,
$N=2$ where both contain certain experimental uncertainties. To explain this data point we take the mass of the DM to be
$M_\Sigma=2.6$ PeV and from \cite{Anchordoqui:2013qsi} we find the 662 days exposure at $E_\nu=1.3$ PeV as 
$A(E_\nu=1.3{\rm PeV})\simeq 10^{14}~{\rm cm^2 ~s ~sr}$. Using these values in the above integral (eq.(\ref{event})), the 
number of events can be easily estimated. Now in the LHS of eq.(\ref{event}) if we set the number of events to be $\sim 2$,
then we get $\tau_\Sigma \simeq (2-3) \times10^{28} ~s~$ which is in
reasonable agreement with the value used in Sec.\ref{sec:bench} and \cite{IceCube-Models}.

\section{Discussion, Summary, and Conclusion}\label{sec:sum}
In this work  we have suggested non-SUSY SO(10) theory with extremely small
 matter parity violation as the underlying origin of decaying Majorana fermion dark matter that manifests as PeV scale high energy neutrino flux at IceCube. 
Since the Dirac neutrino Yukawa couplings in SO(10) are as hierarchical as the
up-quark Yukawa couplings leading to predominantly and naturally
hierarchical RH$\nu$ masses through type-I seesaw we have noted that
equal branching ratio constraint on DM decay to every neutrino flavor can never be achieved with 
one common mixing as proposed in the benchmark model of extended
SM with one common heavy RH$\nu$ mass \cite{Rott:2014}. Despite such
diversities we have shown here that dark matter
 decay operates with
separate distinct mixing  with each of the RH$\nu$s. We have determined these three mixings 
for the first time  by solving the underlying constraint
relations  on DM decay to three different neutrino flavors. The patterns of these three mixings are also found to
depend upon the light neutrino mass hierarchy such as NH, IH or QD.
%------------------------------------------
 In the SM extension \cite{Rott:2014}, the 
RH$\nu$s and the DM fermion $\Sigma_F$ are externally added singlets
with their assumed bare masses having no Higgs origins.
 Whereas in SO(10) models the RH$\nu$ is a member of matter unified
 spinorial representation  ${16}_F$ of odd matter parity, we have identified the decaying DM $\Sigma_F$ to be a
Majorana fermion singlet
of even matter parity contained in ${45}_F$ of SO(10). Whereas the
Higgs origin of RH$\nu$ is well known in SO(10),  we have shown how the
$\Sigma_F$ mass is predicted through matter parity
conserving Higgs Yukawa interaction involving ${210}_H$ that
drives GUT symmetry breaking in both the minimal models suggested here: Model-I and Model-II.

 The direct breaking of SO(10)  to SM predicts type-I seesaw dominance over type-II seesaw through a mild fine tuning of ${126}_H^{\dagger}$ Yukawa coupling  which is also corroborated by the neutrino oscillation data and  the  underlying quark-lepton symmetry of SO(10).\\ 

Deeper theoretical origin of extremely small $N_i-\Sigma_F$ mixings is suggested in
this work to be due to Planck-scale assisted intrinsic matter parity discrete
symmetry  breaking which is  most desired  for the resolution of the
associated cosmological domain wall problem. This spontaneous symmetry breaking origin  is also modeled to occur due to the bounded
vacuum expectation  value $V_{\chi}\le v_{\rm ew}=246$ GeV of a
matter-parity odd real scalar singlet $\chi_S(1,0,1)$. As a result, we predict
  a light Higgs scalar singlet of perturbative mass bound $M_{\chi_S}
\le 860$ GeV accessible to experimental searches at LHC and planned
accelerators. The renormalizable and non-renormalizable values of
different Yukawa couplings of $\chi_S$ underlying the small mixings
have been explicitly derived by us for NH type light neutrino masses and the
method can be used to predict the corresponding sets of values for
other hierarchies like QD and IH. Interestingly, we have also shown how the
mixings can be dynamically generated through a ${\rm dim.}5$ operator but without  using a
Planck-mass fermion singlet $N^{\prime}$. Whereas the SM extended
model \cite{Rott:2014} rests upon the QD type light neutrino masses
and is likely to be severely constrained if other types of hierarchies
such as NH or IH are established in future, our SO(10) ansatz fits all
types of mass hierarchies. Once the light neutrino mass hierarchy is
established, this analysis will fix the associated
Yukawa coupling matrix needed to explain the origin of $N_i-\Sigma_F$ mixings.

%%%%%%%%%%%%%%%%%%%%%%%%%%%%%%%%%%%%%%%%%%%%%%%%%%%%%%%%%%%%%%%%%%%%%%%%%
 Quite interestingly,
we have further shown that  this light Higgs scalar $\chi_S$, besides generating the
spontaneous breaking origin of the
 dark matter 
mixings with $N_i$, also resolves the issue of vacuum stability that persists in
the SM and its extension \cite{Rott:2014}. This  scalar singlet has been
identified as the real part of the complex scalar singlet in
${16}_H^{\dagger}$ and its  ${\cal O} (M_W)$ mass needed for model
predictions has been realized by fine tuning of parameters in the
SO(10) invariant potential. Self-consistency of the model predictions
is guaranteed from the fact that the same fine tuning that gives ${\cal O} (M_W)$
mass to $Re(\chi_S)$ also renders GUT scale mass to the imaginary
part of the scalar singlet which, thus, decouples from contributing to all
relevant physical quantities at all lower scales discussed in this work.
%%%%%%%%%%%%%%%%%%%%%%%%%%%%%%%%%%%%%%%%%%%%%%%%%%%%%%%%%%%%%%%%%%%%%%%%        

We have also shown that this theory of decaying dark matter in SO(10) 
 predicts two different minimal
modifications of the grand desert through Model-I and Model-II which
achieve precision gauge coupling unification with vacuum stability of
the SM in each case which are absent in the bench mark model.
  In Model-I the presence of the scalar component
$\kappa(3,0,8)\subset {210}_H$ of  mass $M_{\kappa}=10^{9.23}$ GeV in the grand
 desert achieves precision unification  at
 $M_U^0=10^{15.224}$ GeV. This  representation ${210}_H$ also
 provides the Higgs origin of the $\Sigma_F$ mass in addition to driving
 the direct breaking of SO(10) to SM. Although only an incomplete and
 limited unification aspect of this 
 grand desert modification as in Model-I was noted earlier in 1993,
 the precision fitting with neutrino oscillation data,  determination of
 $M_{\kappa}$, $\tau_p$, and the model association with decaying dark
 matter dynamics are new including the three different mixings and the
 prediction of the new Higgs scalar singlet $\chi_S$. Furthermore the completion of
 vacuum stability which was  
 absent in the earlier work has been achieved in this work because of its
 decaying dark matter dynamics that predicts light Higgs scalar
 $\chi_S$ with mass $m_2=M_{\chi_S}=177$ GeV for which the perturbative
 upper bound is $M_{\chi_S} < 860$ GeV.

The pattern of precision coupling unification in the minimal Model-II
is the first observation as noted in this work. In this case the new minimal modification of the grand desert is
achieved by supplying the single scalar component $\eta (3,-1/3, 6)
\subset {126}_H$ of mass $M_{\eta}=10^{10.7}$ GeV again with precision
proton lifetime predictions and solution to
vacuum stability emerging from the model explanation of the dynamics
of decaying dark matter. Like Model-I this model also predicts LHC
detectable new light scalar with mass upper bound $M_{\chi_S}\le 860$
GeV. Our vacuum stability resolution predicts a lower mass for this
Higgs scalar $m_2=M_{\chi_S}=177$ GeV.  

Despite two large sized
 representations ${126}_H$ and ${210}_H$, the fermionic representation 
${45}_F$ , and also ${45}_H, {16}^{\dagger}_H$, Model-I and Model-II
 are noted to predict proton lifetime
 prediction up to $\tau_p\simeq 10^{35}$ yrs with reduced threshold
 uncertainties. This value is clearly within the accessible limit of
 ongoing Superkamiokande and Hyperkamiokande experiments.  

Heavy Higgs scalar ( mass $M_{\xi}\sim M_{GUT}$) exchange mechanism
that operates between the DDM and the SM Higgs has been shown to
successfully generate the desired relic density of $\Sigma_F$. Its decay at the galactic center is also found to yield suitable  IceCube
neutrino flux.

In this work we have followed direct spontaneous symmetry breaking of
SO(10)$\to$ SM as explained in Sec.\ref{sec:embed}. Realization of
scalar or fermionic WIMP DM through $SU(5)\times U(1)_X$ intermediate
symmetry at high scale has been discussed while advancing the
application of matter parity in SO(10)
\cite{Kadastik:2009,Hambye:2010,Frig-Ham:2009}. All our results can
apply also through such high scale $SU(5)\times U(1)_X$ intermediate
breaking. An alternative interesting possibility of spontaneous
symmetry breaking of the Abelian subgroup $U(1)_X$ near the
electroweak scale \cite{Khalil}   resulting in an extra low-mass
neutral gauge boson is  beyond the scope of the present work.  
  
In conclusion, we have found that this  non-SUSY SO(10) with its intrinsic
matter parity  is  a self sufficient
theory of decaying dark matter and neutrino physics that predicts all the
particle content, the Higgs or seesaw origin of their masses and
mixings, and the minimal modifications of the grand desert by a single
scalar of intermediate mass for precision gauge coupling unification without the
assistance of any externally imposed stabilising discrete symmetry or fermion
  singlets of SO(10) such as $N^{\prime}$. The  $N_i-\Sigma_F$ mixings are extremely small because of the
underlying intrinsic matter parity of broken SO(10) as gauged discrete symmetry  whose breaking must be assisted by
gravity or the Planck-scale for cosmologically safe acceptable
solutions. Thus, a  Planck scale suppression  of mixings naturally
emerges in both Model-I and Model-II.
 Another factor contributing to the suppression of
this mixings is the matter parity conservation constraint of SM 
that restricts the singlet VEV $V_{\chi}\leq
 v_{\rm ew}$ leading to the experimentally
testable new Higgs scalar mass with perturbative upper bound
$M_{\chi_S}\le 860$ GeV. The resolution of  vacuum instability issue
predicts its actual mass $m_2=M_{\chi_S}=177$ GeV. 
Starting from parity and matter parity invariant SO(10) gauge theory,
in this work we have suggested experimental evidence of  matter parity
nonconservation  at IceCube.\\
Prospects of this model for scalar singlet WIMP dark matter including
the detection possibility of the new Higgs scalar $\chi_S(1,0,1)$
 would be
discussed elsewhere.
%------------------------------------
%%%%%%%%%%%%%%%%%%%%%%%%%%%%%%%%%%%%%%%
%%%%%%%%%%%%%%%%%%%%%%%%%%%%%%%%%%%%%%%

\section{APPENDIX A: Diagonalisation of RH$\nu$ Mass Matrices} \label{sec:VP}
It is to be noted that the light neutrino mixing matrix in PMNS parametrization is given by 
\begin{equation}
 U_{\rm{PMNS}}= \left( \begin{array}{ccc} c_{12} c_{13}&
                      s_{12} c_{13}&
                      s_{13} e^{-i\delta}\cr
-s_{12} c_{23}-c_{12} s_{23} s_{13} e^{i\delta}& c_{12} c_{23}-
s_{12} s_{23} s_{13} e^{i\delta}&
s_{23} c_{13}\cr
s_{12} s_{23} -c_{12} c_{23} s_{13} e^{i\delta}&
-c_{12} s_{23} -s_{12} c_{23} s_{13} e^{i\delta}&
c_{23} c_{13}\cr
\end{array}\right) 
diag(e^{\frac{i \alpha_M}{2}},e^{\frac{i \beta_M}{2}},1) .
\end{equation}
For the sake of simplicity, primarily during the calculation of the light neutrino mass matrix, the 
Majorana phases are assumed to be zero and unphysical phases are also not considered. The 
best fit values of the solar and atmospheric mass squared differences 
\cite{PDG:2016} are used to calculate the light neutrino mass
eigenvalues for different hierarchies
as shown in table \ref{tab_lhnu}.
%----------------------------------------
\begin{table}[!h]
%\caption{}
\caption{Light neutrino mass eigenvalues for different mass hierarchies (taking into account best fit values of the mass squared differences)}
\begin{center}
\begin{tabular}{ |c|c|c|c| } 
\hline
 Mass&  $\hat{m}_{\nu_1}$ (eV) & $\hat{m}_{\nu_2}$ (eV) & $\hat{m}_{\nu_3}$ (eV)\\ 
 ordering& & & \\\hline
 NH &  $0.00127$ & $0.008809$ & $0.049815$ \\ 
 & &  & \\\hline
 QD1 & $0.0630$  & $0.0636$  &$0.08$\\ 
  & &  & \\\hline
 QD2 & $0.1938$  & $0.1940$ & $0.2$\\ 
 & &  &\\\hline
 IH & $0.0476$ & $0.0484$ & $0.00127$\\ 
  &  & & \\\hline
% cell7 & cell8 & cell9 & & &\\ \hline
\end{tabular}
\label{tab_lhnu}
\end{center}
\end{table}
%---------------------------------
Using these mass eigenvalues and best fit values\cite{PDG:2016} of the mixing angles and Dirac CP phase, light neutrino mass matrix $m_\nu$
is calculated. It is then straightforward to calculate the RH$\nu$
mass matrix ($M_N$) using $m_\nu$ and $m_D$  in the u-quark or d-quark diagonal basis.
Further $M_N$ is diagonalised by unitarity $V_P$ matrix. This $V_P$ matrix is written in two parts, one $3\times3$ matrix which includes mixing angles,
Dirac CP phase and unphysical phases) and the other is the  multiplicative diagonal Majorana phase matrix $(P_m)$.
\vspace{1cm}\\
The $V_P$ matrices for different cases are given below
\vspace{.5cm}\\
{\it Case 1: NH with u-quark diagonal basis}
\begin{eqnarray}
V_P  =  \left( 
\begin{array}{ccc}
 -0.513-0.858 i & -0.0023-0.0011 i & (0.8921+8.991i)\times10^{-6}  \\
 -0.0016+0.0019 i & 0.130 -0.991 i & 0.00035 -0.0027 i \\
 (0.17+1.4i)\times10^{-5} & -0.0003-0.0027 i & 0.1217 +0.992 i \\
\end{array}\right) 
\times P_m
\end{eqnarray}
where $P_m= diag  (e^{i\alpha_M/2},e^{i\beta_M/2},1)$ with $\alpha_M=79.83^\circ, \beta_M=-19.1^\circ$ and RH$\nu$ mass eigenvalues are
$\hat{M}_{N_1}=8.9\times10^5$ GeV, $\hat{M}_{N_2}=2.28\times10^9$ GeV, $\hat{M}_{N_3}=1.19\times10^{15}$ GeV.
\vspace{.5cm}\\
{\it Case 2: QD1 with u-quark diagonal basis}
\begin{equation}
V_P=\left(
\begin{array}{ccc}
 -0.992-0.118 i & -0.0003-0.00007 i & (0.1193+1.314 i)\times10^{-6} \\
 -9.38\times10^{-6}+0.0003 i & 0.0944, -0.995 i & 0.0004 -0.004727 i \\
 9.363\times10^{-9}+1.35\times10{-6} i & -2.921\times10^{-6}-0.00042361 i & 0.0068+0.999 i \\
\end{array}
\right) \times P_m
\end{equation}
where $P_m= diag  (e^{i\alpha_M/2},e^{i\beta_M/2},1)$ with $\alpha_M=167.08^\circ, \beta_M=-11.57^\circ$ and RH$\nu$ mass eigenvalues are
$\hat{M}_{N_1}=4.7\times10^4$ GeV, $\hat{M}_{N_2}=9.64\times10^8$ GeV, $\hat{M}_{N_3}=9.22\times10^{13}$ GeV.
\vspace{.5cm}\\
{\it Case 3: QD2 with u-quark diagonal basis}
\begin{equation}
V_P=\left(
\begin{array}{ccc}
 -0.999-0.0285 i & -0.0003-0.0001 i & (0.011 + 2.23i)\times10^{-8} \\
 0.00033 i & 0.3389 -0.940 i & 0.000034 -0.00009 i \\
 7.196\times10^{-9}+1.20i\times10^{-6}  & -0.000100356 i & 0.0059+0.999 i \\
\end{array}
\right) \times P_m
\end{equation}
here $P_m= diag  (e^{i\alpha_M/2},e^{i\beta_M/2},1)$ with $\alpha_M=-171.31^\circ, \beta_M=-40.18^\circ$ and RH$\nu$ mass eigenvalues are
$\hat{M}_{N_1}=1.54\times10^3$ GeV, $\hat{M}_{N_2}=3.47\times10^8$ GeV, $\hat{M}_{N_3}=3.35\times10^{13}$ GeV.
\vspace{.5cm}\\
{\it Case 4: IH with u-quark diagonal basis}
\begin{equation}
V_P=\left(
\begin{array}{ccc}
 (-3.101-0.01281i)\times10^{-6}  & 0.00001 +0.00047 i & 0.0133 -0.999 i \\
 0.0022+0.00147 i & -0.8378-0.5459 i & -0.00038-0.00027 i \\
 0.837 +0.5459 i & 0.00225 +0.00147 i & 0 \\
\end{array}
\right) \times P_m
\end{equation}
here $P_m= diag  (e^{i\alpha_M/2},e^{i\beta_M/2},1)$ with $\alpha_M=113.83^\circ, \beta_M=113.83^\circ$ and RH$\nu$ mass eigenvalues are
$\hat{M}_{N_1}=2.97\times10^{15}$ GeV, $\hat{M}_{N_2}=2.49\times10^9$ GeV, $\hat{M}_{N_3}=6.15\times10^{3}$ GeV.
\vspace{.5cm}\\
%The $V_P$ matrices for different cases are given below
%\vspace{.5cm}\\
{\it Case 5: NH with d-quark diagonal basis}
\begin{equation}
V_P=\left(
\begin{array}{ccc}
 -0.9348-0.2723 i & -0.0282-0.225i & -0.00219+0.00803 i \\
 0.2187\, +0.0637 i & -0.119-0.9655 i & 0.00554\, +0.0386 i \\
 -0.000445-0.00351 i & 0.00499\, +0.0394 i & 0.1254\, +0.9912 i \\
\end{array}
\right)\times P_m 
\end{equation}
where $P_m= diag  (e^{i\alpha_M/2},e^{i\beta_M/2},1)$ with $\alpha_M=-177.08^\circ, \beta_M=-3.6^\circ$ and RH$\nu$ mass eigenvalues are
$\hat{M}_{N_1}=5.85\times10^4$ GeV, $\hat{M}_{N_2}=3.69\times10^9$ GeV, $\hat{M}_{N_3}=1.14\times10^{15}$ GeV.
\vspace{.5cm}\\
{\it Case 6: QD1 with d-quark diagonal basis}
\begin{equation}
V_P=\left(
\begin{array}{ccc}
 -0.9046-0.3615 i & 0.00073 -0.225 i & -0.00322+0.00801 i \\
 0.2095 +0.0840 i & 0.00504 -0.973 i & 0.00053 +0.0403 i \\
 -0.000031-0.0035 i & 0.00036 +0.0411 i & 0.0088 +0.9991 i \\
\end{array}
\right) \times P_m 
\end{equation}
where $P_m= diag  (e^{i\alpha_M/2},e^{i\beta_M/2},1)$ with $\alpha_M=145.5^\circ, \beta_M=-5.3^\circ$ and RH$\nu$ mass eigenvalues are
$\hat{M}_{N_1}=4.71\times10^3$ GeV, $\hat{M}_{N_2}=9.9\times10^8$ GeV, $\hat{M}_{N_3}=9.14\times10^{13}$ GeV.
\vspace{.5cm}\\
{\it Case 7: QD2 with d-quark diagonal basis}
\begin{equation}
V_P=\left(
\begin{array}{ccc}
 0.9044 +0.361 i & 0.00102 -0.225 i & -0.00323+0.0080 i \\
 -0.2094-0.0841 i & 0.00621 -0.9733 i & 0.00048+0.0406 i \\
 0.00002 +0.0035 i & 0.0003 +0.0414 i & 0.0078 +0.999 i \\
\end{array}
\right) \times P_m
\end{equation}
where $P_m= diag  (e^{i\alpha_M/2},e^{i\beta_M/2},1)$ with $\alpha_M=9.27^\circ, \beta_M=-3.35^\circ$ and RH$\nu$ mass eigenvalues are
$\hat{M}_{N_1}=1.55\times10^3$ GeV, $\hat{M}_{N_2}=3.5\times10^8$ GeV, $\hat{M}_{N_3}=3.35\times10^{13}$ GeV.
\vspace{.5cm}\\
{\it Case 8: IH with d-quark diagonal basis}
\begin{equation}
V_P=\left(
\begin{array}{ccc}
 0.00920815\, +0.000266017 i & -0.213401+0.0719151 i & 0.0459181\, -0.973188 i \\
 0.040555\, -0.0143812 i & -0.922805+0.309508 i & -0.0107965+0.225095 i \\
 0.942383\, -0.33163 i & 0.0413766\, -0.0145607 i & 0.00331518\, -0.00116663 i \\
\end{array}
\right)\times P_m 
\end{equation}
here $P_m= diag  (e^{i\alpha_M/2},e^{i\beta_M/2},1)$ with $\alpha_M=-141.25^\circ, \beta_M=-144.9^\circ$ and RH$\nu$ mass eigenvalues are
$\hat{M}_{N_1}=3.177\times10^{15}$ GeV, $\hat{M}_{N_2}=2.23\times10^9$ GeV, $\hat{M}_{N_3}=6.59\times10^{3}$ GeV.

\section{APPENDIX B: Renormalization Group Solutions for Mass Scales  and Threshold Effects}\label{sec:RGE}
%%%%%%%%%%%%%%%%%%%%%%%%%%%%%%%%%%%%%%%%%%%%%%%%%%%%%%%%
%%%%%%%%%%%%%%%%%%%%%%%%%%%%%%%%%%%%%%%%%%%%%%%%%%%%%%%%%%%%%%%%%
%%%%  two loop eqs for gauge couplings and top couplings %%%%%%%%
The RG equations for SM gauge couplings and top quark yukawa coupling at two loop level are given by
\begin{align}
{dy_{top} \over d \ln \mu}= & {1 \over 16\pi^2}\left({9 \over 2}y_{top}^2-{17 \over 12}g_{1Y}^2
 -{9 \over 4}g_{2L}^2-8g_{3C}^2 \right)y_{top} \\ \nonumber
  +  & {1 \over (16\pi^2)^2} [-{23 \over 4}g_{2L}^4-{3 \over 4}g_{2L}^2g_{1Y}^2+{1187 \over 216}g_{1Y}^4 + 9g_{2L}^2g_{3C}^2+{19 \over 9}g_{3C}^2g_{1Y}^2-108g_{3C}^4 
\\ \nonumber
+& \left({225 \over 16}g_{2L}^2+{131 \over 16}g_{1Y}^2+36g_{3C}^2 \right)y_{top}^2+6(-2y_{top}^4-2y_{top}^2\lambda_{\phi}+\lambda_{\phi}^2) ], \\ %\nonumber
 {dg_{1Y} \over d \ln \mu}= & {1 \over 16\pi^2}\left({41 \over 6}g_{1Y}^3\right)+{1 \over (16\pi^2)^2}\left({199 \over 18}g_{1Y}^2+{9 \over 2}g_{2L}^2+{44 \over 3}g_{3C}^2-
{17 \over 6}y_{top}^2\right)g_{1Y}^3,  \\%\nonumber
{dg_{2L} \over d \ln \mu}= & {1 \over 16\pi^2}\left(-{19\over 6}g_{2L}^3\right)+{1 \over (16\pi^2)^2}\left({3 \over 2}g_{1Y}^2+{35 \over 6}g_{2L}^2+12g_{3C}^2-
{3 \over 2}y_{top}^2\right)g_{2L}^3 , \\%\nonumber
{dg_{3C} \over d \ln \mu }= & {1 \over 16\pi^2}\left(-7g_{3C}^3\right)+{1 \over (16\pi^2)^2}\left({11 \over 6}g_{1Y}^2+{9 \over 2}g_{2L}^2-26g_{3C}^2-
2y_{top}^2\right)g_{3C}^3.
\end{align}
\vspace{1cm}\\ 
 The matching formula for different gauge couplings($\alpha^{-1}_i,i=2L,Y,3C$) at the unification scale is given by 
 \be
\alpha^{-1}_i (M_U)=\alpha^{-1}_G-\frac {\lambda_i(M_U)}{12\pi},\label{eq:matching1} 
\ee
 where  $\lambda_i,i=2L,Y, 3C$ are matching functions due to superheavy scalars (S), Majorana fermions (F) and gauge bosons (V),

\ba
\lambda_{i}^S(M_U) & =&\sum_{j}Tr\left(t_{iSj}^2\hat{p}_{Sj}\ln{M_j^S \over M_U}\right),\nonumber \\
 \lambda_{i}^F(M_U) & =&\sum_{k} 4 Tr\left(t_{iFk}^2\ln{M_k^F \over
  M_U}\right), \nonumber \\ 
\lambda_{i}^V(M_U) & =&\sum_{l} Tr\left(t_{iVl}^2\right)
-21\sum_lTr\left(t_{iVl}^2\ln{M_l^V \over M_U}\right),\label{eq:matching2}
\ea
 where $t_{iS}$, $t_{iF}$ and  $t_{iV}$ represent the matrix representations of 
broken generators for scalars, Majorana fermions, and gauge bosons, respectively. The term $\hat{p}_{Sj}$ denotes the projection operator  that removes the Goldstone components from the scalar that contributes to spontaneous symmetry breaking. 

Decomposition of different 
SO(10) representations under $G_{213}$ with respect to their superheavy components are given in Table.{\ref{tab:decomp}}. We use the following notations for the respective components  $10_H\supset H_i$,~~~ $126_H\supset H_i^{\prime}$, ~~~
$16_H \supset H_i^{\prime\prime}$,~~~~$45_H \supset S_i$,~~~$210_H\supset S_i^{\prime}$,~~~~~ and $45_F \supset F_i$.

\begin{table}[h!]
\caption{Superheavy components of  
SO(10) representations under the SM gauge group $G_{213}$ used to
estimate GUT threshold effects in Model-I. In Model-II
$H_5^{\prime}=\eta^(3,-1/3,6)\subset {126}_H$ is excluded as it has lower mass. }
\vskip 0.5cm
\begin{tabular}{| p{14 cm} | }
\hline
 $10_H \supset H_1(1,-1/3,3)+H_2(1,1/3,\bar{3})+ H_3(2,-1/2,1) $ 
 \\ \hline  
$ 126_H  \supset  H_1^{\prime}(1,-1/3,3)+H_2^{\prime}(1,1/3,\bar{3})+H_3^{\prime}(3,1,1)+H_4^{\prime}(3,1/3,\bar{3})+H_5^{\prime}(3,-1/3,\bar{6})+H_6^{\prime}(1,-1/3,3)+H_7^{\prime}(1,-4/3,3)+H_8^{\prime}(1,4/3,6)+H_{9}^{\prime}(1,1/3,6)+H_{10}^{\prime}(1,-2/3,6)+H_{11}^{\prime}(2,1/2,1)+H_{12}^{\prime}(2,-1/2,1)+H_{13}^{\prime}(2,7/6,3)+H_{14}^{\prime}(2,-7/6,\bar{3})+H_{15}^{\prime}(2,1/2,8)+H_{16}^{\prime}(2,-1/2,8)$ 
\\ \hline
 $210_H  \supset S_1^{\prime}(1,2/3,3)+S_2^{\prime}(1,-2/3,\bar{3})+S_3^{\prime}(1,0,8)+S_4^{\prime}(2,1/6,3)+S_5^{\prime}(2,-5/6,3)+S_7^{\prime}(2,-1/6,\bar{3})+S_8^{\prime}(3,0,1)+S_9^{\prime}(3,2/3,3)+S_{10}^{\prime}(3,-2/3,\bar{3})+S_{11}^{\prime}(1,1,1)+S_{12}^{\prime}(1,1,-1)+S_{13}^{\prime}(1,5/3,3)+S_{14}^{\prime}(1,2/3,3)+S_{15}^{\prime}(1,-1/3,3)+S_{16}^{\prime}(1,1/3,\bar{3})+S_{17}^{\prime}(1,-2/3,\bar{3})+S_{18}^{\prime}(1,-5/3,\bar{3})+S_{19}^{\prime}(1,1,8)+S_{20}^{\prime}(1,0,8)+S_{21}^{\prime}(1,-1,8)+S_{22}^{\prime}(2,-1/2,1)+S_{23}^{\prime}(2,-3/2,1)+S_{24}^{\prime}(2,1/6,3)+S_{25}^{\prime}(2,-5/6,3)+S_{26}^{\prime}(2,5/6,6)+S_{27}^{\prime}(2,-1/6,6)+S_{28}^{\prime}(2,3/2,1)+S_{29}^{\prime}(2,5/6,\bar{3})+S_{30}^{\prime}(2,-1/6,\bar{3})+S_{31}^{\prime}(2,1/6,\bar{6})+S_{32}^{\prime}(2,-5/6,\bar{6})+S_{33}^{\prime}(2,1/2,1)$
\\ \hline 
 $16_H \supset H_1^{\prime\prime}(2,1/6,3)+H_2^{\prime\prime}(1,1/3,\bar{3})+H_3^{\prime\prime}(1,-2/6,\bar{3})+H_4^{\prime\prime}(1,1,1)+H_5^{\prime\prime}(2,-1/2,1)$    
\\ \hline 
 $45_F \supset F_1(1,1,1)+ F_2(1,-1,1)+ F_3(3,0,1)+F_4(2,1/6,3)+ F_5(2,-5/6,3)
+ F_6(2,5/6,\bar{3})+ F_7(2,-1/6,\bar{3})+ F_8(1,2/3,3) +
F_9(1,-2/3,\bar{3})
+F_{10}(1,0,8)$ 
\\ \hline  
\end{tabular}
\label{tab:decomp}
\end{table}

Using Table.{\ref{tab:decomp}} we calculate values of matching functions
$\lambda_i(M_U),i=2L,Y,3C$. 

\subsection{Minimal Model-I}

  Analytic formulas for the unification scale and $\kappa$
 mass constrained by gauge coupling unification are 

\begin{align}
\ln {M_U \over M_Z} &= \frac{16\pi}{187\alpha}\left({7\over8}-\frac{10\alpha}
{3\alpha_{3C}}+s_W^2 \right)+\Delta_{I}^U \nonumber\\
\ln {M_{\kappa}  \over M_Z}& = \frac{4\pi}{187\alpha}\left(15+\frac{23\alpha}
{3\alpha_{3C}}-63s_W^2 \right)+\Delta_{I}^{\kappa}  \nonumber\\
{1 \over \alpha_G}& = {3 \over 8\alpha}+{1 \over 187\alpha}\left({347 \over 8}
 +{466\alpha \over 3\alpha_{3C}}-271s_W^2 \right)+\Delta_{I}^{\alpha_G} \label{eq:guform} 
\end{align}

where $s_W^2=\sin^2\theta_W (M_Z)$ and the first term in the above eq.(\ref{eq:guform}) represent one loop contributions. The terms $\Delta_I^i$, $i=U,~\kappa, ~ \alpha_G$ denoting the threshold corrections due to unification scale($M_U$), intermediate scale ($M_{\kappa}$) and GUT fine structure constant({$1 \over \alpha_G$}) are  given by
 
\begin{align}
\Delta_{I}^{\kappa} =\Delta\ln{M_{\kappa}\over M_Z}& ={1\over 561}(-48\lambda_{2L}+25\lambda_Y+23\lambda_{3C}) \nonumber \\
\Delta_{I}^{U} =\Delta\ln{M_U \over M_Z}& ={5\over 3366}(9\lambda_{2L}+7\lambda_Y
-16\lambda_{3C}) \nonumber\\
\Delta_{I}^{\alpha_{G}} =\Delta\left({1 \over \alpha_G}\right)&={1\over 13464\pi}(-945\lambda_{2L}+1135\lambda_Y+932\lambda_{3C}),
\label{eq:theq1} 
\end{align}

Here $a_i$,~~ $b_{ij}$ and  $a^{\prime}_i$,~~~ $b^{\prime}_{ij}$ 
are one loop and two loop beta function coefficients for the range of mass scales $M_Z-M_{\kappa}$ and $M_{\kappa}-M_U$, respectively, they  are given in the Table.{\ref{tab:12lp}}

%==========================================================================     
\begin{table}[h!]
\caption{One loop and two loop beta function coefficients for RG evolution of gauge couplings}
\begin{center} 
\begin{tabular}{| p{2.5cm} | p{2.5cm} | p{4.1cm} |p{4cm}|p{4cm}|}
\hline   Mass scales($\mu$) &~~~~~~${\bf{a_{i}}}$ &~~~~~~~${\bf{b_{ij}}}$
\\ \hline 
 $M_Z < \mu < M_{\kappa} $ & 
  \centering
 $\begin{pmatrix}                    
 \frac{41}{10} \\ \frac{-19}{6} \\ -7 \\
\end{pmatrix}$      
 &   
 $\begin{pmatrix} 
    \frac{199}{50} & \frac{27}{10} & \frac{44}{5} \\
    \frac{9}{10} & \frac{35}{6} &  12 \\
    \frac{11}{10} & \frac{9}{2} & -26 \\
\end{pmatrix}$  
 \\ \hline  
 $M_{\kappa} < \mu < M_U $ & 
  \centering
 $\begin{pmatrix}                    
 \frac{41}{10} \\ -\frac{1}{2} \\ -\frac{11}{2} \\
 \end{pmatrix}$      
 &   
 $\begin{pmatrix} 
    \frac{199}{50} & \frac{27}{10} & \frac{44}{5} \\
    \frac{9}{10} & \frac{161}{2} & 108 \\
    \frac{11}{10} & \frac{45}{2} & 37 \\
   \end{pmatrix}$ 
\\ \hline 
\end{tabular} 
\end{center}
\label{tab:12lp}
\end{table}

Using one loop beta function coefficients from Table \ref{tab:12lp} in  
eq.(\ref{eq:guform}), we get 
\ba 
M_U &=&10^{15.2446}\,{\rm GeV},\nonumber\\ 
M_{\kappa}&=&10^{9.23}\,{\rm GeV},\nonumber\\
\alpha_{G}^{-1} &=&41.77.
\label{eq:RGsol}
\ea
 Contribution due to gauge coupling matching at the GUT scale that occurs even when all superheavy masses are identical to $M_U$ \cite{cmgmp:1985} has been included.  Using  eq.(\ref{eq:theq1}) and values of matching functions,  we estimate the corrections to mass scales due to superheavy masses in partially degenerate case when all the superheavy component masses belonging to
a definite SO(10) representation are degenerate \cite{rnm-mkp:1993,lmpr:1995,mkp:1987},

\begin{align}
\Delta\ln{M_U \over M_Z}& =-0.0196078\eta_{10}-0.037433\eta_{126}
 +0.0322341\eta_{210} \nonumber \\
\Delta\ln{M_{\kappa} \over M_Z}& = 0.05882\eta_{10}-0.160428\eta_{126}
 +1.13815\eta_{210} \nonumber \\
\Delta\left({1 \over \alpha_G}\right)& =0.059293\eta_{10}+1.62758\eta_{126}
+1.72778\eta_{210} ~.\nonumber \\
\end{align}

Maximising the threshold uncertainty in $M_U$ leads to  ~~~~~~~~
\ba
\Delta \ln \left({M_U \over M_Z}\right)&=&\pm 0.089275|\eta_{SH}| ,~~\nonumber\\
\Delta \ln \left({M_{\kappa} \over M_Z}\right)&=& \pm 1.23975|\eta_{SH}| , ~~\nonumber\\
\Delta \left({1 \over \alpha_G}\right)&=&\pm 0.040907|\eta_{SH}|,
\label{eq:maxthsh} 
\ea 
where $\eta_{SH}=\ln ({M_{SH} \over M_U})$ and $(M_{SH}/M_U)= n(1/n)$ with
 plausible allowed values of real number $n=1-10$.  
 
Similarly the threshold effects due to superheavy gauge boson components in ${45}_V$ have
been estimated as shown in eq.(\ref{eq:errors}) and
eq.(\ref{eq:thcoef}) in Sec.7. Threshold effects due to ${45}_F,{45}_H,{54}_H$ and ${16}_H$ are noted to vanish in the minimal model due to the theorem \cite{RNM-PLB:1993}.  
In the case of complete degeneracy in superheavy scalar masses from
all representations, including the threshold effects are
\ba
\Delta \ln \left({M_U \over M_Z}\right)&=&\pm 0.0248|\eta_{SH}| ,~~\nonumber\\
\Delta \ln \left({M_{\kappa} \over M_Z}\right)&=& \pm 1.0365|\eta_{SH}| .\label{eq:mindegthsh}
\ea
Most dominant threshold uncertainty on the unification mass and proton lifetime occurs due to superheavy gauge bosons
\be
{[\Delta \ln \left({M_U \over M_Z}\right)]}_V= -0.9358\eta_V, \label{eq:minthV}
\ee
where $\eta_V=\ln \left(M_V/M_U\right)$. Thus degenerate superheavy gauge boson
masses few times lighter than $M_U$ can cause substantial enhancement
in proton lifetime prediction.

\subsection{Minimal Model-II} 
 Analytic formulas for the two mass scales $M_{\rm U}$ and $M_{\eta}$  are
\begin{align}
\ln {M_U \over M_Z} &=\frac{18\pi}{247\alpha}\left(1+{4 \over 3} s_W^2-4{\alpha \over \alpha_{3C}} \right)+\Delta_{II}^U \nonumber\\
\ln {M_{\eta}  \over M_Z}& =\frac{4\pi}{247\alpha}\left(16+\frac{55}{3}\frac{\alpha}{\alpha_{3C}}-61s_W^2 \right)+\Delta_{II}^{\eta} \nonumber\\
{1 \over \alpha_G}& ={1 \over 494\alpha}\left(241-502s_W^2+{1060 \over 3}{\alpha \over \alpha_{3C}}\right)+\Delta_{II}^{\alpha}\label{eq:guformnm} 
\end{align}
 where $\Delta_{II}^i$, $i=U,~\eta, ~ \alpha_G$ denote  threshold corrections  to unification scale($M_U$), intermediate scale($M_{\eta}$), and inverse GUT fine structure constant({$1 \over \alpha_G$}) 

\begin{align}
\Delta_{II}^U= \Delta\ln{M_U \over M_Z}& ={1\over 494}(5\lambda_Y+7\lambda_{2L}-
12\lambda_{3C} ) \nonumber\\
\Delta_{II}^{\eta}=\Delta\ln{M_{\eta}\over M_Z}& ={5\over 2223}(27\lambda_{2L}-
16\lambda_Y- 11 \lambda_{3C}) \nonumber \\
\Delta_{II}^{\alpha_G}= \Delta\left({1 \over \alpha_G}\right)&={1\over 17784\pi}(-783\lambda_{2L}+1205\lambda_Y+1060\lambda_{3C} )~. \label{eq:thform2}
\end{align}

One loop and two loop beta function coefficients for different 
ranges of mass scales are given in the Table.{\ref{tab:122p}} 

%==========================================================================     
\begin{table}[h!]
\caption{One loop and two loop beta function coefficients for RG evolution of gauge couplings}
\begin{center} 
\begin{tabular}{| p{2.5cm} | p{2.5cm} | p{4.1cm} |p{4cm}|p{4cm}|}
\hline   Mass scales($\mu$) &~~~~~~${\bf{a_{i}}}$ &~~~~~~~${\bf{b_{ij}}}$
\\ \hline 
 $M_Z < \mu < M_{\eta} $ & 
  \centering
 $\begin{pmatrix}                    
 \frac{41}{10} \\ \frac{-19}{6} \\ -7 \\
\end{pmatrix}$      
 &   
 $\begin{pmatrix} 
    \frac{199}{50} & \frac{27}{10} & \frac{44}{5} \\
    \frac{9}{10} & \frac{35}{6} &  12 \\
    \frac{11}{10} & \frac{9}{2} & -26 \\
\end{pmatrix}$  
 \\ \hline  
 $M_{\eta} < \mu < M_{U} $ & 
  \centering
 $\begin{pmatrix}                    
 \frac{45}{10} \\ {5 \over 6} \\ {-9 \over 2 }\\
 \end{pmatrix}$      
 &   
 $\begin{pmatrix} 
    \frac{43}{10} & \frac{63}{10} & \frac{84}{5} \\
    \frac{41}{10} & \frac{707}{6} & 172 \\
    \frac{53}{30} & \frac{129}{2} & \frac{53}{2} \\
   \end{pmatrix}$ 
\\ \hline       
\end{tabular} 
\end{center}
\label{tab:122p}
\end{table}

Using one loop beta function coefficients from Table.{\ref{tab:122p}} in  eq.(\ref{eq:guformnm}), we get respective solutions for mass scales

\ba 
M_U&=&10^{15.2835}\,{\rm GeV}~.\nonumber\\ 
M_{\eta}&=& 10^{10.73} {\rm GeV}, \nonumber\\
 \alpha_{G}^{-1}&=&38.397.
\label{eq:RGsolnm}
\ea

 We estimate  threshold corrections to different mass scales and GUT gauge coupling following procedures similar to model-I

\begin{align}
\Delta\ln{M_U \over M_Z}& =-0.020243\eta_{10}+0.044534\eta_{126}
 -0.0809717\eta_{210} \nonumber \\
\Delta\ln{M_{\eta} \over M_Z}& = -0.039136\eta_{10}-0.7139\eta_{126}
-0.323212\eta_{210} \nonumber \\
\Delta\left({1 \over \alpha_G}\right)& =0.0541256\eta_{10}+1.50961\eta_{126}+
1.17143\eta_{210}. \nonumber \\
\end{align}

Maximising the uncertainty in $M_U$ leads to  ~~~~~~~~
\ba
\Delta \ln \left({M_U \over M_Z}\right)&=&\pm 0.145749|\eta_{SH}| ,~\nonumber\\
\Delta \ln \left({M_{\eta}\over M_Z}\right)&=& \pm 0.3515517|\eta_{SH}| , ~~\nonumber\\
\Delta \left({1 \over \alpha_G}\right)&=&\pm 0.284|\eta_{SH}|,\label{eq:maxthsh} 
\ea 
where $\eta_{SH}=\ln ({M_{SH} \over M_U})$ and $M_{SH}/M_U= n(1/n)$ with
 plausible allowed values of real number $n=1-10$ .
 Neglecting threshold effects due to superheavy gauge
bosons, our estimation gives
for completely degenerate superheavy scalar masses of all SO(10)
representations
\ba
\Delta \ln \left({M_U \over M_Z}\right)&=&\pm 0.05668|\eta_{SH}|, \nonumber\\
\Delta \ln \left({M_{\eta}\over M_Z}\right)&=& \pm 1.07625|\eta_{SH}|
.\label{eq:nonmindegthsh}
\ea 
As in Model-I, the most dominant contribution to the GUT scale and  proton lifetime uncertainties is due to superheavy gauge boson masses
\be
{[\Delta \ln \left({M_U \over M_Z}\right)]}_V= -0.9352\eta_V, \label{eq:minthV}
\ee
where $\eta_V=\ln (M_V/M_U)$. Thus degenerate superheavy gauge boson
masses only few times lighter than $M_U$ can cause substantial enhancement
in proton lifetime prediction. An well known potential of SO(10) for fitting all charged fermion masses \cite{Berto-mass:2006,Joshipura:2011,Altarelli:2011} is beyond the scope of the present work.      

%%%%%%%%%%%%%%%%%%%%%%%%%%%%%%%%%%%%%%%%%%%%%%%%%%%%%%%%%%%%%

\section{ACKNOWLEDGMENT}
The authors thank  Rabindra N. Mohapatra for useful suggestions.
M. K. P. acknowledges financial support under the project
SB/S2/HEP-011/2013 from the Department of Science and Technology,
Government of India. He also thanks International Centre of
Theoretical Sciences, Bengaluru for invitation to the international conference
``Candles of Darkness 2017'' and discussions with Urjit Yajnik. 
For financial support from Siksha 'O' Anusandhan (SOA),
Deemed to be University, M. C. acknowledges  a 
Post-Doctoral fellowship and B.S. a Ph. D. research
fellowship. 
%\bibliography{mybibfile}
%\vspace {1cm} 

%-----------------------------------------------------
%-----------------------------------------------------

\end{document}